\documentclass[a4paper,fleqn,usenatbib]{mnras}

\usepackage{newtxtext,newtxmath}

\usepackage[T1]{fontenc}
\usepackage{ae,aecompl}
\usepackage{mathrsfs}

\usepackage{graphicx}	
\usepackage{amsmath}	
\usepackage[flushleft]{threeparttable}	
\UseRawInputEncoding
\usepackage{color}
\usepackage{todonotes}
\usepackage{mathtools, cuted}
\usepackage{stfloats}
\usepackage{acronym}
\usepackage{booktabs}
\usepackage{multirow}
\usepackage{float}
\usepackage{subfig}
\usepackage{soul}
\usepackage{makecell}
\usepackage{colortbl}
\acrodef{CMDs}{colour-magnitude diagrams}
\acrodef{CVs}{Cataclysmic variables}
\acrodef{FoV}{field-of-view}
\acrodef{IFU}{integral field unit}
\acrodef{SED}{spectral energy distribution}
\acrodef{NMAD}{Normalized Median Absolute Deviation}

\definecolor{lime}{HTML}{A6CE39}
\DeclareRobustCommand{\orcidicon}{
\begin{tikzpicture}
\draw[lime, fill=lime] (0,0)
circle [radius=0.16]
node[white] {{\fontfamily{qag}\selectfont \tiny ID}};
\draw[white, fill=white] (-0.0625,0.095)
circle [radius=0.007];
\end{tikzpicture}
\hspace{-2mm}
}

\foreach \x in {A, ..., Z}{\expandafter\xdef\csname orcid\x\endcsname{\noexpand\href{https://orcid.org/\csname orcidauthor\x\endcsname}
{\noexpand\orcidicon}}
}

\newcolumntype{P}[1]{>{\centering\arraybackslash}p{#1}}

\newlength\stextwidth

\definecolor{darkgreen}{rgb}{0.09, 0.45, 0.27}

\makeatletter 

\title[Discovering stars, quasars, and galaxies in S-PLUS]{On the discovery of stars, quasars, and galaxies in the Southern Hemisphere with S-PLUS DR2}
\author[L. Nakazono et al.]{L. Nakazono$^{1\orcidL{}}$ \thanks{Email: lilianne.nakazono@usp.br}, C. Mendes de Oliveira$^{1}$, N. S. T. Hirata$^{2}$, S. Jeram$^{3}$ \and C. Queiroz$^{4}$, Stephen S. Eikenberry$^{3}$, A. H. Gonzalez$^{3}$, R. Abramo$^{4}$, R. Overzier$^{5}$, \and M. Espadoto$^{2}$, A. Martinazzo$^{2}$,  L. Sampedro$^{1}$, F. R. Herpich$^{1\orcidF{}}$,  F. Almeida-Fernandes$^{1}$, \and A. Werle$^{1, 6}$, C. E. Barbosa$^{1}$, L. Sodr\'e Jr.$^{1}$, E. V. Lima$^{1}$,  M. L. Buzzo$^{1}$, A. Cortesi$^{7}$, \and K. Men\'endez-Delmestre$^{7}$,  S. Akras$^{8\orcidC{}}$,  Alvaro Alvarez-Candal$^{5,9,10}$,\and  A. R. Lopes$
^{5}$,  E. Telles$^5$,  W. Schoenell$^{11}$,  A. Kanaan$^{12}$, T. Ribeiro$^{13}$ 
\\
$^{1}$
Instituto de Astronomia, Geof\'isica e Ci\^encias Atmosf\'ericas da U. de S\~{a}o Paulo, Cidade Universit\'aria, 
05508-900, S\~{a}o Paulo, SP, Brazil \\
$^{2}$ 
Departamento de Ci\^{e}ncia da Computa\c{c}\~{a}o, Instituto de Matem\'{a}tica e Estat\'{i}stica da USP,
Cidade Universit\'{a}ria, 05508-090, S\~ao Paulo, SP, Brazil\\
$^{3}$ Department of Astronomy, University of Florida, 211 Bryant Space Center, Gainesville, FL 32611, USA\\
$^{4}$ Departamento de F\'isica Matem\'atica, Instituto de F\'isica, Universidade de S\~{a}o Paulo, SP, Rua do Mat\~{a}o 1371, S\~{a}o Paulo, Brazil\\
$^{5}$
Observat\'orio Nacional / MCTIC, Rua General Jos\'e Cristino 77, Rio de Janeiro, RJ, 20921-400, Brazil \\
The remaining institutions are at the end of the paper.\\
}

\date{Accepted 2021 June 18. Received 2021 June 16; in original form 2020 July 29}

\pubyear{2021}

\begin{document}
\label{firstpage}
\pagerange{\pageref{firstpage}--\pageref{lastpage}}
\maketitle

\begin{abstract}
This paper provides a catalogue of stars, quasars, and galaxies for the Southern Photometric Local
Universe Survey Data Release 2 (S-PLUS DR2) in the Stripe 82 region. We show that a 12-band filter system (5 Sloan-like and 7 narrow bands) allows better performance for object classification than the usual analysis based solely on broad bands (regardless of infrared information). Moreover, we show that our classification is robust against missing values. Using spectroscopically confirmed sources retrieved from the Sloan Digital Sky Survey DR16 and DR14Q, we train a random forest classifier with the 12 S-PLUS magnitudes + 4 morphological features. A second random forest classifier is trained with the addition of the W1 (3.4 $\mu$m) and W2 (4.6 $\mu$m) magnitudes from the Wide-field Infrared Survey Explorer (WISE). Forty-four percent of our catalogue have WISE counterparts and are provided with classification from both models. We achieve 95.76\% (52.47\%) of quasar purity, 95.88\% (92.24\%) of quasar completeness, 99.44\% (98.17\%) of star purity, 98.22\% (78.56\%) of star completeness, 98.04\% (81.39\%) of galaxy purity, and 98.8\% (85.37\%) of galaxy completeness for the first (second) classifier, for which the metrics were calculated on objects with (without) WISE counterpart. A total of 2 926 787 objects that are not in our spectroscopic sample were labelled, obtaining 335 956 quasars, 1 347 340 stars, and 1 243 391 galaxies. From those, 7.4\%, 76.0\%, and 58.4\% were classified with probabilities above 80\%. The catalogue with classification and probabilities for Stripe 82 S-PLUS DR2 is available for download.
\end{abstract}

\section{Introduction}
\label{sec:introduction}
The next generation of astronomical surveys will map large volumes of the sky, collecting data for millions, and even billions of objects. For instance, GAIA DR2 \citep{2018A&A...616A...1G} contains measurements of 1.7 billion sources; the Vera C. Rubin Observtory Legacy Survey of Space and Time (LSST; \citealt{Ivezi__2019}) aims to provide 32 trillion observations of 40 billions objects; the Javalambre-Physics of the Accelerated Universe Astrophysical Survey (J-PAS; \citealp{2014arXiv1403.5237B}, \citealt{miniJPAS}) will cover at least 8\,000 deg$^2$ with 54 overlapping narrow bands, 2 medium-bands extending to the UV and the near-infrared, and completemented by the $u$, $g$, $r$, $i$ Sloan Digital Sky Survey (SDSS) broad-band filters; the Javalambre Photometric Local Universe Survey (J-PLUS; \citealp{2019A&A...622A.176C}) will cover the same area of the Northern Sky as J-PAS with the Javalambre optical filter system consisting of 7 narrow bands and 5 SDSS-like bands. Regarding the Southern Sky, the Southern Photometric Local Universe Survey (S-PLUS; \citealp{2019arXiv190701567M}) will cover $\sim$9\,300 deg$^2$ using the same 12 optical bands as J-PLUS. With such amount of data, object classification is essential for all these surveys, requiring an efficient and automated methodology. 

Generally, galaxies and stars can be easily separated due to their different morphologies (e.g. \citealp{Pimbblet_2001}, \citealp{Moore_2006}), while quasars can be confused with stars at any magnitude range, {especially when quasar targets are selected in broad-band colour-colour diagrams (see e.g. Fig. 14 from \citealt{Richards2002})}. In the low signal-to-noise regime, however, galaxies and point-like sources can also be difficult to separate. The performance of the object classification can be improved when spectroscopy is available, since it brings more information than the imaging in just a few optical bands. However, given that complete spectroscopic surveys are more limited in area and magnitude than photometric surveys, many techniques have been developed to perform star/quasar separation with photometric data, such as:
colour-colour cuts (e.g. \citealp{Wu_2010}, \citealp{2012AJ....144...49W}, \citealp{2017ApJ...851...13S}, \citealp{Paris2018}); proper motion criteria (e.g. \citealp{Guo_2018}, \citealp{2018A&A...615L...8H}); Bayesian Statistics (e.g. \citealp{Kirkpatrick_2011}, \citealp{2015ApJ...811...95P}, \citealp{2017AJ....154..269Y}), and Machine Learning algorithms (e.g. \citealp{2012MNRAS.425.2599P},  \citealp{Carrasco_2015}, \citealp{2019MNRAS.485.4539J}). For star/galaxy separation, methods that rely on morphological features of the photometric images are widely used to distinguish the classes (e.g.  \citealp{costaduarte2019splus}, \citealp{L_pez_Sanjuan_2019} and \citealp{Baqui2021}, for S-PLUS, J-PLUS and miniJPAS, respectively). Few works have tackled both problems at once in a 3-class (star/quasar/galaxy) separation (e.g. \citealp{2019arXiv190910963C}, \citealp{Yang_2017}, \citealp{Kurcz_2016}, \citealp{Ball_2006}). \citet{Brescia_2015} have shown, using SDSS magnitudes, that a 3-class separation improved accuracy in comparison with a 2-step approach that consisted of a galaxy/star and a star/quasar classification.

\begin{figure}
\includegraphics[trim= 0 0 1.5cm 1.5cm, clip, width=\columnwidth]{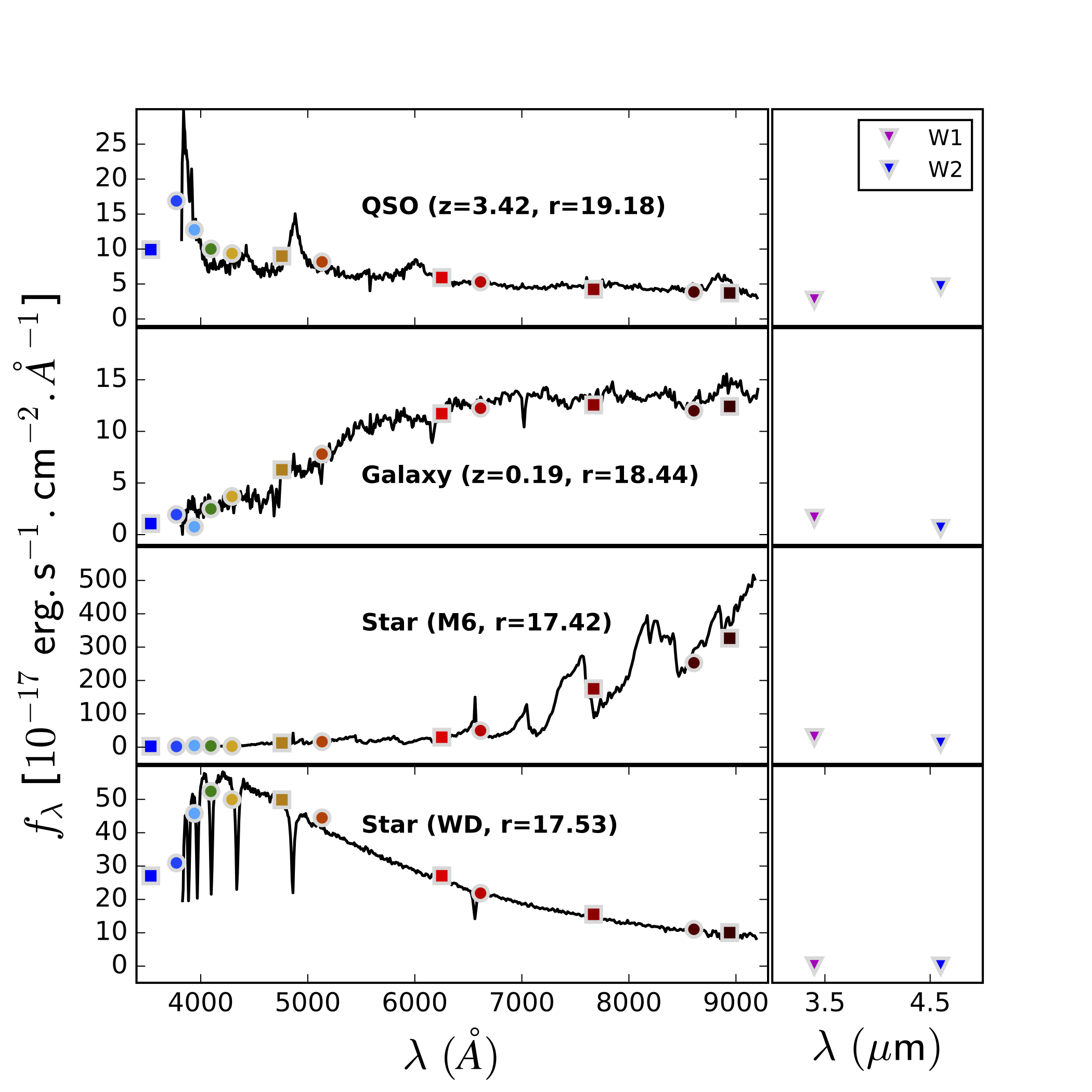}\\
\includegraphics[trim= 0 0 0 .5cm 0, clip, width=\columnwidth]{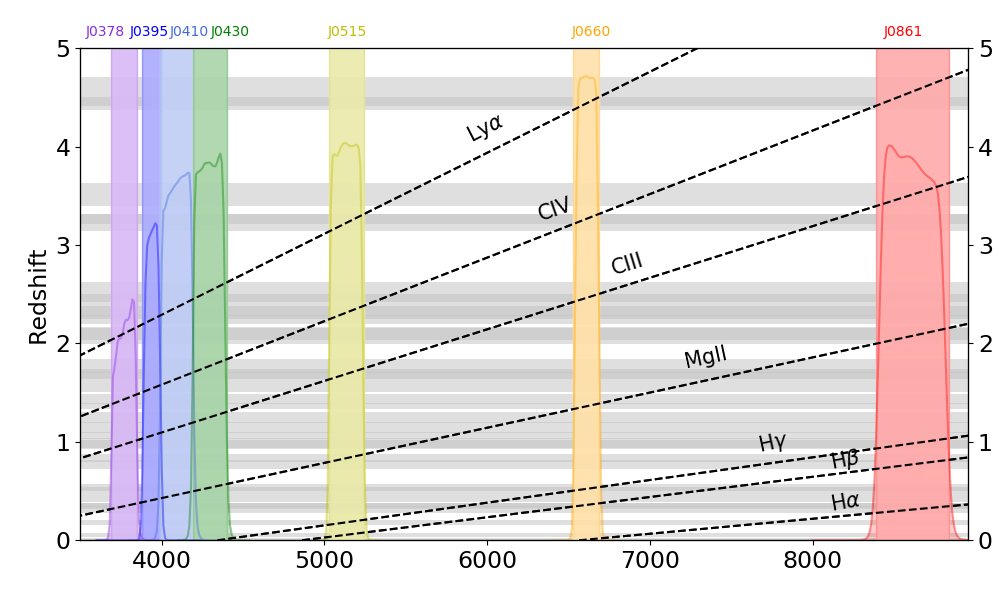}
  \caption{Top: Examples of spectra and their fluxes observed with the S-PLUS photometric system. The S-PLUS broad- and narrow-band filters are illustrated by coloured squares and circles, respectively. The triangles indicate the fluxes observed with WISE. From top to bottom: a quasar at redshift 3.42, a galaxy at redshift 0.19, an M6 star and a white dwarf. Bottom: Detection of the main emission lines of a quasar at different redshifts. Only the narrow-band filter transmission curves are shown in colours (with an arbitrary scale) and their names are indicated on the upper part of the plot. The grey shaded regions correspond to the expected redshifts (y-scale) of the quasars that we can detect with the S-PLUS narrow-band filters. In the regions with no horizontal grey bands, we do not expect to detect any quasar emission lines at the corresponding redshift. Light shades of gray indicate one detected line, whereas the strongest shades of grey indicate that at least two lines can be detected at the same redshift. For instance, the Ly$\alpha$ and CIV emission lines are mutually visible for quasars at $z = 3.2$, whereas the H$\gamma$ and the CIII emission lines are mutually visible for quasars at $z=1$.}
    \label{fig:filter_z}
\end{figure}

 Object classification in photometric surveys is much improved if there are images available in a number of different filters, covering a wide range of spectral energy distribution, resembling a low-resolution spectrum. In this sense, combining optical with infrared information $-$ usually from the Wide-field Infrared Survey Explorer (WISE; \citealp{2010AJ....140.1868W}) $-$  greatly improves object classification, especially for finding quasars (e.g. \citealp{Bovy_2012}, \citealp{2012AJ....144...49W}, \citealp{2017ApJ...851...13S}). This is due to the dust surrounding the nucleus of quasars that absorbs light and re-radiates in the near-infrared (NIR) and mid-infrared, together with stellar NIR emission of the host galaxies (see \citealp{Hernan} and references therein). Narrow-band surveys $-$ such as J-PAS, J-PLUS and S-PLUS $-$ can also play an important role on finding quasars, given that the broad-emission features of the quasars can be easily identified when they coincide with the narrow bands, as illustrated in Fig. \ref{fig:filter_z}. In the top panel, we show the spectra of a quasar, a galaxy, an M6 star, and a white dwarf with their respective observed fluxes in each of the 12 S-PLUS filters and WISE bands. In the bottom panel, we show the redshift ranges where the main quasar emission lines (Ly$\alpha$, CIV, CIII, MgII, H$\gamma$, H$\beta$, H$\alpha$) can be observed for each of the seven S-PLUS narrow-band filters.With the J-PAS unique set of 54 narrow-band filters, one will not only detect the quasar photospectrum but will immediately obtain a high-accuracy photometric redshift measurement.

In this era of Big Data, we face the need of migrating from more traditional techniques (e.g. colour-colour cuts), to computational, statistical and mathematical methods, which is corroborated by the growth of works in many astronomical fields with artificial intelligence and machine learning applications (see review from \citealt{2020WDMKD..10.1349F}). In this work we adopted a supervised learning approach, given that a sufficient number of spectroscopically confirmed stars, quasars and galaxies are available in the observed S-PLUS fields of Stripe 82 region (see Section \ref{sec:description}). We have chosen to test Random Forest and Support Vector Machines, as both are non-parametric models and, therefore, require fewer assumptions about the data (such as probability distribution). A good understanding of the S-PLUS data (as described in Section \ref{sec:visual}) is required to best construct our evaluation and model fitting strategy (see Section \ref{sec:analysis}). In this process, we also want to statistically: (1) assess the potential of the 7 narrow-band filters in comparison with analyses based only on the broad-band magnitudes, which is novelty and (2) reaffirm that the infrared bands and morphological parameters improve the classification performances.We discuss the comparison of different models through cross-validation in Section \ref{sec:comparison}, from which we select the best classifiers (one for objects with WISE counterpart, and another for objects without it). We perform a statistical hypothesis test (Wilcoxon signed-rank) in order to compare the performances with confidence level of 90\%, which are shown in Appendix \ref{sec:stat}. The expected performances of the chosen models are calculated on a separate test dataset and are shown in Section \ref{sec:desc_final}.
We finally use our validated random forest models to classify the $\sim$3 million photometric sources from the Stripe 82 region in S-PLUS DR2. We describe the details of this catalogue containing object classification and probabilities in Section \ref{sec:catalogue}. Unless specified otherwise, all magnitudes here are presented in the AB system. In the Data Availability section we give the links to download the datasets used in this work and the access to our ADQL queries. All appendices are only available as supplementary online material.

\section{Database}
\label{sec:description}

We use photometric catalogues from S-PLUS and WISE surveys, which are further introduced in \S\ref{ssec:info}. In order to perform a supervised-learning approach, we retrieved spectroscopically confirmed stars, quasars and galaxies from SDSS. We established a few selection criteria on these samples, as further described in \S\ref{ssec:sample}, aiming to avoid spurious sources for training and validating the algorithms. 
\subsection{Surveys}
\label{ssec:info}
\subsubsection{The Southern Photometric Local Universe Survey (S-PLUS)}
\label{ss:splus}
S-PLUS is a photometric survey performed by the T80-South robotic telescope, located at Cerro Tololo, Chile. By the end of 2023, it will have covered $\sim$9300 deg$^2$ of the Southern Sky in 12 optical filters.

S-PLUS uses the Javalambre 12-band system, which includes the narrow-band filters J0378, J0395, J0410, J0430, J0515, J0660 and J0861, centered on [OII], Ca H+K, H$\gamma$, G-band, Mgb triplet, H$\alpha$ and Ca triplet features, respectively. The broad-band filters are the SDSS-like $u$, $g$, $r$, $i$, $z$. These filters are similar to the SDSS filter system, except for the u band, which differs significantly (see \citealp{2019A&A...622A.176C}). { The information of their effective wavelength, width and photometric depth are available in Table \ref{tab:filters}. The photometric depth for narrow bands is typically 20.0 mag, except for H$\alpha$ that is 21.1 mag. For broad bands, the depth is typically 21 mag, except for $z$ band that is 20.1 mag.}

\begin{table}
\caption{Characteristics of the S-PLUS 12-bands filter system. The effective wavelength and width information were taken from \citealt{2019arXiv190701567M}. The depths were taken from  \citealt{2021arXiv210400020A}.}

\label{tab:filters}
\begin{threeparttable}
\begin{tabular}{ccccl}
\toprule
Filter & \begin{tabular}[c]{@{}l@{}}Effective \\ Wavelength\tnote{*}    (\AA{})\end{tabular} & \begin{tabular}[c]{@{}l@{}}Width\tnote{**}    \\ (\AA{})\end{tabular} & \begin{tabular}[c]{@{}l@{}}Depth for  \\ S/N > 3 (mag)\end{tabular} & Comments     \\

\midrule
u           & 3536                           & 352             & 21.0       & Javalambre u \\
J0378       & 3770                           & 151             & 20.4       & {[}OII{]}    \\
J0395       & 3940                           & 103             & 19.9       & Ca H+K       \\
J0410       & 4094                           & 201             & 20.0       & H$\gamma$    \\
J0430       & 4292                           & 201             & 20.0       & G-band       \\
g           & 4751                           & 1545            & 21.3       & SDSS-like g  \\
J0515       & 5133                           & 207             & 20.2       & Mgb Triplet  \\
r           & 6258                           & 1465            & 21.3       & SDSS-like r  \\
J0660       & 6614                           & 147             & 21.1       & H$\alpha$    \\
i           & 7690                           & 1506            & 20.9       & SDSS-like i  \\
J0861       & 8611                           & 408             & 19.9       & Ca Triplet   \\
z           & 8831                           & 1182            & 20.1       & SDSS-like z \\
\bottomrule
\end{tabular}
\begin{tablenotes}[para,flushleft]
\item[*] Wavelength corresponding to the middle point at half maximum.
\item[**] Width at half maximum.
\end{tablenotes}
\end{threeparttable}
\end{table}

 {For this work, we used data from S-PLUS DR2 in the Stripe 82 area} ($0^\text{o}<$ RA $ <60^\text{o}$; $300^\text{o}<$ RA $ <360^\text{o}$ and $-1.4^\text{o}<$ Dec $<+1.4^\text{o}$), covering $\sim$336 deg$^2$ close to the Celestial Equator.  {To construct our catalogue with class labels  and probabilities, we consider a subset of 3\,076\,191 sources from this area that are within $13<r\leq22$ and $\texttt{PhotoFlag\_r}==0$, which will be further discussed in Section \ref{sec:catalogue}. To build our training sample for our classifiers, we establish more specific criteria for each spectroscopic class (see \S\ref{ssec:sample}).}   {The photometry was obtained with SExtractor \citep{1996A&AS..117..393B} as described in section 4.5 in \citet{2019arXiv190701567M} using the following parameters: \texttt{detection\_threshold} of 1.1$\sigma$ and a minimum of 4 pixels for a object to be detected.}     {Photometry was done with SExtractor dual mode using a detection image built from the sum of $g$, $r$, $i$, $z$ bands.}  {For this work, we only considered the \texttt{ISO} magnitudes, in AB system. We use a subscript (e.g. $r_\text{SDSS}$) whenever we refer to the modelmag magnitudes from SDSS in order to distinguish from the S-PLUS broad-band filters.} 

\subsubsection{Wide-field Infrared Survey Explorer (WISE)}
\label{ssec:wise}
The WISE mapped the entire sky with a 40-cm telescope in orbit of the Earth. It has observed the sky in infrared using $W1$, $W2$, $W3$ and $W4$ bands centered at 3.4, 4.6, 12 and 22 $\mu$m, respectively. The Vega magnitudes can be converted to AB magnitudes using the following relation $m_{ab} = m_{vega} + \Delta{m}$, where $\Delta{m}$ = 2.699 and 3.339 for $W1$ and $W2$ bands, respectively.  {Using different magnitude systems do not interfere in our analyses. Thus, we use the Vega magnitudes to train our models}. For this work, we are only using $W1$ and $W2$ magnitudes, whose limiting Vega magnitudes are 16.5 and 15.5 respectively. We disregard $W3$ and $W4$, whose limiting Vega magnitudes are 11.2 and 7.9, respectively, due to their shallowness. For this work we are using the AllWISE catalog \citep{2013yCat.2328....0C}.  

\subsection{Spectroscopic sample selection} 
\label{ssec:sample}

\begin{table*}
\caption{Description of the spectroscopic samples from the cross-match between S-PLUS DR2, SDSS DR14Q (quasars), SDSS DR16 (stars and galaxies) and AllWISE catalogues.  Selection criteria are $\texttt{zWarning}=0$ for galaxies, and $r>13$ for stars. For all classes, the criteria are $\texttt{PhotoFlag}\_r=0$ and an upper limit of 22 magnitudes in $r$. After the selection criteria, we applied a 4-sigma clipping of the difference between $r$ magnitudes from SDSS and S-PLUS.}
\label{tab:number_sample}
\begin{tabular}{lcccccc}
\toprule
       & Total matches & \begin{tabular}[c]{@{}c@{}}Number of lost objects \\  from criteria\end{tabular} & \begin{tabular}[c]{@{}c@{}}Number of lost objects \\ from 4-sigma clipping\end{tabular} & \texttt{with\_WISE} sample & \texttt{no\_WISE} sample & Final sample \\ \midrule
QSO    & 28\,275         & 9\,217                                                                             & 69                                                                                     & 15\,538  &  3\,451  &  18\,989    \\
STAR   & 78\,217         & 22\,464                                                                            & 1\,702                                                                                  &  39\,595    &  14\,456 &  54\,051  \\
GALAXY & 154\,529         & 87\,66                                                                            & 951                                                                                    &  60\,059  & 5\,553 &   65\,612  \\ \bottomrule
\end{tabular}
\end{table*}

Among many wide-field spectroscopic surveys, the SDSS has provided the largest number of discovered quasars (QSOs). Recently, the quasar catalog from DR14 (DR14Q; \citealp{Paris2018}) was released with 526\,356 quasars over 9\,376 deg$^2$, including new discoveries from SDSS-IV extended Baryon Oscillation Spectroscopic Survey (eBOSS; \citealp{2016AJ....151...44D}) and identified quasars from SDSS-I/II (DR7Q; \citealp{2010AJ....139.2360S}) and SDSS-III programs (DR12Q; \citealp{2017A&A...597A..79P}). The DR14Q is provided with the cross-matched information from the AllWISE catalogue for sources within 2$\arcsec$, { with 2\% rate of false positives \citep{Paris2018}}. We cross-matched  {Stripe 82 S-PLUS DR2} with DR14Q with a matching radius of 1$\arcsec$ to retrieve the quasar sample. We selected all spectroscopically confirmed stars and galaxies in Stripe 82 from the SDSS DR16.
We cross-matched the catalogues from SDSS DR16 and Stripe 82 S-PLUS DR2 within 1$\arcsec$, that was also cross-matched with ALLWISE catalogue within 2$\arcsec$. 

 {Note that there is a difference in resolution between WISE and S-PLUS, which in principle could represent a source of duplicate WISE assignments (one S-PLUS object could be assigned to more than one WISE source). We try to minimize this problem by not allowing the same WISE source to be cross-matched more than once. Here we are assuming that most of the emission from the WISE match is coming from the closest counterpart, in angular separation, to the S-PLUS source. If we were to allow multiple matches, 47\,448 WISE sources would be attributed multiple times within a total of 1\,459\,524  S-PLUS DR2 Stripe 82 objects with WISE counterpart (47.44\% of the total photometric sample). Those would be mostly cross-matched with two different close sources of S-PLUS, had we allowed this to happen. This exercise was done in order to have an idea of the magnitude of the problem, although it is not solved here (the only solution would be to have higher resolution images in the infrared).}

Table \ref{tab:number_sample} shows the total number of quasars, stars and galaxies retrieved from SDSS. There are 62 S-PLUS observations that were cross-matched twice as both quasar and galaxy. We decided to keep these duplicates as this number will not affect our models. The following selection criteria were applied in order to avoid any contamination from spurious sources in our samples:

\begin{itemize}
    \item $\texttt{PhotoFlag\_r} = 0$  {and $r \leq 22$} for all classes: this output from \texttt{SExtractor} allows us to select only the objects with reliable photometry from the S-PLUS images;
    \item $\text{zWarning} = 0$ for galaxies: this flag is retrieved from SDSS catalogue. Any non-zero values indicate problems with the pipeline fit to the SDSS spectrum;
    \item $r>13$ for stars: this condition was set from a visual inspection on the S-PLUS images, to avoid saturated images.
\end{itemize}

In the S-PLUS catalog, non-detected bands have magnitudes set to 99. These missing values were not excluded from our analysis or treated in any special way, as one of our goals is to build a model that is robust against missing data.
Some sources have a discrepant magnitude in S-PLUS when compared to SDSS, even considering the filtering conditions listed above. This can be explained by either intrinsic variable sources or problems in the image reduction in either or both surveys. To be conservative and select the purest samples possible, an additional 4-sigma clipping was applied on the difference between SDSS and S-PLUS magnitudes ($r_\text{SDSS} - r$) per $r$ bin of size 0.5. The number of lost objects in these processes and the size of the final samples are described in Table \ref{tab:number_sample}.

\section{Description of the spectroscopic samples}
\label{sec:visual}

\begin{figure}
  \includegraphics[trim=0.5cm 0.5cm 1.5cm 1.5cm, clip, width=\columnwidth]{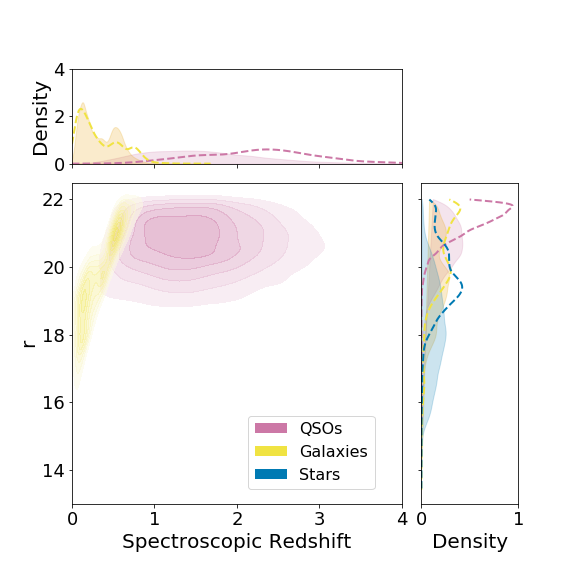}
  \caption{Distribution of the spectroscopic redshift (z) and r magnitude of the 12\,543 known quasars and 60\,234 known galaxies that were observed with S-PLUS and WISE (from \texttt{with\_WISE} sample). The stars, which lie at $z=0$, are not plotted. The spectroscopic redshift of all quasars and galaxies were retrieved from SDSS. We show the marginal distributions for the \texttt{no\_WISE} sample in dashed curves, and for the \texttt{with\_WISE} sample in solid curves.}
    \label{fig:dist_r_z}
\end{figure}

\begin{figure}
  \includegraphics[width=\columnwidth]{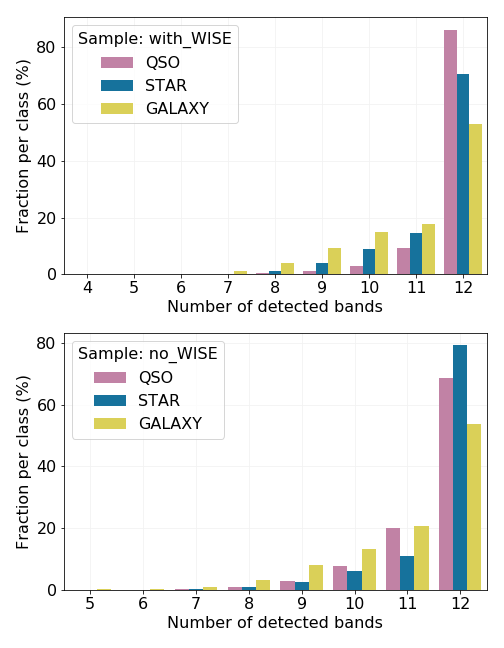}
  \caption{Fraction of galaxies (yellow), quasars (pink) and stars (blue) from \texttt{with\_WISE} (top panel) and \texttt{no\_WISE} (bottom panel) samples that were detected in a given number of bands out of the 12 S-PLUS bands.}
    \label{fig:nDet_auto}
\end{figure}

 {In order to evaluate the inclusion of WISE data for classifying objects, it is important to distinguish between two complementary subsets of our spectroscopic sample, which will be called as \texttt{with\_WISE} and \texttt{no\_WISE}. These two subsets contain objects with and without WISE magnitudes (W1 and W2), respectively.}
The \texttt{no\_WISE} sample includes sources with null information on the error or signal-to-noise for W1 or W2, containing a total of 3\,451 QSOs (18.17\% out of 18\,989), 14\,456 stars (26.74\% out of 54\,051) and 5\,553 galaxies (8.46\% out of 65\,612). The total number of sources in each of these samples are shown in Table \ref{tab:number_sample}. Figure \ref{fig:dist_r_z} shows the relation between redshift and $r$ magnitude of the 15\,538 known quasars and 60\,059 known galaxies in the \texttt{with\_WISE} sample. For the 39\,595 stars in the \texttt{with\_WISE} sample, we only show the magnitude distribution. In this figure, we also show the marginal distributions for both \texttt{with\_WISE} and \texttt{no\_WISE} samples, shown in filled and dashed curves, respectively. We note that the \texttt{with\_WISE} sample is distributed in brighter magnitudes compared to \texttt{no\_WISE}.

 {For a classifier that does not include the WISE magnitudes as features, this model would be better generalized if trained with more data (with\_WISE $+$ no\_WISE). However, the comparison of performances between classifiers with and without WISE magnitudes should be done on the same dataset (in this case, with\_WISE) to avoid any bias due to different distributions as seen in Fig. \ref{fig:dist_r_z}.} 

The majority of the sources in our samples were detected in all twelve bands of S-PLUS. All sources were detected in at least 3 filters. However, we note that there are some missing bands and these are mostly in the bluer region of the spectra, where the S/N is typically lower. In Fig. \ref{fig:nDet_auto}, we show the fraction of QSOs, galaxies and stars detected in a certain number of the S-PLUS filters. From the total number of sources in each class in the \texttt{with\_WISE} (\texttt{no\_WISE}) sample, 86.2\% (68.6\%)  QSOs, 52.9\% (53.6\%) galaxies and 70.7\% (79.4\%)  stars were detected in all 12 bands. When we consider objects that have been detected in at least 9 bands, the fractions are: 99.5\% (99.1\%), 94.9\% (95.4\%) and 98.4\% (98.6\%), respectively.
\begin{figure}

  \includegraphics[trim=0.3cm 0.5cm 0.5cm 0, clip, width=\columnwidth]{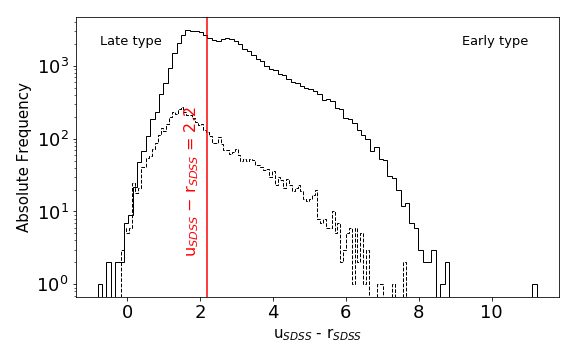}
  \caption{Distribution of u$_\text{SDSS} -$ r$_\text{SDSS}$ of the spectroscopically confirmed galaxies with or without WISE counterparts. The red vertical line shows the empirical relation u$_\text{SDSS} -$ r$_\text{SDSS} = 2.2$ that separates early- and late-type galaxies \citep{Strateva_2001}.}
    \label{fig:early_late}
\end{figure}

The target selection of SDSS quasar surveys has relatively low completeness at $z =2.2-3.5$, due to overlapping of quasars and the stellar locus in the five-band optical colour space (\citealp{2006AJ....131.2766R}; \citealp{2013ApJ...773...14R}). Moreover, \cite{2017ApJ...851...13S} showed that the survey has systematically missed many bright ($m_i$ < 18.5) quasars at ${z} > 3$.  {We also have shown in Fig. \ref{fig:filter_z} that some quasars at specific redshifts will not have any emission lines detected in a narrow band filter (we further discuss this in \S\ref{ssec:bias}).} Therefore, our sample is not representative of all quasars.

We show the distribution of early- and late-type galaxies in Fig. \ref{fig:early_late}, using the empirical relation from \cite{Strateva_2001}. The \texttt{with\_WISE} (\texttt{no\_WISE}) sample is composed of 60.7\% (33.7\%) early-type and 39.3\% (66.3\%) late-type galaxies.
Both \texttt{with\_WISE} and \texttt{no\_WISE} samples have a majority of A, F, G, K, M stars, according to SDSS classification. The \texttt{with\_WISE} sample has: 37.6\% M-type, 27.6\% K-type, 24.7\% F-type, 7.3\% G-type and 2.2\% A-type stars. The remaining 0.6\% correspond to cataclysmic variables, sub-dwarfs, white-dwarfs, carbon-, O-, B-, L-, and T-type stars. The \texttt{no\_WISE} sample has: 39.4\% F-type, 17.2\% K-type, 16.5\% A-type, 8.5\% M-type, 7.6\% G-type, 4.8\% white-dwarfs, 1.6\% B-type stars. Cataclysmic variables, sub-dwarfs, carbon-, L-, T-, and O-type stars represent the remaining 4.4\% out of the \texttt{no\_WISE} sample.

\begin{figure*}
\centering
\subfloat[12 S-PLUS bands]{\includegraphics[trim=0.1cm 0 0.1cm 0.1cm, clip, width=0.5\textwidth]{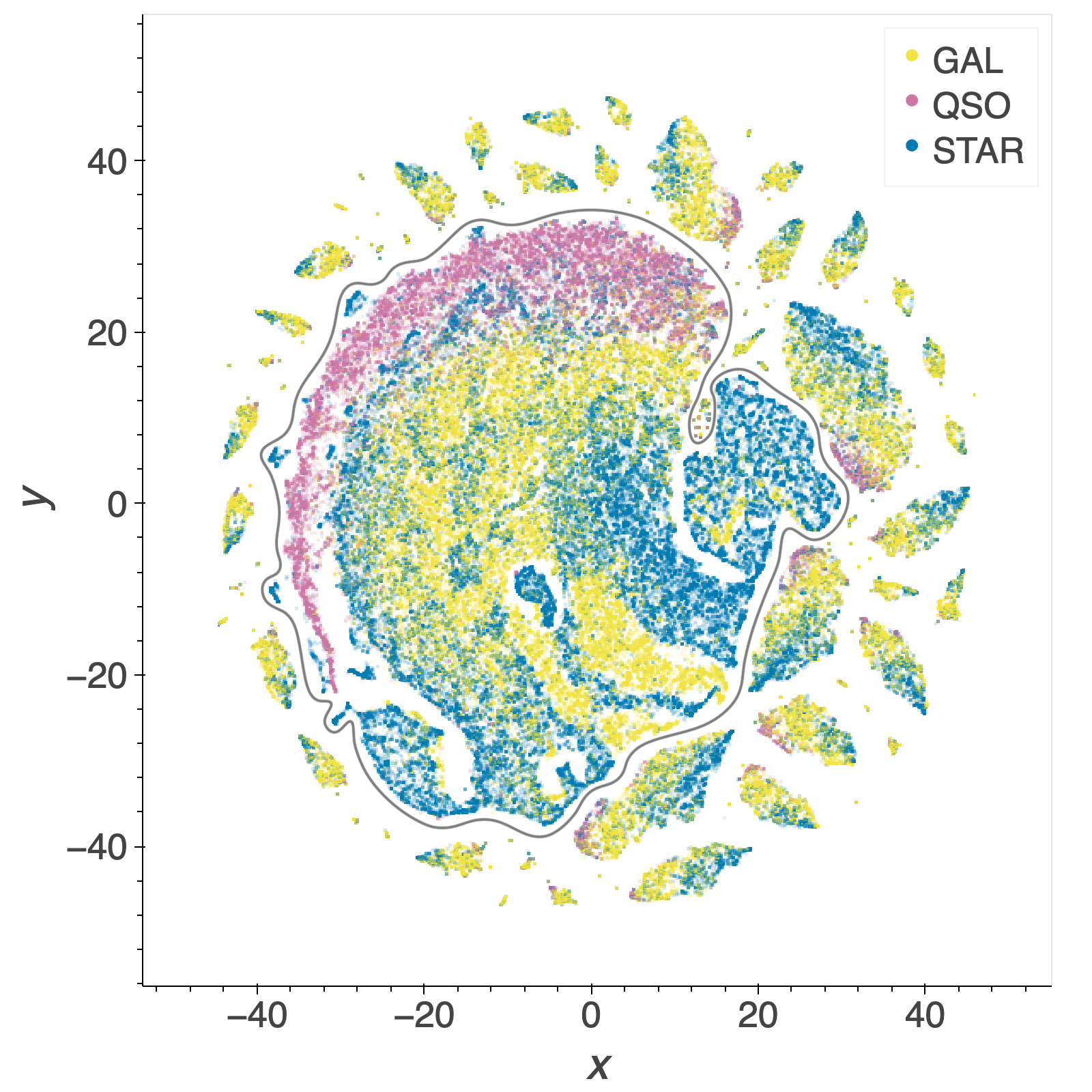}}
\subfloat[12 S-PLUS bands + 4 morphological features]{\includegraphics[trim=0.1cm 0 0.1cm 0.1cm, clip, width=0.5\textwidth]{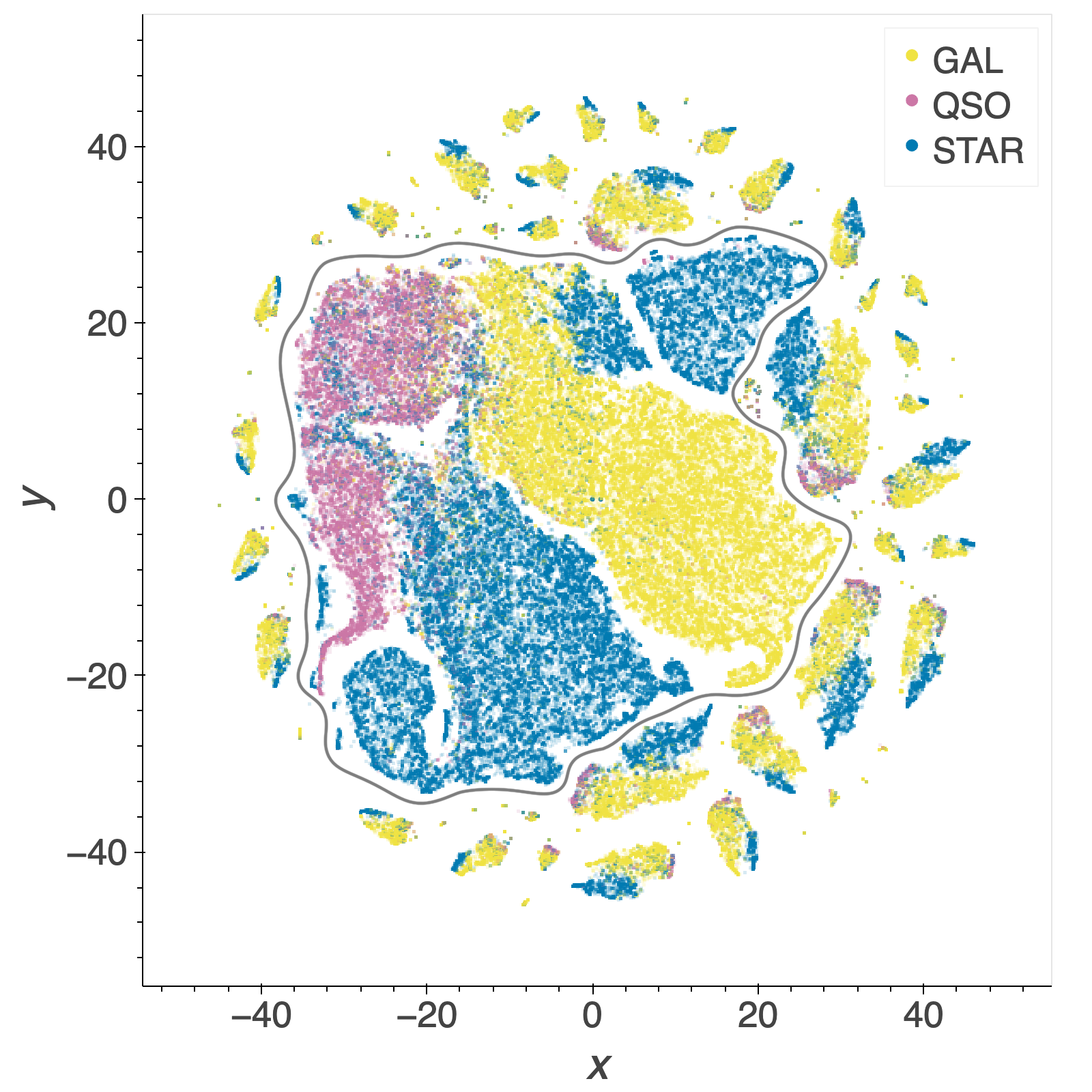}} \\
\subfloat[12 S-PLUS bands + 4 morphological features + 2 WISE bands]{\includegraphics[trim=0.1cm 0 0.1cm 0.1cm, clip,width=0.5\textwidth]{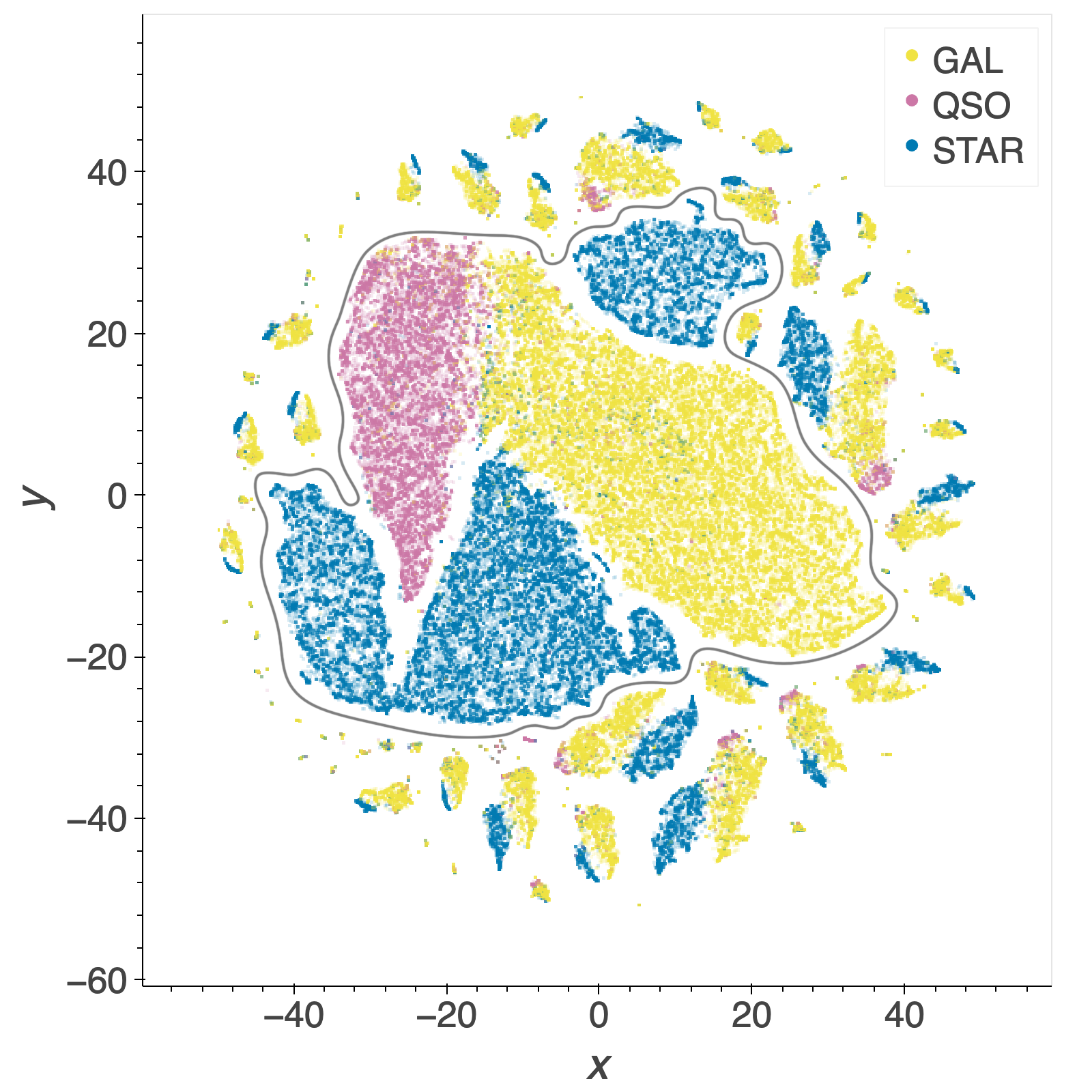}}
    \caption{Two-dimensional projection of the (a) 12 S-PLUS bands; (b) 12 S-PLUS bands + 4 morphological features; and (c) 12 S-PLUS bands + 4 morphological features + 2 WISE bands, where these four morphological features are: \texttt{FWHM}, \texttt{A}, \texttt{B} and \texttt{KRON\_RADIUS}. The projection of the features was done using t-SNE with \texttt{perplexity} $= 30$. A total of 18\,989 quasars (pink), 54\,051 stars (blue), and 65\,612 galaxies (yellow) are shown in (a) and (b). In (c), 15\,538 quasars, 39\,595 stars and 60\,059 galaxies are shown. The contour in the three plots was hand-drawn to roughly separate the cases with and without missing bands. The small islands outside the contours are related to objects with at least one missing value in the S-PLUS bands. These figures show that, despite the missing-band values, the three classes are well clustered in their corresponding loci. Panel (b) clearly shows less confusion between stars and galaxies due to the inclusion of the four morphological features. The confusion between stars and quasars is diminished when the WISE features are included, as shown in panel (c).} 
    \label{fig:tsne}
\end{figure*}

In Fig. \ref{fig:tsne} we show a 2-dimensional projection of the  {(a) 12 S-PLUS bands; (b) 12 S-PLUS bands + 4 morphological features; and (c) 12 S-PLUS bands + 4 morphological features + 2 WISE bands} from our  {spectroscopic} sample of quasars, stars and galaxies.  {These morphological features are: full width at half maximum, major semi-axis, minor semi-axis and Kron radius.} The projection was performed by the t-Distributed Stochastic Neighbour Embedding (t-SNE) algorithm \citep{vanDerMaaten2008}, which is a nonlinear dimensionality reduction technique. The chosen hyper-parameters for the t-SNE run are shown in Appendix \ref{sec:hyper}. The stars and galaxies seem to be better discriminated when we include the morphological parameters as seen in Fig. \ref{fig:tsne}(b), compared to (a). When including WISE features, there is less confusion between stars and quasars, as seen in Fig. \ref{fig:tsne}(c). For this last plot, only objects with WISE counterparts are being shown. The objects detected in all 12 bands are concentrated in the centre of all figures, whereas the objects with at least one missing band are in the small  {islands} surrounding the centre, outside the contour.
From the projection, we can see that the observed stars, quasars and galaxies are generally concentrated in their own loci, even for the small islands outside of the contours.  This indicates that, in spite of missing data, objects of the same class
tend to cluster, thus suggesting that classification of those objects is possible. We therefore decided to keep sources with missing-band values  {(mag = 99)} in our classification strategy based on machine learning.  {In Section \ref{ssec:bias} we discuss in more detail if the missing band-values are affecting our performances. }

\section{Methodology based on supervised-learning algorithms}
\label{sec:analysis}
\begin{figure*}
\centering
  \includegraphics[trim=1cm 0cm 0.5cm 0cm, clip, width=\textwidth]{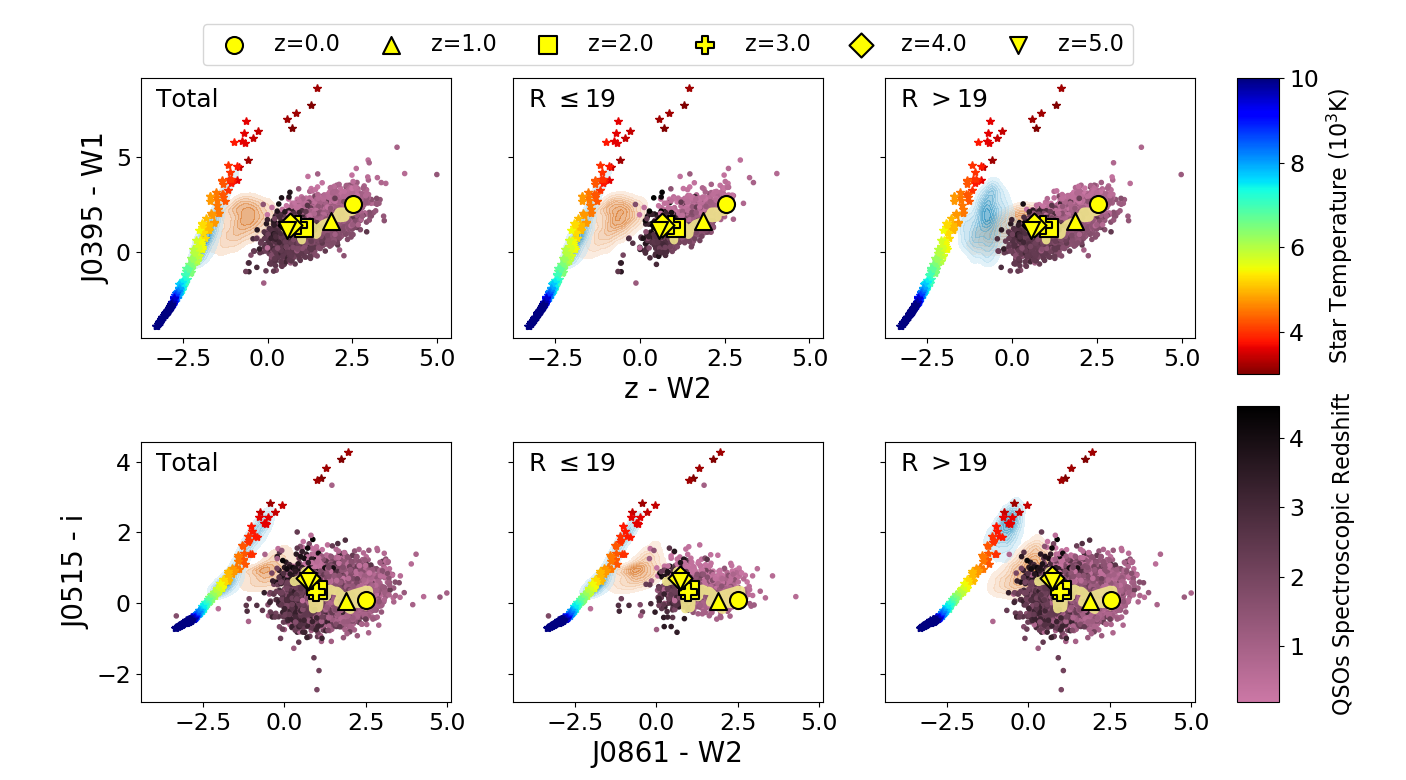}
  \caption{Colour-colour diagrams containing observed stars (blue contours), galaxies (orange contours) and quasars (points coloured by their spectroscopic redshift, according to the bottom right colour bar) with good photometric flag (\texttt{PhotoFlag\_R} $=$ 0) and signal-to-noise greater than 3 for each considered filter. These diagrams include simulated colours for a stellar template library \citep{PaulaCoelho}, indicated by star-shaped points coloured by their temperature, according to the upper right colour bar. Colour simulations of a quasar template \citep{Hernan} at different redshifts are shown as a yellow track. We also indicate with yellow symbols the colours at specific redshifts $z =$ \{0,1,2,3,4,5\}. Upper panels: $J0395 - W1$ versus $z - W2$. Bottom panels: $J0515 - i$ versus $J0861 - W2$.  From left to right: without limiting magnitude; including only $r \leq 19$ observed objects; including only $r > 19$ observed objects.
  }
    \label{fig:data_color}
\end{figure*}

 {Initially, in order to facilitate the data visualization, we created colour-colour diagrams following the star/quasar separation methodology from Jeram et al. (in preparation)}. We use simulated AB magnitudes of a theoretical stellar library which covers 3\,000$\,K \leqslant T_{\text{eff}} \leqslant 25\,000$\,K from 130\,nm to 100\,$\mu$m  \citep{PaulaCoelho} and a single QSO template in the range 0.1 - 11\,$\mu$m \citep{Hernan}. The quasar template was, then, shifted from $z =0 - 5$ in steps of 0.25. From the simulated data, we were able to select the combination of colours that provide the best separation of stars and quasars. We then included in the plot the known stars, galaxies and QSOs from \texttt{the with\_WISE sample}. These are shown in Fig. \ref{fig:data_color}, where we selected the following best colour combinations: $\text{J0395} - W1$ versus $z - W2$ and $\text{J0515} - i$ versus $\text{J0861} - W2$. These diagrams show a good separation between stars and QSOs, for both observed and simulated data. In both colour spaces, the observed galaxies lie between stars and QSOs. We can see from the panels for $r > 19$ that the overlap of the three regions is greater at fainter magnitudes. For $r < 19$, however, the overlap is minimum. The figure also displays a smooth gradient for the spectroscopic redshift (retrieved from SDSS) of the known QSOs, following the trend from low to high redshift, from right to left, along the tracks. In section 5.4 of \cite{2019arXiv190701567M}, preliminary results using the methodology above are given.  {While this method works well, we have the caveat of classifying only objects with signal-to-noise higher than 3 in each of the mentioned filters. Moreover, we have seen that our selected colour-colour diagram underperforms at $r$ magnitudes higher than 19. Therefore, we decided for a machine learning approach using more from the available information (12-band magnitudes and morphological features).  With such approach, patterns in our dataset can be detected even in a high dimensional space, allowing better performances. However, these patterns cannot be easily described or interpreted. As it will be shown in Section \ref{sec:comparison} and in following sections, methods based on machine learning turn out to yield robust results.}

In this work, two different supervised learning algorithms for classification are tested: Support Vector Machine (SVM) and Random Forest (RF). We implemented both algorithms from the \texttt{scikit-learn} (\texttt{sklearn}) library \citep{scikit-learn} in \textsc{Python}. In \S \ref{ssec:svm} and \S \ref{ssec:rf}, we briefly explain SVM and RF, respectively, and the choice of the hyper-parameters. To compare the performance of both methods, we used the cross-validation strategy described in \S \ref{ssec:crossval}, considering different feature spaces described in \S \ref{ssec:feat} and evaluated each experiment with the metrics described in \S \ref{ssec:metric}.  {We arbitrarily fixed the seed for the random number generator as \texttt{random\_state} $= 2$, in order to provide reproducible results.}

\subsection{Introduction to Support Vector Machine (SVM)} 
\label{ssec:svm}

SVM is a supervised machine learning algorithm, first introduced for binary classification (for more details, see \citealp{Cortes1995}). For linearly separable cases, it consists in finding a hyper-plane with maximum margin that separates the two classes of objects. Positive objects should stay on one side and negative ones on the opposite side with respect to the hyper-plane and no object should lie within the margin. The objects on the margin are the support vectors. As the two classes may not be linearly separable, the soft-margin formulation allows violation of the margin, and the optimal solution is one that maximizes margin at the same time it minimizes classification error. There is a parameter C that controls the trade-off between these two factors. Small C corresponds to larger margins, but possibly with more classification errors while large C tends to generate smaller margins and fewer classification errors (which might lead to over-fitting, however). To account for non-linearity, the kernel trick is used to fit the hyper-plane in a higher dimensional space using non-linear transformations \citep{10.5555/559923}. 


The SVM can be extended to a multi-class classification, decomposing it into binary sub-classifications. We adopted the one-against-one strategy  {(\texttt{decision\_function\_shape} = "ovo")}, that consists in building one SVM for each combination of pairs. 
For this work we chose the radial basis function kernel  {(\texttt{kernel} = "rbf")} and other parameters that we did not mention were set using default values  {(see Appendix \ref{sec:hyper})}.

\subsection{Introduction to Random Forest (RF)} 
\label{ssec:rf}

Random Forest is a method based on multiple decision trees and commonly used for classification and regression problems (for more details, see \citealp{Breiman2001}). In a decision tree, the feature space is split successively into smaller sub-spaces. At each splitting step a node is added to the tree, and branches for the corresponding sub-spaces are created. This splitting process is successively applied on each branch subspace until the training objects contained in the resulting subspace is "pure" according to some purity criteria (e.g., contains only or predominantly objects of same class or value). Nodes with no splitting are the leaf nodes. The final prediction is done by traversing the tree from the root to a leaf node, following thus a single set of decisions.
In a random forest, multiple trees are constructed. Each tree is grown independently of one another, from a sample with replacement of the training set. This procedure is known as bootstrap aggregating (or bagging). Each node is determined from a randomly chosen subset of the input features, instead of all of them.  The class prediction is then given by the majority vote from all trees, resulting in predictions with reduced variance. Such strategy is robust against over-fitting, due to the Strong Law of Large Numbers. Probabilities for each class are estimated as the number of votes per total trees. In order to calculate the features importances, we use the implementation of \texttt{scikit-learn} that takes the Gini Importance approach \citep{reason:BreFriOlsSto84a}, also known as Mean Decrease in Impurity. It is defined as the total decrease in node impurity weighted by the fraction of the sample that reaches that node and then averaged over all trees of the ensemble. \cite{oshiro_2012} suggest using a range between 64 and 128 trees in a forest, as larger numbers of trees increase the computational cost and do not bring significant performance gain. 
 {In a first attempt}, we fixed a total of 100 trees  {(\texttt{n\_estimators} = 100)}, with the square root of the total number of features being sampled per node (\texttt{max\_features} = "auto"). Parameters not described here were set using default values (see Appendix \ref{sec:hyper}). 

\subsection{Cross-validation strategy}
\label{ssec:crossval}

 The goal of doing cross-validation \citep{10.2307/2336116} is to find the best configuration for classifying objects in terms of features, algorithms and their hyper-parameters. For that, we only use the \texttt{with\_WISE} dataset, allowing us to make direct comparisons between different feature space configurations. The \texttt{no\_WISE} dataset will be considered in further analyses.
 
We sampled 75\% out of  {\texttt{with\_WISE}}  {(\texttt{no\_WISE})} as our training set and 25\% as our testing set in a stratified way, i.e. maintaining the same initial proportion  {1:2.8:3.4 (1:4.2:1.6) of quasars, stars, and galaxies, respectively}. We sampled the training and testing sets  using \texttt{sklearn.train\_test\_split} with a fixed $\texttt{random\_state} = 2$.

\begin{figure*}
\includegraphics[width=1\textwidth]{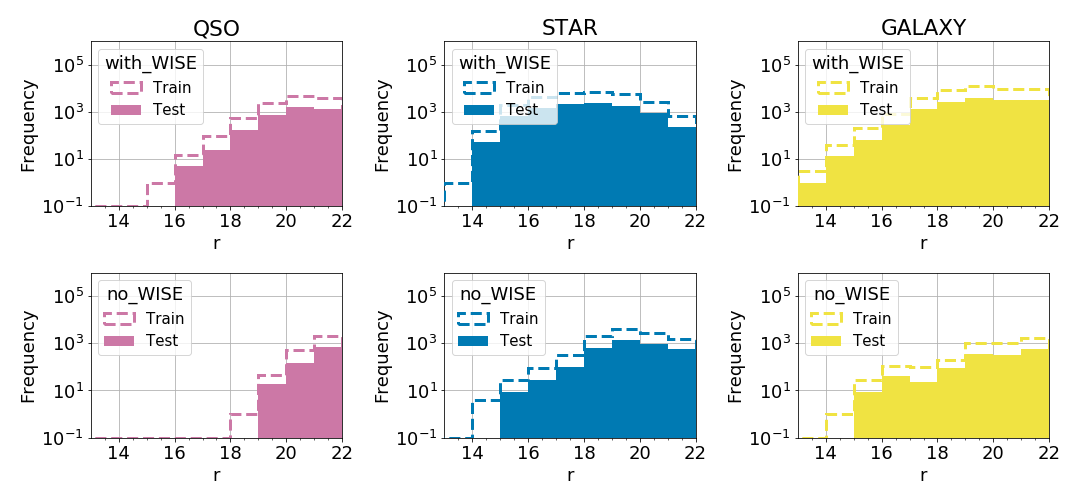}
  \caption{Distribution of r magnitudes for 15\,538 quasars (top left), 39\,595 stars (top middle), and 60\,059 galaxies (top right) that have WISE counterpart. In the bottom panels we show the distribution for objects without WISE counterpart (3\,451 quasars, 14\,456 stars, and 5\,553 galaxies). All panels shows the distributions for the train and test sets that correspond to 75\% and 25\% of the data, respectively.}
    \label{fig:test_dist}
\end{figure*}

 In Fig. \ref{fig:test_dist} we show the distribution of $r$ magnitude for each class in the \texttt{with\_WISE} testing sample.  
 {Our strategy consisted in two steps: model selection by cross-validation with the training set, and evaluation of the selected model with the testing set}. We trained different combinations of model settings (e.g. algorithm and feature space). We will further refer to each unique combination of settings as an "experiment". These experiments were evaluated in the validation step through k-fold cross-validation, using \texttt{sklearn.model\_selection.StratifiedKFold}, as follows:

\begin{enumerate}
    \item Divide the training set into $k$ folds of equal size in a stratified way
    \item Train the model with $k-$1 out of the $k$ folds, and validate the experiment with the remaining fold
    \item Repeat until each fold is used for validation exactly once.
\end{enumerate}
 
We fixed $k = 5$, which resulted in five different model fittings for each experiment. In order to evaluate them, we calculated an average of the metrics (further described in \S\ref{ssec:metric}), thus reducing variance.

From the k-fold cross-validation we can select the best settings. Afterwards, the final step consisted in training the chosen experiments using all folds and using the test set to assess the performance of the trained model. This final step yields more realistic metrics, as the testing set has not been used to train the algorithm at any point. 

\subsection{Feature spaces} 
\label{ssec:feat}

Several works have used the five SDSS broad-bands for either star/quasar or star/quasar/galaxy separation (e.g. \citealp{Brescia_2015}, \citealp{2015ApJ...811...95P}). We want to evaluate the advantage of the 7 S-PLUS narrow-bands against using only the 5 broad-bands. Thus, we tested the following feature spaces: 

\begin{enumerate}
    \item 12 S-PLUS bands 
    \item 12 S-PLUS bands + 2 WISE bands
    \item 5 S-PLUS broad-bands
    \item 5 S-PLUS broad-bands + 2 WISE bands
\end{enumerate}

As fixed hyper-parameters are considered for SVM and RF  {during the feature and algorithm evaluation}, we ended up with 8 unique experiments (4 feature spaces $\times$ 2 algorithms).
After selecting the best feature spaces among the ones listed above  {and the best algorithm}, we also want to evaluate the improvements in our classification when adding morphological features, such as Full Width at Half Maximum (\texttt{FWHM}), major semi-axis (\texttt{A}), minor semi-axis (\texttt{B}) and Kron Radius (\texttt{KRON\_RADIUS} in DR2 catalogue; \citealt{1980ApJS...43..305K}). As the FWHM is variable field-to-field, we use the value normalized to the detection image seeing (\texttt{FWHM\_n} in DR2 catalogue, but hereafter we will mention it simply as \texttt{FWHM}). 

For further references, we define the following naming system to refer to each experiment:

\begin{equation*}
\text{Algorithm} \_ \alpha S + 2W + 4M,
\end{equation*}

where "Algorithm" are RF or SVM, "$\alpha$S" refers to the number of features from S-PLUS bands (i.e. 12 or 5), "2W" refers to the inclusion of $W1$ and $W2$ magnitudes from WISE, and "4M" refers to the four morphological features mentioned above. Any term is omitted if the corresponding features were not included in the experiment. For instance, RF\_5S+2W is the model trained with Random Forest, 5 S-PLUS broad bands and 2 WISE bands, without including the morphological features.

\subsection{Performance Metrics} 
\label{ssec:metric}
\begin{figure}
\includegraphics[width=0.45\textwidth]{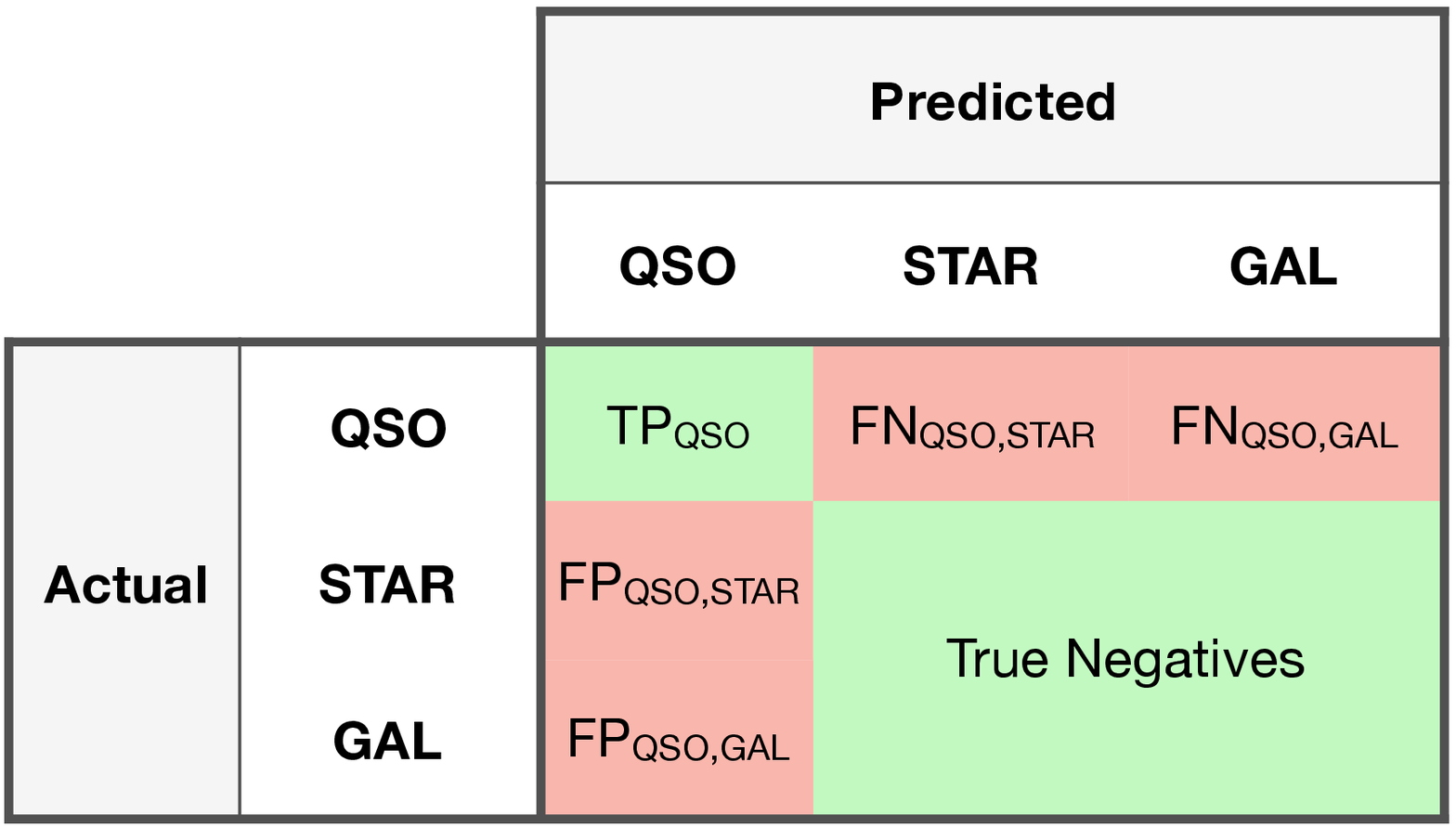}
  \caption{Example of a 3-classes confusion matrix. ${TP}_i$, or True Positives, is the amount of objects of class $i$ that were classified correctly, while ${FP}_{i,j}$, or False Positives, is the number of objects of class $j$ that were misclassified as class $i$. In this example we show the True Positives, True Negatives and False Positives in respect to QSOs.}
    \label{tab:conf}
\end{figure}

Our dataset is imbalanced, as the number of stars and galaxies in our spectroscopic sample are much larger than the number of quasars. In that case, accuracy is not a good performance measurement, due to bias towards the classes with higher frequencies. We will consider other metrics based on the confusion matrix (Fig. \ref{tab:conf}) as further described.

Considering $i,j \in \mathcal{Y}$, we can define precision ($\mathnormal{P}$) and recall ($\mathnormal{R}$) as follows:

\begin{equation}
\label{eq:precision}
P_i = \frac{(TP)_i}{ (TP)_{i} + \sum\limits_{j \ne i} (FP)_{i,j}},
\end{equation}

\begin{equation}
\label{eq:recall}
R_i = \frac{(TP)_i}{ (TP)_{i} + \sum\limits_{j \ne i} (FN)_{i,j}},
\end{equation}

\noindent
 {where $\mathnormal{TP}$, $\mathnormal{FP}$, and $\mathnormal{FN}$ denotes the true positives, false positives and false negatives, respectively.} In other words, $\mathnormal{P}_i$ is the percentage of objects assigned as class $i$ that are indeed class $i$, while $\mathnormal{R}_i$ is the percentage of class $i$ objects that were classified as class $i$. In astronomy, precision is also known as purity, and recall is also known as completeness or efficiency.
We can also define another metric that evaluates $\mathnormal{R}$ and $\mathnormal{P}$ in a single value, which is the harmonic mean of $\mathnormal{P}_i$ and $\mathnormal{R}_i$ called F-measure ($\mathnormal{F}_i$):

\begin{equation}
F_i = 2\Big(P_i^{-1}+R_i^{-1}\Big)^{-1}  = 2 \times \frac{P_iR_i}{P_i+R_i}.
\end{equation}

Note that $\mathnormal{P}_i$, $\mathnormal{R}_i$ and $\mathnormal{F}_i$  $\in$ [0,1]. A perfect classification would yield $\mathnormal{P}_{i}$, $\mathnormal{R}_{i}$ and, consequently, $\mathnormal{F}_{i}$ to be 1. For an overall evaluation of the model performance (i.e. considering the classification of quasars, stars and galaxies altogether), we define the macro-averaged F-measure ($\bar{\mathnormal{F}}$) as follows:

\begin{equation}
\bar{F} = \frac{1}{N}{\sum\limits_{\mathcal{Y}}F_i} ,
\end{equation}

\noindent
where $\mathnormal{N} = 3$ in our case. This definition differs from the micro-averaged F-measure that first computes an overall $\mathnormal{P}$ and $\mathnormal{R}$ and then calculates the harmonic mean. The micro-average is very sensitive to the imbalance of the classes towards the one with higher frequency, while macro-average will give more importance to the class with smallest frequency.

\section{Model validation}
\label{sec:comparison} 

In this section we show the performance metrics from the cross-validation for each of the experiments (Table \ref{tab:svm_rf_results}). We will refer to each experiment by the naming system previously described in \S\ref{ssec:feat}. From the cross-validation results, we first compare the overall performances from SVM and RF (see \S\ref{ssec:algo}). Then we discuss the feature selection by comparing the performances from different feature spaces within the same algorithm (see \S\ref{ssec:feat_e}). All comparisons were statistically performed by a Wilcoxon signed-rank test that are fully described in Appendix \ref{sec:stat}. Throughout this section, we only consider an improvement on the performance of a given classifier if there is a statistically significant increase at 90\% confidence level (i.e. p-values $<$ 0.1), when comparing two experiments.  {All measured p-values are shown in Appendix \ref{sec:stat}.}   {In \S\ref{ssec:dataset} we show the cross-validation results for the experiment without WISE features when we consider an increased dataset. Finally, in \S\ref{ssec:tuning} we perform a randomized grid search to look for better hyper-parameter values.}  {In this section we select two models that achieve the best performance: one that includes WISE features and another without WISE features. The expected performance of the chosen models will be shown in Section \ref{sec:desc_final}. We then provide classifications for unlabelled sources in Stripe 82 S-PLUS DR2 using both classifiers, which will be further discussed in Section \ref{sec:catalogue}.}

\subsection{Algorithm evaluation}
\label{ssec:algo}
First, we evaluated which algorithm gave us the best performance in overall terms. To answer this question, we only compared the macro-averaged F-score within each feature space. Table \ref{tab:svm_rf_results} shows that, in this setup, RF provides a better performance in all feature spaces compared to SVM. These differences are statistically significant (see Appendix \ref{sec:stat}). Moreover, we can see that RF generally requires  {less than half} of the SVM computational time for the same feature space. Thus, we will only consider results from RF for the further comparisons.

\subsection{Feature evaluation}
\label{ssec:feat_e}

\begin{table*}
\centering
\caption{Cross-validation results. Fit time and macro-averaged F-score are shown for each model. Besides that, the metrics precision, recall and F-score are shown for each class. We highlight the highest observed metrics among all experiments in bold. Experiments in bold were chosen for further improvements (see Table \ref{tab:morph_2}). Note that these results were based only on the with\_WISE dataset.}
\label{tab:svm_rf_results}
\begin{tabular}{@{}lP{2cm}cccccc@{}}
\arrayrulecolor{gray}
\toprule
EXPERIMENT  & Fit time (s) &  Prediction time (s) & $\bar{\text{F}}$ (\%) & CLASS  & Precision (\%) & Recall (\%) & F (\%) \\ 
 
\midrule

\multirow{3}{*}{\textbf{RF\_12S}}  & \multirow{3}{*}{ 37.61 $\pm$ 1.26 } & \multirow{3}{*}{0.426 $\pm$ 0.042}  &\multirow{3}{*}{ 92.17 $\pm$ 0.3 } & QSO & 91.99 $\pm$ 0.41 & 89.31 $\pm$ 0.97 & 90.63 $\pm$ 0.52 \\
&&&& STAR & 95.55 $\pm$ 0.46 & 89.59 $\pm$ 0.44 & 92.47 $\pm$ 0.31 \\
&&&& GAL & 91.26 $\pm$ 0.41 & 95.7 $\pm$ 0.29 & 93.42 $\pm$ 0.2 \\

\midrule
\multirow{3}{*}{\textbf{RF\_12S+2W}} & \multirow{3}{*}{ 36.56 $\pm$ 4.39 } & \multirow{3}{*}{0.354 $\pm$ 0.010}  & \multirow{3}{*}{ \textbf{96.85 $\pm$ 0.05} } & QSO & \textbf{95.48 $\pm$ 0.12} & 94.83 $\pm$ 0.26 & \textbf{95.15 $\pm$ 0.12} \\
&&&& STAR & \textbf{98.61 $\pm$ 0.08} & \textbf{97.06 $\pm$ 0.25} & \textbf{97.83 $\pm$ 0.15} \\
&&&& GAL & \textbf{96.97 $\pm$ 0.15} & \textbf{98.14 $\pm$ 0.08} & \textbf{97.55 $\pm$ 0.08} \\

\midrule

\multirow{3}{*}{RF\_5S} & \multirow{3}{*}{ 23.99 $\pm$ 0.2 } & \multirow{3}{*}{0.431 $\pm$ 0.012} & \multirow{3}{*}{ 90.19 $\pm$ 0.25 } & QSO & 90.58 $\pm$ 0.46 & 88.4 $\pm$ 0.7 & 89.48 $\pm$ 0.53 \\
&&&& STAR & 91.93 $\pm$ 0.73 & 87.68 $\pm$ 0.53 & 89.75 $\pm$ 0.26 \\
&&&& GAL & 89.74 $\pm$ 0.35 & 93.03 $\pm$ 0.45 & 91.36 $\pm$ 0.14 \\

\midrule

\multirow{3}{*}{RF\_5S+2W} & \multirow{3}{*}{ 21.77 $\pm$ 0.2 } & \multirow{3}{*}{0.335 $\pm$ 0.010} &\multirow{3}{*}{ 96.62 $\pm$ 0.13 } & QSO & 94.98 $\pm$ 0.2 & \textbf{95.08 $\pm$ 0.32} & 95.03 $\pm$ 0.18 \\
&&&& STAR & 98.14 $\pm$ 0.12 & 96.93 $\pm$ 0.34 & 97.53 $\pm$ 0.18 \\
&&&& GAL & 96.92 $\pm$ 0.17 & 97.68 $\pm$ 0.16 & 97.3 $\pm$ 0.12 \\

\midrule

\multirow{3}{*}{SVM\_12S}  & \multirow{3}{*}{ 142.49 $\pm$ 7.64 } & \multirow{3}{*}{81.29 $\pm$ 0.537} &\multirow{3}{*}{ 76.96 $\pm$ 0.24 } & QSO & 90.89 $\pm$ 0.13 & 77.53 $\pm$ 1.17 & 83.67 $\pm$ 0.7 \\
&&&& STAR & 76.17 $\pm$ 0.39 & 61.36 $\pm$ 0.4 & 67.96 $\pm$ 0.36 \\
&&&& GAL & 73.59 $\pm$ 0.25 & 85.82 $\pm$ 0.18 & 79.24 $\pm$ 0.13 \\

\midrule

\multirow{3}{*}{SVM\_12S+2W}  & \multirow{3}{*}{ 127.48 $\pm$ 13.89 } & \multirow{3}{*}{68.81 $\pm$ 2.50} & \multirow{3}{*}{ 89.7 $\pm$ 0.18 } & QSO & 95.46 $\pm$ 0.31 & 83.56 $\pm$ 1.0 & 89.11 $\pm$ 0.63 \\
&&&& STAR & 94.39 $\pm$ 0.34 & 84.06 $\pm$ 0.48 & 88.93 $\pm$ 0.37 \\
&&&& GAL & 86.77 $\pm$ 0.2 & 95.83 $\pm$ 0.24 & 91.07 $\pm$ 0.16 \\

\midrule

\multirow{3}{*}{SVM\_5S} & \multirow{3}{*}{ 100.13 $\pm$ 1.38 } & \multirow{3}{*}{74.91 $\pm$ 0.11} & \multirow{3}{*}{ 78.47 $\pm$ 0.3 } & QSO & 89.46 $\pm$ 0.54 & 82.87 $\pm$ 0.84 & 86.04 $\pm$ 0.6 \\
&&&& STAR & 74.6 $\pm$ 0.4 & 65.98 $\pm$ 0.48 & 70.02 $\pm$ 0.24 \\
&&&& GAL & 75.91 $\pm$ 0.23 & 83.14 $\pm$ 0.38 & 79.36 $\pm$ 0.19 \\

\midrule

\multirow{3}{*}{SVM\_5S+2W} & \multirow{3}{*}{ 54.71 $\pm$ 0.62 } &  \multirow{3}{*}{39.62 $\pm$ 0.24} &\multirow{3}{*}{ 95.11 $\pm$ 0.15 } & QSO & 95.37 $\pm$ 0.26 & 91.43 $\pm$ 0.45 & 93.36 $\pm$ 0.32 \\
&&&& STAR & 95.8 $\pm$ 0.15 & 96.4 $\pm$ 0.21 & 96.1 $\pm$ 0.08 \\
&&&& GAL & 95.57 $\pm$ 0.16 & 96.2 $\pm$ 0.09 & 95.88 $\pm$ 0.09 \\
\bottomrule
\end{tabular}
\end{table*}

One aim of this work is to assess if the narrow-bands --- which is the major S-PLUS advantage  --- improve the classification. Then we compared RF\_12S versus RF\_5S, i.e. when WISE magnitudes are not being considered, and RF\_12S+2W versus RF\_5S+2W, when considering WISE magnitudes. With addition of WISE photometry, the narrow-bands do not significantly increase the performance of quasar  {recall and galaxy precision} with 90\% confidence. This is possibly due to the fact that WISE plays a more important role on distinguishing these objects. For stars, all metrics have been improved.  {The macro-averaged F-score does significantly improve for RF\_12S+2W.} On the other hand, the S-PLUS narrow-bands significantly increase the precision and recall of all classes when training RF without WISE magnitudes. This means that, for sources without a WISE counterpart, the 12-band magnitude system is able to provide a more reliable star/quasar/galaxy separation in comparison to classifications based solely on broad-band filters. This is important to note, given that  {43.81\% of our photometric sample from} S-PLUS DR2 sources do not have WISE detection. Nevertheless, we evaluated the inclusion of WISE magnitudes by comparing RF\_12S versus RF\_12S+2W, and RF\_5S versus RF\_5S+2W {, where all metrics were statistically improved. This confirms the importance of WISE magnitudes for the classification.
}

 {Comparing our findings with results from other works is not straightforward as we would be comparing metrics that are calculated on completely different validation datasets. Nevertheless, it can guide us through possible interpretations.
For instance, \citet{Brescia_2015} achieves $F_\text{QSO}$ = 88.65\%, $F_\text{STAR}$ = 89.95\%, and $F_\text{GAL}$ = 96.32\% with their best three-class experiment using only broad-bands (\texttt{psfMag} and \texttt{magModel}) from SDSS DR10. These results would be more related to our RF\_5S experiment. We can possibly assume that the morphological aspect may be covered by using both \texttt{psfMag} and \texttt{magModel} in their models, which could explain a $F_{GAL}$ better than ours. On the other hand, they find a similar confusion between quasars and stars, which could be explained by the fact that their model also does not consider WISE features. This similarity is reassuring, since different samples and photometry are used and SDSS is about 1 magnitude deeper than S-PLUS. For instance, the SDSS photometric depth in $r$ is 22.7 mag, whereas the S-PLUS depth is 21.3 mag. The exception is the $z$ band for which the depths are comparable (see SDSS DR16 website\footnote{https://www.sdss.org/dr16/imaging/other\_info/} and table 4 from \citealt{2021arXiv210400020A}). \citet{2019arXiv190910963C} use a method that is more similar to ours by also applying a random forest algorithm on SDSS and WISE features, albeit they include $W3$ and $W4$ from WISE and they use all-sky data. For their analyses based only on SDSS magnitudes, they achieve $F_\text{QSO}$ = 88.1\%, $F_\text{STAR}$ = 92.7\%, and $F_\text{GAL}$ = 97.0\%. Our RF\_5S experiment shows similar behaviour for quasars and worse performance for stars and galaxies. It is not clear what could be the reason for this difference in performance, although percentages can be misleading performance indicators for smaller samples, which is our case.  Including WISE features, they achieve $F_\text{QSO}$ = 94.2\%, $F_\text{STAR}$ = 97.4\%, and $F_\text{GAL}$ = 98.8\%. Our RF\_5S+2M performances seem to be in agreement with the results from  \citet{2019arXiv190910963C}, although our galaxy F-score is slightly worse.}

For further evaluation we only consider the best feature spaces with and without WISE using a RF algorithm: 12 S-PLUS bands (RF\_12S) and 12 S-PLUS bands + 2 WISE bands (RF\_12S+2W).  {For those, we made new experiments including morphological information as features, which can distinguish resolved galaxies from unresolved stars (and quasars) as shown in 
\cite{costaduarte2019splus} for the S-PLUS data.} The results including four morphological parameters (\texttt{FWHM}, \texttt{A}, \texttt{B}, and \texttt{KRON\_RADIUS}) statistically improved the star/quasar/galaxy classification for both experiments, as expected (see two top blocks from Table \ref{tab:morph_2}).  {The only exception is the quasar precision when WISE features are included. For that, the difference on quasar classification precision is very small (from 95.48\% to 95.54\%) for the classifier trained with WISE features. Morphological features would likely help decreasing the number of galaxies being misclassified as quasars or stars. Therefore, the small improvement in quasar precision could be due to the WISE features providing enough information to separate galaxies and quasars.}  

\begin{figure}
  \includegraphics[width=\columnwidth]{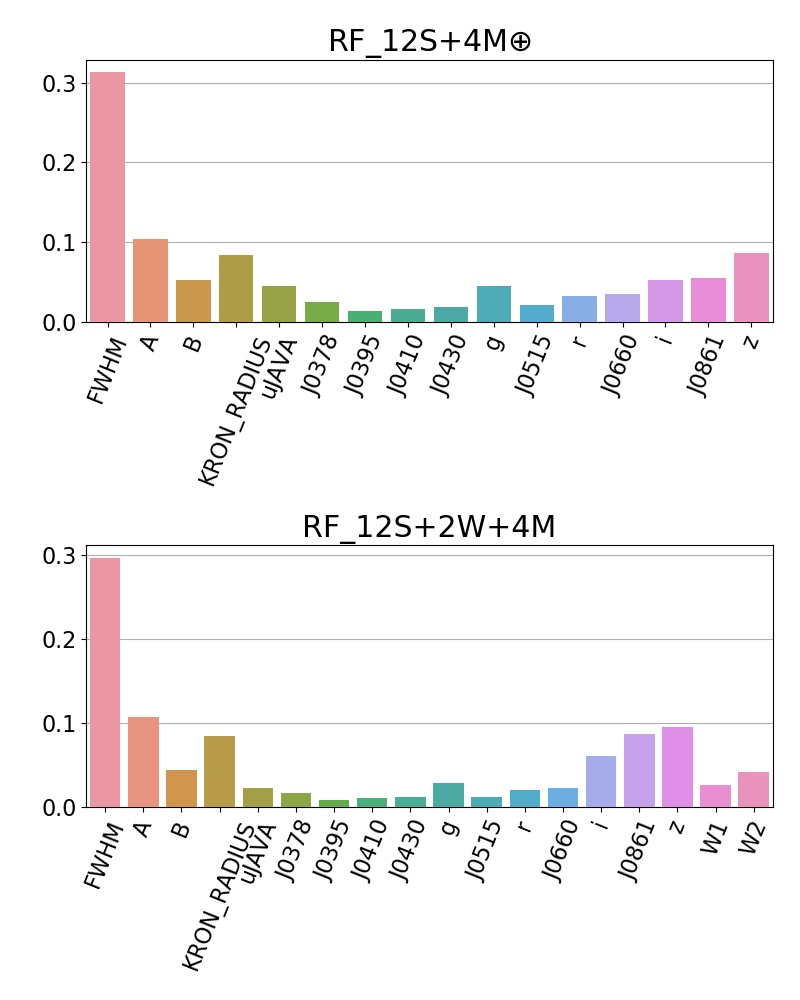}
  \caption{Feature importance in fraction for: 12 S-PLUS filters + 4 Morphological features (top panel) and 12 S-PLUS filters +  2 WISE bands + Morphological features (bottom panel), indicating which features were most important for the classification.}
    \label{fig:feat_imp}
\end{figure}

Comparing RF\_12S+2W+4M versus RF\_12S+4M, the results when considering WISE bands remain statistically superior over RF\_12S+4M.  {Moreover, the metrics had greatly improved from RF\_12S to RF\_12S+4M, with a percentage increase of: 2.28\% for quasars, 5.7\% for stars, and 4.56\% for galaxies, in terms of F-score. From RF\_12S+2W to RF\_12S+2W+4M the percentage increase for F-score is: 0.36\% for quasars, 0.84\% for stars, and 0.74\% for galaxies.} Thus, morphological information plays an important role mainly for the classification when WISE features are not available. In Fig. \ref{fig:feat_imp}, we show the estimated feature importance rankings, where we can see that the \texttt{FWHM}, followed by \texttt{A}, is the most useful information for both classifiers. This reassures the importance of morphological features in our models. However, it is not clear from our analyses if these features are able to distinguish complex structures (such as interacting galaxies). This could be further investigated in future work, as it could potentially help to avoid confusion in the bright end of saturated stars being classified as galaxies, allowing us to build a classification catalogue with more sources. This confusion is not present within our spectroscopic sample results due to our selection criteria (\texttt{PhotoFlag\_r}$=0$). Along with morphology, the reddest S-PLUS filters ($i$, J0861 and $z$) seem to be more relevant for classification than the bluest ones, { being also more relevant than $W1$ and $W2$ features.}


\begin{table*}
\centering
\caption{Cross-validation results. Fit time, prediction time and macro-averaged F-score are shown for each model. Besides that, the metrics precision, recall and F-score are shown for each class.  {Experiments in bold were selected to be the final models. The symbol $\oplus$ refers to an increased dataset (no\_WISE + with\_WISE) for which the algorithm is being validated. Note that they are not comparable with experiments without the $\oplus$ symbol that are validated only on with\_WISE dataset. The $\star$ symbol refers to experiments that were done with the hyper-parameters defined from a randomized grid search (see \S\ref{ssec:tuning}}). }
\label{tab:morph_2}
\begin{tabular}{@{}lP{2cm}cccccc@{}}
\arrayrulecolor{gray}
\toprule
EXPERIMENT  & Fit time (s) & Prediction time (s) & $\bar{\text{F}}$ (\%) & CLASS  & Precision (\%) & Recall (\%) & F (\%) \\ 
 
\midrule

\multirow{3}{*}{RF\_12S+4M} & \multirow{3}{*}{ 53.01 $\pm$ 3.68 } &  \multirow{3}{*}{0.296 $\pm$ 0.008} & \multirow{3}{*}{ 96.04 $\pm$ 0.1 } & QSO & 93.11 $\pm$ 0.27 & 92.29 $\pm$ 0.36 & 92.7 $\pm$ 0.27 \\
&&&& STAR & 98.45 $\pm$ 0.14 & 97.04 $\pm$ 0.08 & 97.74 $\pm$ 0.06 \\
&&&& GAL & 97.12 $\pm$ 0.14 & 98.25 $\pm$ 0.07 & 97.68 $\pm$ 0.07 \\

\midrule
\multirow{3}{*}{\textbf{RF\_12S+2W+4M}} & \multirow{3}{*}{ 53.62 $\pm$ 2.77 }&  \multirow{3}{*}{0.273 $\pm$ 0.002} & \multirow{3}{*}{ 97.47 $\pm$ 0.12 } & QSO & 95.54 $\pm$ 0.24 & 95.45 $\pm$ 0.59 & 95.49 $\pm$ 0.29 \\
&&&& STAR & 99.32 $\pm$ 0.12 & 97.98 $\pm$ 0.04 & 98.65 $\pm$ 0.08 \\
&&&& GAL & 97.82 $\pm$ 0.12 & 98.71 $\pm$ 0.07 & 98.27 $\pm$ 0.05 \\

\midrule

\multirow{3}{*} {\textbf{RF\_12S+4M$\oplus $}}  & \multirow{3}{*}{ 57.73 $\pm$ 1.17 } & \multirow{3}{*}{0.416 $\pm$ 0.013} & \multirow{3}{*}{ 91.8 $\pm$ 0.24 }  & QSO & 82.58 $\pm$ 0.5 & 86.8 $\pm$ 0.85 & 84.64 $\pm$ 0.55 \\
&&&& STAR & 96.15 $\pm$ 0.12 & 92.62 $\pm$ 0.25 & 94.35 $\pm$ 0.15 \\
&&&& GAL & 95.68 $\pm$ 0.31 & 97.15 $\pm$ 0.15 & 96.41 $\pm$ 0.14 \\

\midrule

\multirow{3}{*}{RF\_12S+4M$\oplus\star$}  & \multirow{3}{*}{ 241.7 $\pm$ 5.77  } & \multirow{3}{*}{1.53 $\pm$ 0.03} &  \multirow{3}{*}{ 91.78 $\pm$ 0.22 } & QSO & 82.65 $\pm$ 0.39 & 86.71 $\pm$ 0.92 & 84.63 $\pm$ 0.44 \\
&&&& STAR & 96.14 $\pm$ 0.14 & 92.56 $\pm$ 0.25 & 94.32 $\pm$ 0.16 \\
&&&& GAL & 95.61 $\pm$ 0.38 & 97.19 $\pm$ 0.16 & 96.39 $\pm$ 0.15 \\

\midrule

\multirow{3}{*}{RF\_12S+2W+4M$\star$}  & \multirow{3}{*}{ 144.55 $\pm$ 5.45 } & \multirow{3}{*}{1.02  $\pm$ 0.55} & \multirow{3}{*}{ 97.48 $\pm$ 0.11 } & QSO & 95.58 $\pm$ 0.15 & 95.48 $\pm$ 0.47 & 95.53 $\pm$ 0.24 \\
&&&& STAR & 99.31 $\pm$ 0.11 & 97.98 $\pm$ 0.05 & 98.64 $\pm$ 0.07 \\
&&&& GAL & 97.82 $\pm$ 0.09 & 98.72 $\pm$ 0.07 & 98.27 $\pm$ 0.06 \\

\bottomrule
\end{tabular}
\end{table*}

\subsection{Increasing the dataset for the classifier without WISE features}
\label{ssec:dataset}
 {The analyses presented in \S\ref{ssec:algo} and \S\ref{ssec:feat_e} were done on the same training dataset from \texttt{with\_WISE} sample in order to obtain comparable results. However, the classifier that does not use WISE features as input (i.e. RF\_12S+4M) is therefore biased towards objects with WISE counterpart. One can see the differences in distribution between \texttt{with\_WISE} and \texttt{no\_WISE} samples in Fig. \ref{fig:dist_r_z}. Thus, we should evaluate how RF\_12S+4M performs with an increased training dataset consisting of \texttt{with\_WISE} and \texttt{no\_WISE} datasets. To distinguish between the two experiments, we will call this one trained with more data as RF\_12S+4M$\oplus$. 
The increased dataset for RF\_12S+4M$\oplus$ now contains 18\,989 quasars, 54\,054 stars, and 65\,612 galaxies, where 75\% of each class is used for training. The results for the cross-validation considering the increased dataset are presented in the middle block from Table \ref{tab:morph_2}.  Although the results for RF\_12S+4M$\oplus$ seems to be poorer than RF\_12S+4M, one should note that these results are no longer comparable as they are based on different datasets. In \S\ref{ssec:final} we detail how RF\_12S+4M$\oplus$ and RF\_12S+4M perform in both \texttt{no\_WISE} and \texttt{with\_WISE} testing sets for a better understanding of this model.    }

\subsection{Tuning models with randomized grid search}
\label{ssec:tuning}
 {We attempted to tune RF\_12S+4M$\oplus$ and RF\_12S+2W+4M by employing a 3-fold randomized grid search using  \texttt{RandomizedSearchCV} from the \texttt{Python} library \texttt{scikit-learn}. We detail in Table \ref{tab:gridsearch} the evaluated hyper-parameters, the list of values to be validated for each hyper-parameter, and the best values for RF\_12S+4M$\oplus$ and RF\_12S+2W+4M. Besides \texttt{n\_estimators}, \texttt{min\_samples\_split}, and \texttt{max\_depth}, all other hyper-parameters found by the grid search are the same default values that we used in our previous analyses. The cross-validation results are shown in the two bottom blocks from Table \ref{tab:morph_2}, denoted by $\star$ symbols. From these results, we can see that our metrics changed very slightly and they are not worth the increase in computational time. For RF\_12S+4M$\oplus$, the average prediction time is 0.416 seconds\footnote{Computations were done with a Macbook Pro Touchbar 13-inch 2017 with 8 GB RAM and a Kaby Lake 3.1 GHz Dual-Core Intel Core i5.}, considering \texttt{n\_estimators} = 100 and approximately 27\,730 sources. We can roughly estimate an order of $\sim$85 million sources in the total area of S-PLUS ($\sim$9300 deg$^2$) within $13<r<22$ and $\texttt{PhotoFlag}\_r=0$. For N $\sim85$ million objects and \texttt{n\_estimators} = 100, the prediction time would be approximately 21.25 minutes. For the RF\_12S+4M$\oplus\star$ with \texttt{n\_estimators} = 400  and fixing $\text{N} \sim 85$ million, we estimate $\sim$3.53 days to complete the predictions. Therefore, we maintain the same hyper-parameters used to train RF\_12S+2W+4M and RF\_12S+4M$\oplus$} that we described previously in \S\ref{ssec:rf}.

\begin{table*}
\caption{ {List of hyper-parameters and values evaluated through a randomized grid search for the \texttt{RandomForestClassifier}. The best hyper-parameters are shown for RF\_12S+4M$\oplus$ and RF\_12S+2W+4M and their performances are shown in Table \ref{tab:morph_2} denoted by star symbols.} }
\label{tab:gridsearch}
\begin{tabular}{@{}llcc@{}}
\toprule
Hyper-parameter      & Values                             & \multicolumn{1}{l}{Best values for RF\_12S+4M$\oplus$} & \multicolumn{1}{l}{Best values for RF\_12S+2W+4M} \\ \midrule
n\_estimators       & 100, 200, 300, 400                            & 400                                            & 300                                               \\
min\_samples\_split & 2, 5, None                                     & 5                                              & 5                                                 \\
min\_samples\_leaf  & 1, 2, None                                      & 1                                              & 1                                                 \\
max\_features       & sqrt, 5, 10, 15                                         & sqrt                                           & sqrt                                              \\
max\_depth          & 10, 20, 30, 40, 50, 60, 70, 80, 90, 100, 110, None & 30                                           & 70                                                \\
bootstrap           & True, False                                        & False                                          & False                                             \\ \bottomrule

\end{tabular}
\end{table*}

\section{Description of the final models}
\label{sec:desc_final}

We have so far been selecting the best algorithm (see \S\ref{ssec:algo}) and the best feature spaces (see \S\ref{ssec:feat_e}) for our classification problem through k-fold cross-validation.  {The statistical comparisons were based only on the \texttt{with\_WISE} sample, in order to minimize possible bias effect due to differences in the features distributions. The experiment that achieved the best performance is a random forest algorithm trained with 12 S-PLUS bands + 2 WISE bands and 4 morphological features (RF\_12S+2W+4M). However, the S-PLUS catalogue has a significant fraction of objects with no WISE counterparts. Therefore it is desirable to have another classifier that can be used on those objects. Thus, we have also selected the experiment with best performance among the ones without WISE features, which is a random forest algorithm trained with 12 S-PLUS bands + 4 morphological features (RF\_12S+4M$\oplus$). For RF\_12S+4M$\oplus$ we use an increased sample that includes \texttt{no\_WISE}, in order to better generalize our model.}

 {For each of the two selected models we show the overall expected performance, and the performance in terms of magnitude and probability threshold (see \S\ref{ssec:final}). In \S\ref{ssec:bias} we discuss and investigate some sources of bias that are possibly affecting our models.}

\subsection{Performance on test dataset}
\label{ssec:final}

\begin{table*}
\caption{ {Performance on the test set in terms of precision, recall and F-score ($F$) for each class. We also show the overall performance in terms of macro-averaged F-score ($\bar{F}$). For the model including 12 S-PLUS bands, 2 WISE bands, and 4 morphological features (RF\_12S+2W+4M) we show the metrics for the test set based on with\_WISE. For the classifier trained with 12 S-PLUS bands and 4 morphological features (RF\_12S+4M$\oplus$) we show the metrics separately for with\_WISE and no\_WISE, besides for the entire test set. The \texttt{with\_WISE} testing set has 3\,884 quasars, 9\,899 stars, and 15\,015 galaxies. The \texttt{no\_WISE} testing set has 863 quasars, 1\,388 stars, and 3\,614 galaxies. }}
\label{tab:final_model}
\begin{tabular}{@{}lcccccccc@{}}

\toprule
MODEL                          & Fit time (s)        & Test set                    & Prediction time (s)               & $\bar{\text{F}}$ (\%) & CLASS & Precision (\%) & Recall (\%) & F (\%) \\ \midrule
\multirow{3}{*}{RF\_12S+2W+4M} & \multirow{3}{*}{39.07} & \multirow{3}{*}{with\_WISE}    & \multirow{3}{*}{0.476}          & \multirow{3}{*}{97.69}      & QSO   & 95.76          & 95.88       & 95.82        \\
                               &                        &                        &                                        &                             & STAR  & 99.44          & 98.22       & 98.83        \\
                               &                        &                        &                                        &                             & GAL   & 98.04          & 98.8        & 98.42        \\
\midrule
\multirow{9}{*}{RF\_12S+4M$\oplus$}   & \multirow{9}{*}{48.91} &  \multirow{3}{*}{with\_WISE}            & \multirow{3}{*}{0.438} & \multirow{3}{*}{95.04}      & QSO   & 95.68          & 86.1        & 90.64        \\
                               &                        &                        &                                        &                             & STAR  & 95.59          & 97.97       & 96.77        \\
                               
                               &                        &                        &                                        &                             & GAL   & 97.26          & 98.19       & 97.72        \\
                               \cmidrule(l){3-9} 
                               &                       & \multirow{3}{*}{no\_WISE}       & \multirow{3}{*}{0.279}         & \multirow{3}{*}{79.17}      & QSO   & 52.47          & 92.24       & 66.89        \\
                               &                        &                        &                                        &                             & STAR  & 98.17          & 78.56       & 87.27        \\
                               &                        &                        &                                        &                             & GAL   & 81.39          & 85.37       & 83.33        \\
                               \cmidrule(l){3-9} 
                               &                        &  \multirow{3}{*}{with\_WISE + no\_WISE}  &\multirow{3}{*}{1.925}  & \multirow{3}{*}{91.92}      & QSO   & 82.6           & 87.21       & 84.84        \\
                               &                        &                        &                                        &                             & STAR  & 96.16          & 92.78       & 94.44        \\
                               &                        &                        &                                        &                             & GAL   & 95.87          & 97.1        & 96.48        \\

\bottomrule
\end{tabular}
\end{table*}

 {Considering classifier RF\_12S+2W+4M, we achieved a macro-averaged F-score of 97.69\% on the \texttt{with\_WISE}  testing set. In terms of quasar classification, we have 138 galaxies (from a total of 15\,015) and 27 stars (from a total of 9\,899) being incorrectly classified as quasars ($P_{\text{QSO}}$ = 95.76\%). Meanwhile, 147 quasars were classified as galaxies and 13 quasars were classified as stars ($R_{\text{QSO}}$ = 95.88\%), from a total of 3\,884 quasars in the testing set. In terms of star classification, we obtained $P_{\text{STAR}}$ = 99.44\% and $R_{\text{STAR}}$ =  98.22\%. For galaxies, we obtained $P_{\text{GAL}}$ =  98.04\% and $R_{\text{GAL}}$ = 98.8\%. Specifically, 149 stars were classified as galaxies and 42 galaxies were classified as stars. For classifier RF\_12S+4M$\oplus$ and considering the same \texttt{with\_WISE} testing set, we achieved a macro-averaged F-score of 95.04\%. We have 123 galaxies and 28 stars being misclassified as quasars ($P_{\text{QSO}}$ = 95.68\%), similarly to RF\_12S+2W+4M. We have 298 and 242 quasars being classified as stars and galaxies, respectively ($R_{\text{QSO}}$ = 86.1\%), which is 10.2\% decrease compared to the same metric from RF\_12S+2W+4M. For stars, we obtained $P_{\text{STAR}}$ = 95.59\% and $R_{\text{STAR}}$ =  97.97\%. For galaxies, we obtained $P_{\text{GAL}}$ =  97.26\% and $R_{\text{GAL}}$ = 98.19\%. Specifically, 173 stars were classified as galaxies and 149 galaxies were classified as stars. Note that the scores for RF\_12S+2W+4M in the \texttt{with\_WISE} testing set are mostly better or similar to its corresponding results in the cross-validation shown in Table \ref{tab:morph_2}. This is also true for the testing results on \texttt{with\_WISE}+\texttt{no\_WISE} from RF\_12S+4M$\oplus$ and its corresponding cross-validation results shown in Table \ref{tab:morph_2}. The fact that we do not see any of the two models performing too differently from their corresponding results in cross-validation indicates no evidence of over- or under-fitting.}

\begin{figure}
  \includegraphics[width=\columnwidth]{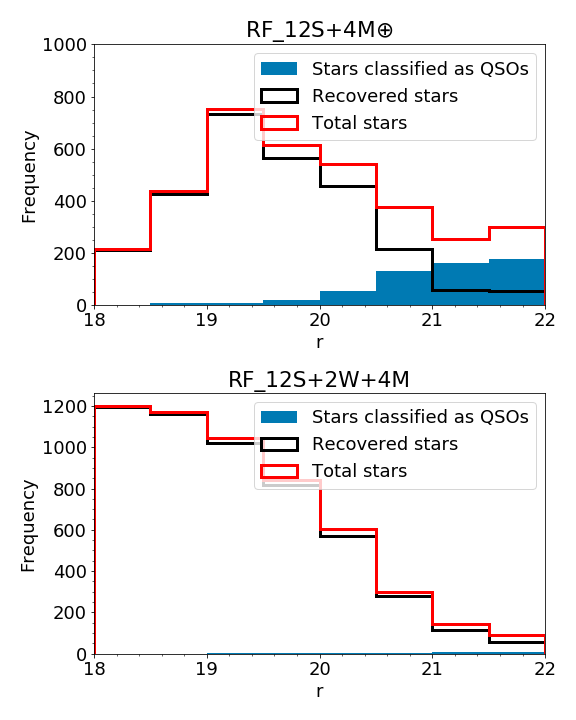}
  \caption{Magnitude distribution in $r$ band for the stars being misclassified as quasars (blue), for the stars being correctly classified as stars (black), and for the total stars in the corresponding test set (red). In the top panel we show the distributions for RF\_12S+4M$\oplus$, and, in the bottom panel, for RF\_12S+2W+4M.}
    \label{fig:missing_stars}
\end{figure}

 {The testing (Table \ref{tab:final_model}) and validation (Table \ref{tab:morph_2}) results for RF\_12S+2W+4M and RF\_12S+4M$\oplus$ considering \texttt{with\_WISE} sample show that using the classification from the former is preferable over the latter whenever the WISE information is available. Otherwise, one can use the classification from RF\_12S+4M$\oplus$. Therefore, it is desirable to know how RF\_12S+4M$\oplus$ performs when no WISE counterpart is found. The results shown in Table \ref{tab:final_model} for the \texttt{no\_WISE} testing set shows a great performance for star precision ($P_{STAR}$ = 98.17\%) and quasar recall ($R_{QSO}$ = 98.17\%); an acceptable performance for star recall ($R_{STAR}$ = 78.56\%), galaxy precision ($P_{GAL}$ = 81.39\%), and galaxy recall ($R_{GAL}$ = 85.37\%); and a very poor quasar precision ($P_{QSO}$ = 52.47\%). The results do not improve in overall if we consider other training dataset (e.g. only no\_WISE or only with\_WISE) for the classifier RF\_12S+4M$\oplus$. The quasar precision is due to 546 stars being mis-classified as quasars (9.0\% are white dwarfs). Checking the $r$ distribution of these stars (Fig. \ref{fig:missing_stars}, top panel), we see that most of them are in the faint end. From that we can understand that RF\_12S+4M$\oplus$ is typically classifying faint point-like sources as quasars and it can indicate a possible bias. Depending on one's science purposes, one should consider an lower upper-limit magnitude than 22. On the other hand, RF\_12S+2W+4M does not present the same issue (Fig. \ref{fig:missing_stars}, bottom panel), thus one can consider going down to 22 magnitudes with this classifier. }

\begin{figure*}
    \centering
    \includegraphics[trim=0.5cm 0.5cm 0.5cm 0.2cm, clip, width=\textwidth]{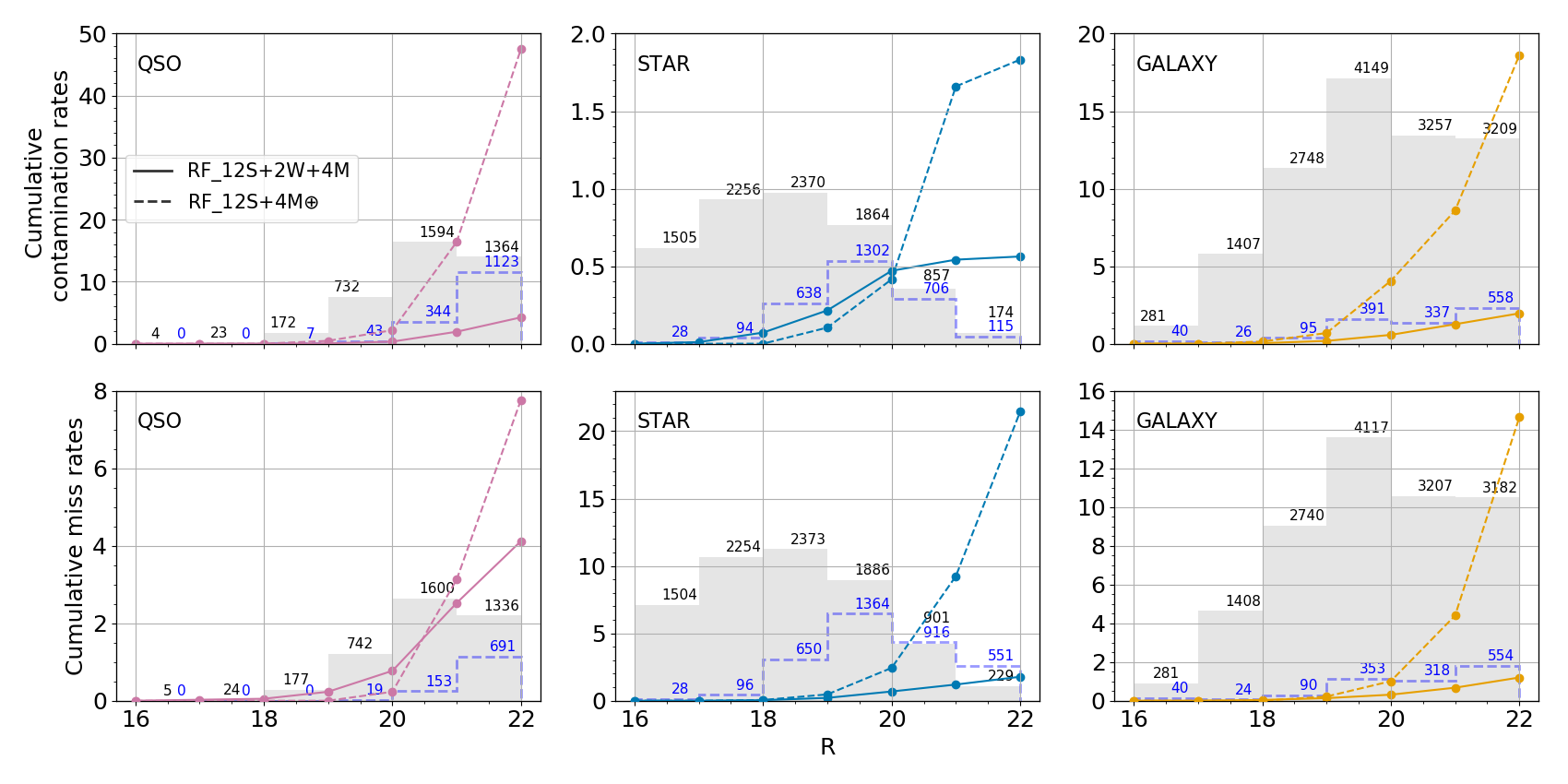}
    \caption{Cumulative contamination (top panels) and miss rates (bottom panels) of quasars (left panels), stars (middle panels), and galaxies (right panels). The models referring to RF\_12S+4M$\oplus$ (12 S-PLUS bands + Morphology) and RF\_12S+2W+4M (12 S-PLUS bands + 2 WISE bands + Morphology) in a Random Forest algorithm are represented by a solid and dashed curves, respectively. In the top panels, we show the $r$ magnitude distribution of all objects classified as quasars (left panel), stars (middle panel) or galaxies (right panel) in the corresponding test set for RF\_12S+4M$\oplus$ (dashed blue) and for RF\_12S+2W+4M (grey). For RF\_12S+4M$\oplus$ we consider only the \texttt{no\_WISE} testing set. In the bottom panels, we show the r magnitude distribution of the known sources in these test sets. These distributions are not described by the y-axis, however the scale of the histograms is the same for all panels. The numbers on top of each histogram bar are the absolute frequency for each magnitude bin. The cumulative contamination and miss rates are relative to the total counts (i.e. sum of all frequencies of the corresponding distribution). One should avoid using the results here presented to compare the two classifiers as they are calculated on different datasets.}
    \label{fig:miss_mag}
\end{figure*}

 { One can also use the information present in Fig. \ref{fig:miss_mag} to choose an upper-limit magnitude.} We show in the top panels of Fig. \ref{fig:miss_mag} the cumulative percentage of contamination (i.e. 1 $-$ $P_i$, also known as false discovery rate) in the classification per model, per class and per magnitude in $r$. In each of these panels, we also show the corresponding distribution of $r$ magnitudes for all the sources in the testing sets classified either as quasar (top left), star (top middle) or galaxy (top right). The contamination rates of quasars, stars and galaxies are 0.05\% (0.46\%), 0.21\% (0.10\%) and 0.18\% (0.07\%) for RF\_12S+2W+4M (RF\_12S+2W+4M$\oplus$) at magnitude $r = 19$, respectively.  {Down to magnitude $r = 21$, these rates are: 1.93\% (16.41\%) for quasars, 0.54\% (1.66\%) for stars, and 1.25\% (0.86\%) for galaxies.}   {The rates for RF\_12S+2W+4M} are in agreement with the colour-colour diagram analyses from \cite{2019arXiv190701567M}.
 In the bottom panels of Fig. \ref{fig:miss_mag} we show the cumulative miss rates (i.e. 1 $-$ $R_i$, { also known as false negative rate}) per model, per class and per magnitude in $r$. In each of these panels, we also show the corresponding distribution of $r$ magnitudes for the total number of the spectroscopically confirmed quasars (bottom left), stars (bottom middle) and galaxies (bottom right). For RF\_12S+2W+4M (RF\_12S+4M$\oplus$) and $r < 19$, the cumulative miss rates are 0.23\% (no percentage is available), 0.21\% (0.47\%) and 0.13\% (0.22\%) for quasars, stars and galaxies, respectively.  {Down to magnitude $r = 21$, these rates are: 1.93\% (3.13\%) for quasars, 0.54\% (9.24\%) for stars, and 1.25\% (4.39\%) for galaxies. Note that the contamination and miss rates are less than 10\% for all classes and for both models (except quasar contamination for RF\_12S+4M$\oplus$) down to magnitude $r = 21$, which is within the photometric depth of the S-PLUS $r$ band.}
 {Moreover,} the classification of quasars, stars, and galaxies  {from RF\_12S+2W+4M} are still reasonably good down to magnitude $r=22$.

In the star/galaxy separation using RF on 12 S-PLUS bands + Morphology from \cite{costaduarte2019splus}, the obtained accuracy (i.e. fraction of correctly classified objects per total objects) is 95\%. We reached an star/galaxy accuracy (i.e. number of correctly classified stars and galaxies per total number of stars and galaxies in our test sample) of 95.15\% when also applying RF on 12 S-PLUS bands + Morphology  {on the \texttt{with\_WISE}+\texttt{no\_WISE} testing set}.  {For the \texttt{with\_WISE} and \texttt{no\_WISE} testing sets separately, the accuracy is 98.1\% and 79.09\%, respectively.}  Moreover, our star/galaxy accuracy reaches 98.59\% when WISE is considered.

 From the \texttt{with\_WISE} test sample, 28\,149 out of 28\,798 (97.75\%) sources were equally labelled using either RF\_12S+4M$\oplus$ or RF\_12S+2W+4M classifiers, representing: 85.94\% out of 3\,884 labelled quasars, 97.73\% out of 9\,899 labelled stars, and 98\% out of 15\,015 labelled galaxies. From these 28\,149 sources, 98.5\% were classified correctly. From the remaining 649 sources (2.25\% out of 29\,109) that were not equally labelled, 555 (85.52\% out of 649) and 58 (8.94\% out of 649) were correctly classified by RF\_12S+2W+4M and RF\_12S+4M$\oplus$, respectively. There is a total of 154 quasars, 152 stars and 152 galaxies that were misclassified by both RF\_12S+4M$\oplus$ and RF\_12S+2W+4M (independent of being equally labelled by both classifiers), representing only 1.59\% out of 28\,798 sources.

\begin{figure*} 
\centering
\includegraphics[trim=0 0.5cm 0.5cm 0.5cm, clip, width=\textwidth]{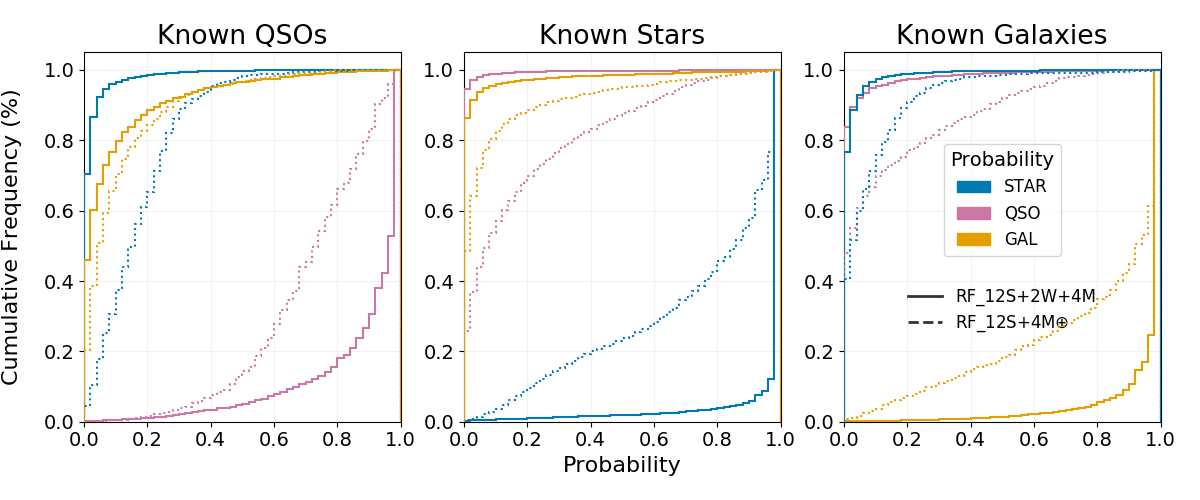}
  
    \caption{From the RF\_12S+2W+4M (RF\_12S+4M$\oplus$) test sample, we show the cumulative distribution of the estimated probabilities for the spectroscopically confirmed: 3\,884 (8\,63)  QSOs, in the left panels; 9\,899 (1\,388) stars, in the middle panels; 15\,015 (3\,614) galaxies, in right panels. The cumulative distributions of the probability of an object being a quasar, star or galaxy are shown in pink, blue and yellow curves, respectively. 
    Results from classifier RF\_12S+2W+4M, referring to 12 S-PLUS bands + 2 WISE bands + Morphology, are shown in solid lines, whereas RF\_12S+4M$\oplus$, referring to 12 S-PLUS bands + Morphology, are shown in dotted lines. One should avoid using the results here presented to compare the two classifiers as they are calculated on different datasets.} 
    \label{fig:prob_wise}
\end{figure*}

Along with the classification, probabilities of being a quasar, star or galaxy were also assigned for each source. We expect that known quasars will present high probabilities of being a quasar, and low probabilities of being a star or a galaxy, as the three probabilities must sum up to 1. The same idea can be applied to the known stars and galaxies. Therefore, we can check the probabilities for each class of our spectroscopic test set. In Fig. \ref{fig:prob_wise} we show the probabilities distributions for each class  {in the \texttt{with\_WISE} (\texttt{no\_WISE}) test set for RF\_12S+2W+4M (RF\_12S+4M$\oplus$)}. We show the distribution in cumulative frequency for the probability of each quasar (left panel), star (middle panel) and galaxy (right panel) being quasar (pink curve), star (blue curve) or galaxy (yellow curve). Most of the known sources  {from \texttt{with\_WISE}} are being correctly classified with high probabilities by the classifier RF\_12S+2W+4M, as shown by the high peaks near probability 1 in each panel. The classifier RF\_12S+4M$\oplus$ returns less certain probabilities for \texttt{no\_WISE} sources, as seen by their broader distribution. The plot shows that the classifier RF\_12S+4M$\oplus$ is mostly confusing the known quasars as stars  {(vice-versa) and known galaxies as quasars}.

\begin{figure}
  \includegraphics[trim=0.5cm 0.5cm 0.5cm 0.5cm, clip, width=\columnwidth]{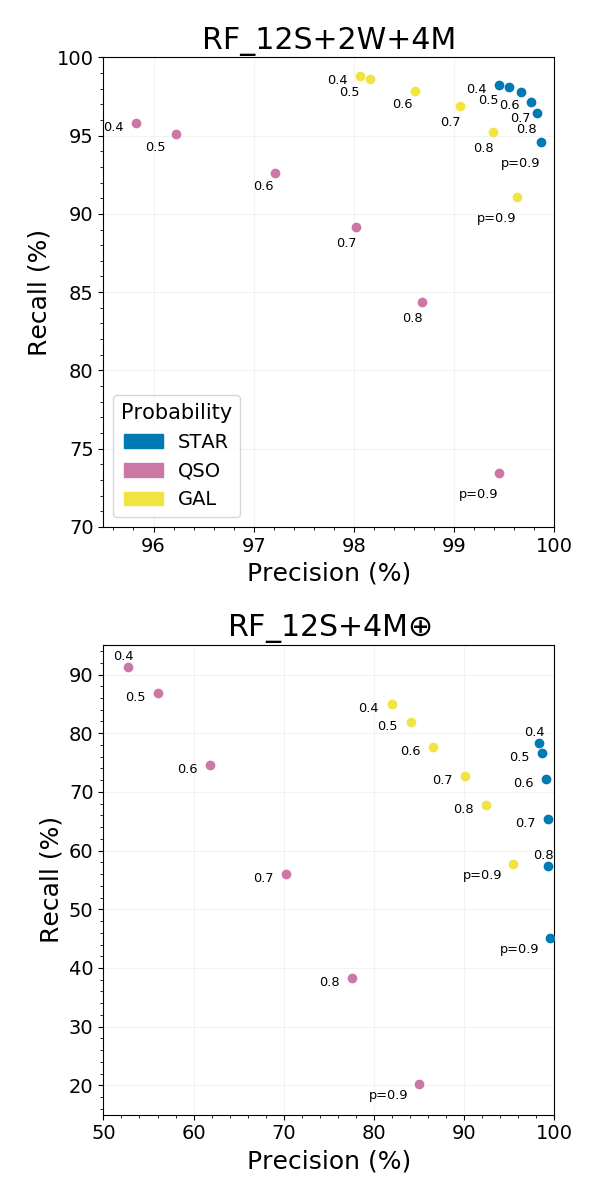}
    \caption{Relation between recall and precision that were calculated in a subset of objects that follows the condition $P_i(X=i)>p$, where $X$ is the event of an source be classified as class $i$, $i \in  \mathcal{Y}$ = \{\text{QSO}, \text{STAR}, \text{GAL}\}, and $p$ is the threshold shown below the points. Results for the classifiers RF\_12S+4M$\oplus$ (\texttt{no\_WISE} test set) and RF\_12S+2W+4M (\texttt{with\_WISE} test set) are shown bottom and top panels, respectively. The results for known stars, quasars, and galaxies are shown in blue, pink, and yellow, respectively. One should avoid using the results here presented to compare the two classifiers as they are calculated on different datasets.}
    \label{fig:prob_threshold}
\end{figure}

 In Fig. \ref{fig:prob_threshold} we show the relation between precision and recall when a threshold $p$ in probability is set for both classifiers. These metrics were calculated using Eq. \ref{eq:precision} and \ref{eq:recall} on a subset of objects from the test sample that follows the condition $P_i(X = i) > p$ for each corresponding class, where $X$ denotes the event of a source be classified as class $i$, $i \in \mathcal{Y}$ = \{\text{QSO}, \text{STAR}, \text{GAL}\}.  {Here again we consider the \texttt{with\_WISE} test set for RF\_12S+2W+4M and the \texttt{no\_WISE} test set for RF\_12S+4M$\oplus$.} As illustrated in the figure, using a higher probability thresholds results in higher precision and lower recall.

Note that the precision, recall and F-score metrics shown in Table \ref{tab:final_model} can be overestimated for the photometric sample, which is likely fainter. It is important to stress that a supervised learning approach can only recognize the patterns learned from the training sample, meaning that any bias in our spectroscopic sample will be reproduced in our classification. In other words, the metrics from Table \ref{tab:final_model} are only reliable for the part of the photometric data within the same feature space distribution of the spectroscopic sample. On the other hand, \cite{2019arXiv190910963C} have shown that the random forest probabilities are correlated to F-score, thus one can consider them as a reliability measurement for the classification. Therefore, we recommend that one selects the quasars, stars or galaxies from our catalogue using a probability threshold that best suits one's science by analysing Fig. \ref{fig:prob_wise} and Fig. \ref{fig:prob_threshold}.

\subsection{Bias}
\label{ssec:bias}
 {Here we discuss the possible bias sources that can be affecting our models. One important aspect to be noted is that our models are not ideal for finding underrepresented sources (e.g. high-redshift quasars) from our spectroscopic samples. For objects that are naturally rare, supplementary techniques would be necessary to properly find them. For non-rare objects that were still underrepresented in our spectroscopic samples, the solution is to increase the spectroscopic observations of these objects. Albeit there are available data from other spectroscopic surveys that could be used in our models, the construction of a better training dataset demands a detailed study, which is not in the scope of this work.} 

\begin{figure}
  \includegraphics[trim=0.3cm 0.7cm 0.5cm 0.7cm, clip, width=\columnwidth]{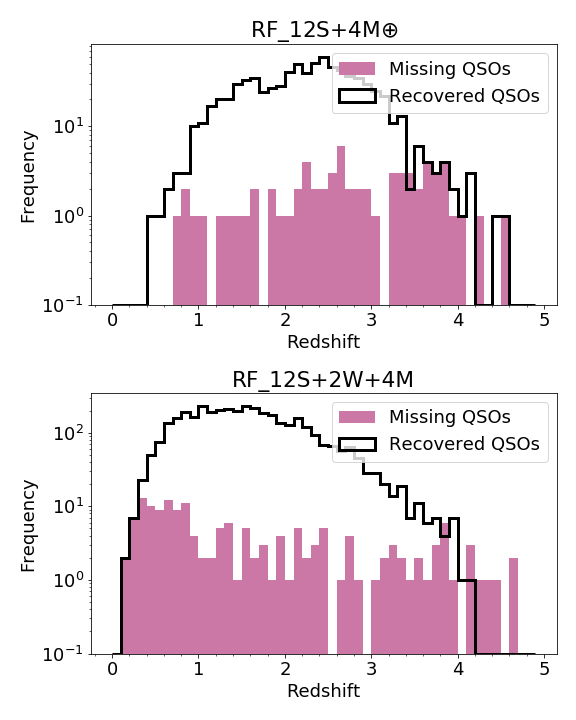}
  \caption{Redshift distribution of the missing and recovered quasars from the corresponding test set for RF\_12S+4M$\oplus$ (top panel) and RF\_12S+2W+4M (bottom panel).}
    \label{fig:qso_bias}
\end{figure}

 {As shown in Fig. \ref{fig:filter_z}, quasars at some specific redshifts will not have emission lines detected in any of the S-PLUS narrow bands, which could possibly cause a bias effect. We can investigate that by checking the redshift distribution of the quasars that are not being recovered by our classifications. Figure \ref{fig:qso_bias} shows that we may have problems classifying quasars at redshift higher than four for both RF\_12S+4M$\oplus$ (top panel) and RF\_12S+2W+4M (bottom panel), although this is due to under-representation in our spectroscopic sample. At lower redshifts, it does not seem that the non-detection of quasar emission lines in any of S-PLUS narrow bands are particularly affecting the classification.}

 {The model RF\_12S+4M$\oplus$ is probably biased towards faint point-like sources being quasars, as previously discussed in \S\ref{ssec:final}. This means that faint stars will be often classified as quasars. On the other hand, we retrieve most of the known quasars. This problem can perhaps be solved after gathering more spectroscopic confirmed sources to train the classifier. This will be possible extending this work for areas beyond Stripe 82 and also using data from future releases. Meanwhile, one can avoid  this problem by restricting the source selection in an upper magnitude below 22 for objects without WISE counterpart.}

 {One could argue that the absence of magnitude measurements (missing bands) in any of the 12 S-PLUS filters could add bias to our models. We have shown in Fig. \ref{fig:tsne} that the missing bands are likely not affecting our classification, as our data seems to be fully separable within the considered features. In order to better investigate that, we can compare the performances between a model trained with a subset of our spectroscopic data that have information in all bands, and another model trained with the same subset but with simulated missing values. To simulate the missing values, we are computing the empirical proportions of each combination of missing bands from our entire dataset. We then use those proportions to sample the sources that will have the corresponding observed magnitudes replaced by a missing value (i.e. 99). Therefore, we can run a 5-fold cross-validation similarly to what was done in Section \ref{sec:comparison}. Moreover, we can see how each model performs on the data that originally have missing-band values, which were not included in those models. The results for this analyses are shown in Appendix \ref{sec:mb}. We see that there is no loss in performance due to the presence of missing-band values in any of our S-PLUS features. Therefore, there is no evidence to conclude that our models are being biased by the inclusion of features with missing-band values, which is in agreement with the t-SNE visualisation. We furthermore argue that missing-band values are usually a physically important information (e.g. the Lyman break) and excluding these objects from our models are not desirable.}

\section{NEW CATALOGUE OF STARS, QUASARS AND GALAXIES FOR S-PLUS DR1}
\label{sec:catalogue}

 In order to build our catalogue, we  re-trained the classifiers using all spectroscopically confirmed objects from SDSS. The final model performance shown in Table \ref{tab:final_model} can change for these new models trained with the entire dataset and we cannot quantitatively account for this. Nevertheless, we would expect very similar performances as the training and testing sets were randomly drawn from the same distribution and their class were ratios maintained.  
 Therefore, the classifier RF\_12S+2W+4M to be used on sets of unknown objects was trained with the entire \texttt{with\_WISE} set, whereas RF\_12S+4M$\oplus$ was trained with all of our spectroscopic sample. Their expected performances were described in the last section. 

The catalogue with the classification for sources in Stripe 82 from S-PLUS DR2 can be found on \url{https://splus.cloud/files/SPLUS_SQGTool_DR2_S82_v1.0.0.csv}.  {In this catalogue, there are 3\,076\,191 objects that are within the selection criteria we used to build our models, i.e. \texttt{Photo\_Flag\_r}$ = 0$, $13<r<22$, from a total of $\sim$11M in the S-PLUS DR2 for the Stripe 82 region. A total of 1\,347\,637 objects (43.81\%) have WISE counterpart and were provided with predictions from RF\_12S+2W+4M. On the other hand, we provide predictions for all sources using the classifier RF\_12S+4M$\oplus$. These numbers include sources that were in our spectroscopic training sample. The columns in our catalogue are fully described in Table \ref{tab:columns} from Appendix \ref{sec:appendix_catalogue}.}

 {From the analyses shown in Section \ref{sec:comparison}, one should consider using the classification from RF\_12S+2W+4M whenever the WISE information is available. Therefore, for the discussion here presented we consider the classification from RF\_12S+4M$\oplus$ for sources without WISE counterpart, and the classification from RF\_12S+2W+4M for sources with WISE counterpart.}
 {For the remaining sources from Stripe 82 (and also for other S-PLUS fields from DR2 and future releases), one can predict their classification using the \texttt{SQGClass} from  \texttt{splusdata.vacs} in \texttt{Python}. The documentation containing instructions for the \texttt{SQGClass} usage can be found on \url{https://splus.cloud/}. The classification for all sources in DR2 is also available thorugh a query service \url{https://splus.cloud/query} as star\_galaxy\_quasar table under dr2\_vacs schema. }  {We recommend that one should consider using the same (or more restrict) selection criteria beforehand.} Moreover, we stress that one must consciously filter out sources that are not appropriate for your scientific goals, which might include sources with high photometric errors and/or bad photometry flags. 

\begin{figure*}
    \includegraphics[width=1\textwidth]{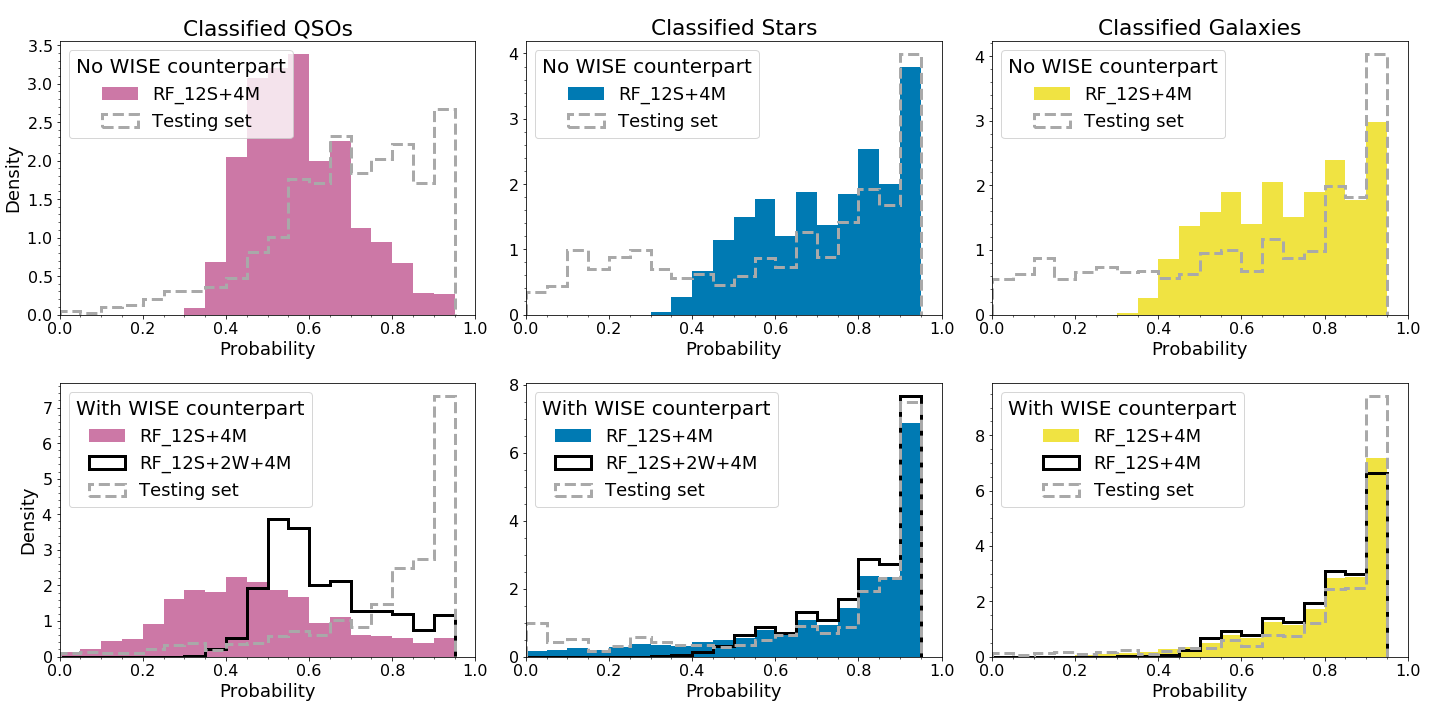}
  \caption{Histograms show the corresponding probabilities distribution of classified quasars (left panels), stars (middle panels), and galaxies (right panels) that were not part of our spectroscopic training sample. Top panels shows the probabilities distribution of sources without WISE counterpart. Bottom panels are for the remaining sources with WISE counterpart. For those, we show the probabilities distribution from both classifiers, but we only consider the classified objects from RF\_12S+2W+4M. We can see from bottom-left panel that many objects classified as quasars by RF\_12S+2W+4M (black) are not being classified as such by RF\_12S+4M$\oplus$. For each panel we also show the probability distribution from their corresponding test set (gray).}
    \label{fig:prob_cat}
\end{figure*}

\begin{figure*}
\includegraphics[width=1\textwidth]{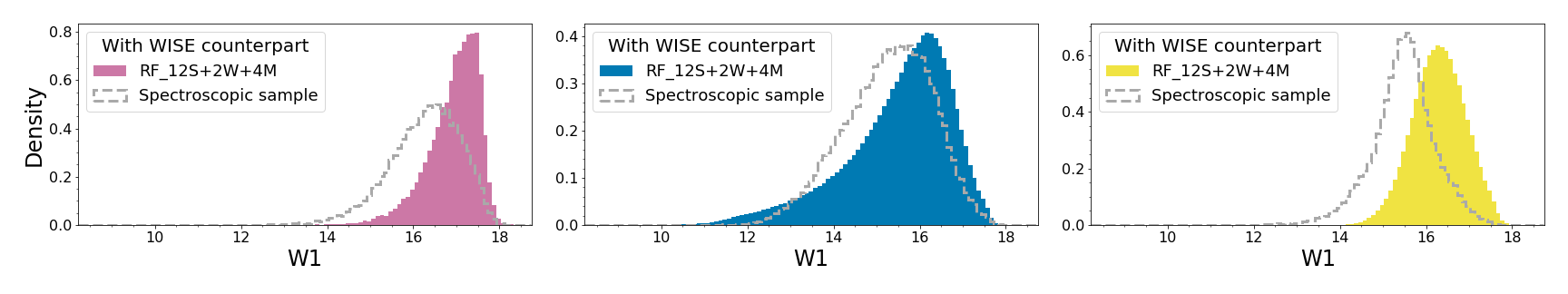}\\
    \caption{Distribution  of $W1$ for the sources we classify as quasars (pink), stars (blue), or galaxies (yellow). We also show in gray the distribution for the spectroscopically confirmed objects. }
    \label{fig:dist_W1}

\end{figure*}

 Hereafter we only consider the 2\,926\,787 sources that were not in our spectroscopic training sample. Based on the results from our classification, our catalogue for the Stripe 82 region has: 335\,956 quasars, 1\,347\,340 stars, and 1\,243\,391 galaxies. From these numbers, 24\,299 quasars, 669\,759 stars, and 530\,893 galaxies have WISE counterpart and were classified with RF\_12S+2W+4M. In Fig. \ref{fig:prob_cat} we show the distribution of the estimated probabilities for each class in our catalogue. For sources without WISE counterpart (top panels), 19\,485 quasars, 399\,159 stars, and 309\,338 galaxies are being classified with very confident probabilities ($> 0.8$). For sources with WISE counterpart (bottom panels), we have the information from both RF\_12S+2W+4M and RF\_12S+4M$\oplus$, but we only show the distribution from objects classified from RF\_12S+2W+4M. We can see that we have a more prominent peak around very confident probabilities than seen for sources without WISE counterpart (being more evident for stars and galaxies), i.e. the WISE features are likely making our objects easier to be classified with higher certainty. We have 5\,352 (2\,169) quasars, 624\,689 (610\,715) stars, and 417\,476 (426\,895) galaxies being classified with very confident probabilities from RF\_12S+2W+4M (RF\_12S+4M$\oplus$) for sources with WISE counterpart. On the other hand, a total of 149 quasars, 1\,204 stars, and 41 galaxies classified by RF\_12S+2W+4M are being assigned with low probability ($<0.3$) by RF\_12S+4M$\oplus$. In each panel of Fig. \ref{fig:prob_cat} we show the corresponding probability distribution of the test sample, previously shown in Fig. \ref{fig:prob_wise}. From these distributions we see a fairly good agreement, except for quasars. The fact that our catalogue has fewer quasars with very confident probabilities compared to our test sample is possibly due to a high quasar completeness within our spectroscopic sample. Moreover, we show in Fig. \ref{fig:dist_W1} the $W1$ magnitude distribution for both our classified sources and our spectroscopic training sample, where we can see that our objects are approximately 1 magnitude fainter than the spectroscopic sample. Therefore, as our performances worsens with increasing magnitude (see Fig. \ref{fig:miss_mag}), we can assume that our expected performances may be overestimated. In Fig. \ref{fig:dist_feat_classification} from Appendix \ref{sec:appendix_catalogue} we also show the distribution for FWHM and $r$ magnitude.

\begin{figure}
  
  \includegraphics[width=0.45\textwidth, trim={0 0 2.5cm 1cm}, clip]{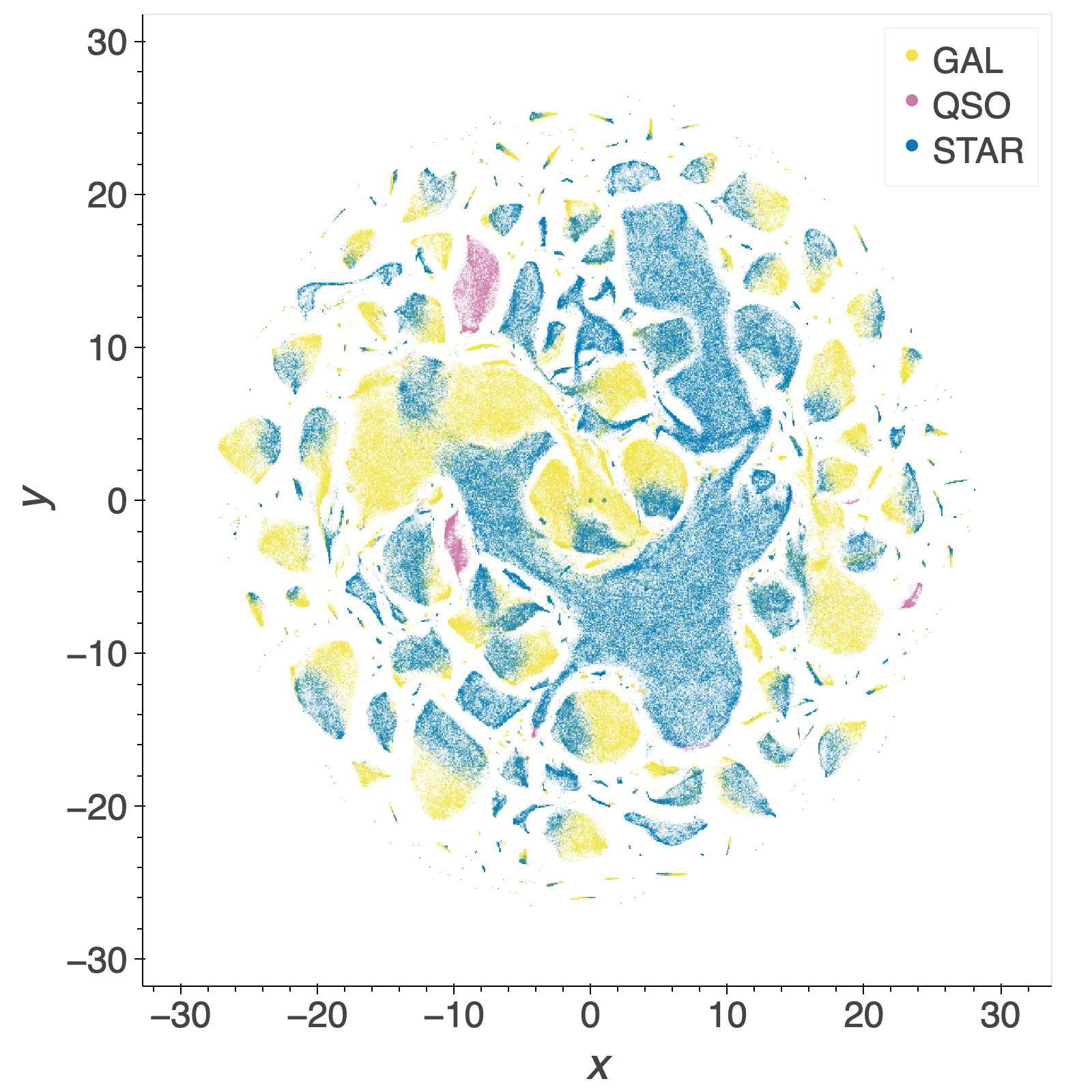}
  \caption{Two-dimensional projection performed by t-SNE for the objects in our catalogue that were not in our spectroscopic sample for the 12 S-PLUS bands + 4 morphological features. Classes were obtained from our model RF\_12S+4M$\oplus$. Only objects with classification with very confident probabilities ($>80\%$) are shown.}

    \label{fig:tsne_80}
\end{figure}
 We show in Fig. \ref{fig:tsne_80} a t-SNE visualisation of the sources from our catalogue that are being classified as galaxies, stars, or quasars with very confident probabilities ($>80\%$). This plot does not include any source with spectroscopic information and we only show objects without WISE counterpart.  We can conclude from this plot that we are able to identify patterns with RF, otherwise, our classes would not appear clustered. We cannot conclude, however, that we are labelling these patterns correctly by only looking at the t-SNE plots.

Figure \ref{fig:tsne_80} shows how the labels from the RF classifier are related to an unsupervised clustering using t-SNE algorithm, which is an independent process from RF. In this figure we show sources from our catalogue that are being classified as galaxies, stars, or quasars with very confident probabilities ($>80\%$). This plot does not include any source with spectroscopic information and we only show objects without a WISE counterpart. The t-SNE plot itself does not prove that we are labelling these objects correctly, nevertheless, it would be an indication of problems in the classification if the colours from RF labels were randomly dispersed throughout the t-SNE clusters. Although the t-SNE algorithm is totally different from the RF process, the figure clearly shows that the spatial clusters are in agreement with the RF labels. In addition, in the great majority of the cases, the three classes do not significantly overlap and most of the groups show a clear separation between the labels from RF and quasars are clearly in distinct islands, in most cases. 

\begin{figure*}
    \includegraphics[width=1\textwidth]{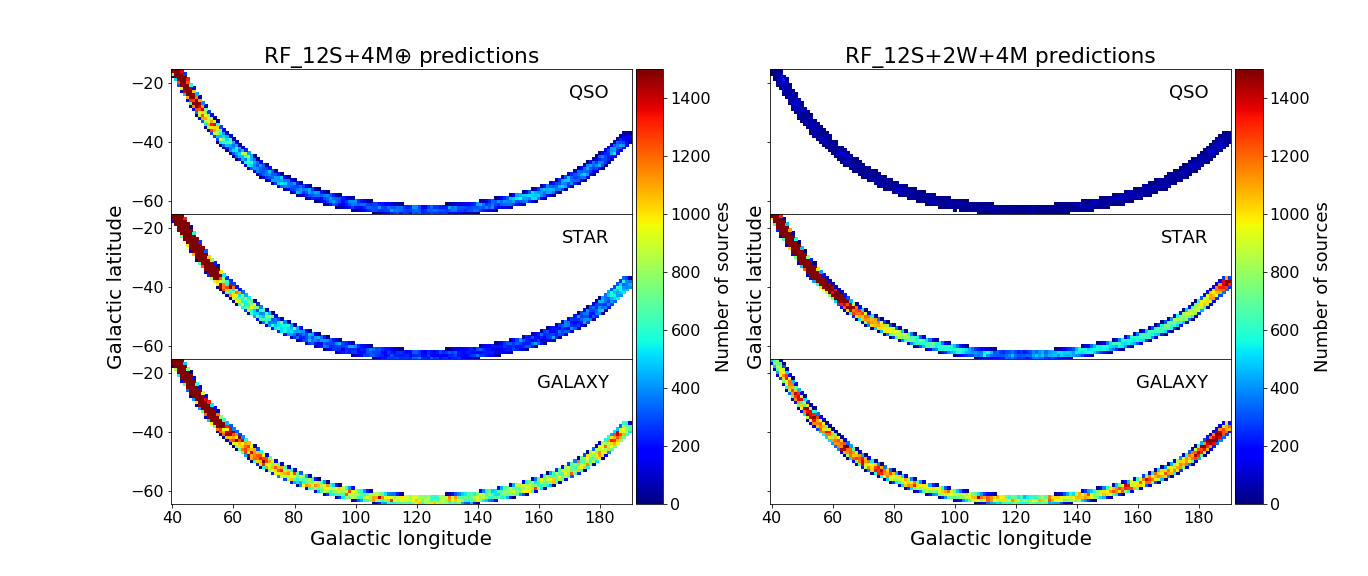}
  \caption{Number counts of total objects from Stripe 82 S-PLUS DR2 being classified as quasars (top panels), stars (middle panels) and galaxies (bottom panels) per galactic coordinates. Each pixel corresponds to an area of 1x1 deg$^2$. Results for the classifiers RF\_12S+4M$\oplus$ and RF\_12S+2W+4M are shown in left and right panels, respectively. We set a top limit of 1\,500 counts in order to provide a better visualization of the higher longitudes. Thus, the dark red regions at lower longitudes should not be interpreted as homogeneous. We recommend caution when using our classification for crowded regions (e.g. regions close to the Galactic Plane), due to the high contamination of stars. }
    \label{fig:bxl}
\end{figure*}

The S-PLUS DR2 fields in the interval 20h < RA < 21h (300$^{\text{o}}$ < RA < 320$^{\text{o}}$) are extremely populated by stars, given they are in the Galactic Plane, i.e. |b| < 30. Thus, we might have greater number counts of stars being confused as quasars or galaxies compared to less populated regions. This can also happens in regions around resolved stellar clusters. Figure \ref{fig:bxl} shows the total number of objects from Stripe 82 S-PLUS DR2  being classified as quasar, star or galaxy with both of our classifiers as a function of Galactic coordinates. As the number counts for galaxies and quasars are expected to be homogeneous in the spatial distribution, it is clear from Fig. \ref{fig:bxl} that there is a high absolute number of contaminated sources in the quasar and galaxy samples at lower galactic latitudes for classifier RF\_12S+4M$\oplus$. In contrast, this behaviour is alleviated when WISE magnitudes are taken into consideration, as seen from the right panel of Fig. \ref{fig:bxl}.

\begin{figure*}
  \includegraphics[width=1\textwidth]{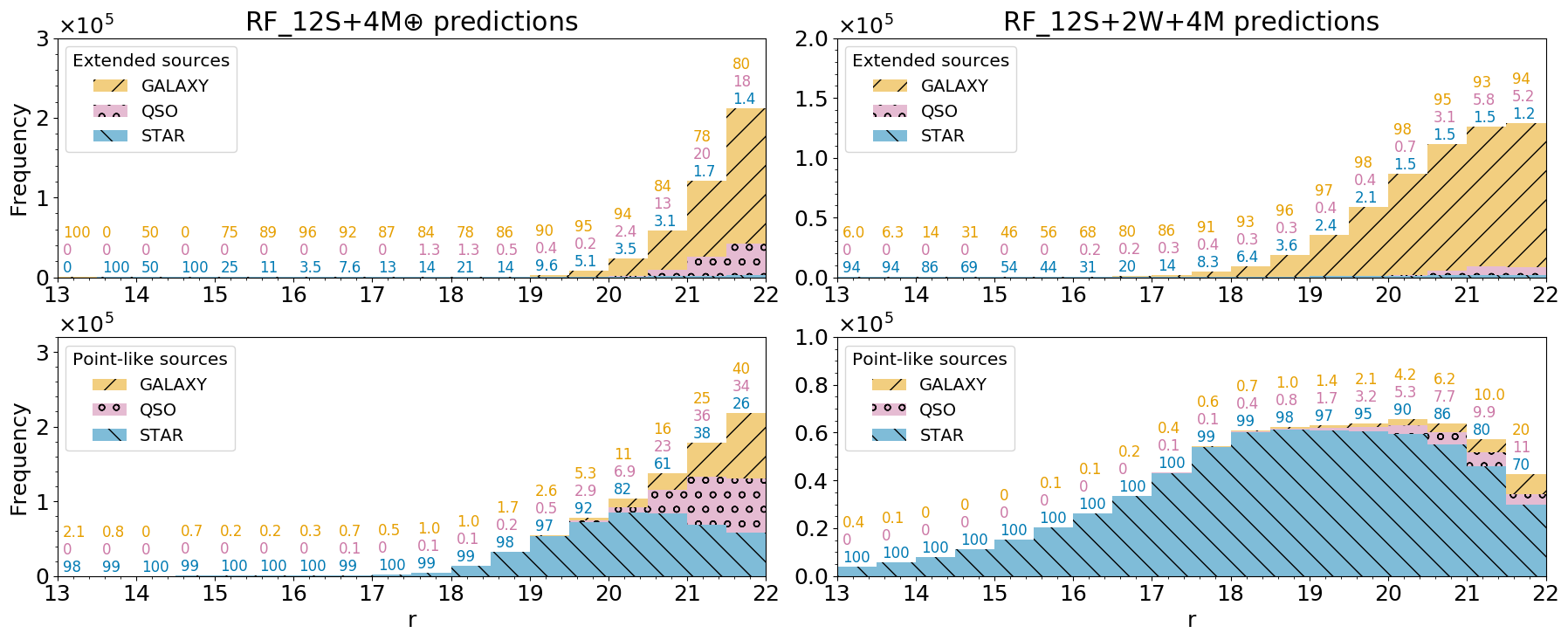}
  \caption{The stacked histograms show the frequencies of extended and point-like sources (in the top and bottom panels, respectively) that were classified as stars, quasars or galaxies with classifier RF
  \_12S+4M$\oplus$ (left panels) and RF\_12S+2W+4M (right panels). The photometric classification of extended and point-like sources were retrieved from SDSS DR16. The numbers on the top of each histogram bar refer to the percentage of spectroscopically confirmed galaxies (orange), quasars (pink) and stars (blue) in each bin. In total, 0.78\% (1.97\%) out of the extended sources are being classified as quasars (most at the faint end) and 1.35\% (0.83\%) are being classified as stars with RF\_12S+2W+4M (RF\_12S+4M$\oplus$). On the other hand,we are classifying 1.52\% (4.91\%) out of the point-like sources as galaxies. Assuming that the extended sources are likely to be galaxies and the point-like sources are likely to be quasars or stars, these results are in fairly good agreement with the estimated contamination and miss rates on the galaxy sample that were shown in Fig.
 \ref{fig:miss_mag}.  } 
    \label{fig:morph_class}
\end{figure*}

The SDSS has a photometric pipeline  that performs a star/galaxy separation based on the difference of the composite model magnitudes (\texttt{cmodelMag}) and PSF magnitudes (\texttt{psfMag}) with 95\% confidence level to r$_{SDSS}$ = 21 \citep{2002AJ....123..485S}. The sources are classified as extended source (\texttt{type} $== 3$) if they satisfies the relation: \texttt{psfMag} $-$ \texttt{cmodelMag} > 0.145. Otherwise, they will be classified as point-like source (\texttt{type} $== 6$). We compared this information with our obtained results, assuming that extended sources are likely to be galaxies, whereas point-like sources are likely to be stars or quasars.  In Fig. \ref{fig:morph_class} we show the distribution of all extended (top panels) and point-like sources (bottom panels) for our two classifiers down to magnitude $r = 22$. For the classifier RF\_12S+2W+4M (RF\_12S+4M$\oplus$), 0.78\% (1.97\%) out of the extended sources are being classified as quasars (most at the faint end) and 1.35\% (0.83\%) are being classified as stars. On the other hand, we are classifying 1.52\% (4.91\%) out of the point-like sources as galaxies.

There are 195 galaxies in Stripe 82 S-PLUS DR2 that were not in our training sample, but were classified by Galaxy Zoo DECaLS \citep{walmsley2021galaxy}. From those, only three galaxies were classified incorrectly as stars by RF\_12S+4M$\oplus$ and one galaxy were classified incorrectly as quasar by RF\_12S+2W+4M.  

\section{Summary and Conclusions}
\label{sec:conc}

In the context where many large photometric surveys are emerging, we need an efficient, accurate, optimized and automated technique for classifying objects.  We compared two supervised learning methods based on Support Vector Machines (SVM) and Random Forest (RF) algorithms for the star/quasar/galaxy classification of S-PLUS DR2 for the Stripe 82 region. 
From our analysis we conclude that S-PLUS has a great advantage in distinguishing these three classes due to its Javalambre filter system, consisting of 7 narrow and 5 broad bands.  {Our classification is robust against missing bands, as first indicated by the t-SNE results and further evidenced from an analysis of simulated missing data.} We were also capable of improving the star/galaxy classification with respect to \cite{costaduarte2019splus}, extending the analysis to $r = 22$. Here we summarize our main results from the cross-validation:

\begin{itemize}
    \item  {Algorithm selection:} we compare the performance of RF and SVM through macro-averaged F-score and we concluded that the best results came from RF. It also generally required less computational time to train the classifier. 
    \item  {Feature space evaluation:} we tested the performance of four feature spaces:  12 S-PLUS bands (12S);  12 S-PLUS bands + 2 WISE bands (12S+2W);  5 S-PLUS broad-bands (5S); and  5 S-PLUS broad-bands + 2 WISE bands (5S+2W). In order to assess the importance of the S-PLUS narrow bands, we compared RF\_12S versus RF\_5S and RF\_12S+2W versus RF\_5S+2W. We concluded, with 90\% confidence, that the classification is improved when narrow-band information is available along with broad-band magnitudes. We also compared RF\_12S versus RF\_12S+2W and RF\_5S versus RF\_5S+2W, confirming the importance of WISE bands in the classification. From these results, we selected RF\_12S and RF\_12S+2W for new tests including morphological parameters (\texttt{FWHM}, \texttt{A}, \texttt{B} and \texttt{KRON\_RADIUS}), which we further referred to as RF\_12S+4M and RF\_12S+2W+4M, respectively. All metrics were statistically improved with the addition of morphological information in our models.  {In order to provide a more generalized model, we increase the dataset for RF\_12S+4M by including sources without WISE counterpart, which is now denoted by RF\_12S+4M$\oplus$.}
    \item  {Final model performances:} the expected performances were calculated  {from two unseen test samples: one that only contains objects with WISE counterpart, and another that only contains objects with no WISE information}. Considering the former, we calculated the overall F-score, which is the harmonic mean of precision and recall, macro-averaged over all three classes as 97.69\% (95.04\%) for RF\_12S+2W+4M (RF\_12S+4M$\oplus$) . In terms of quasar classification, we achieved 95.76\% (95.68\%) for precision/purity and 95.88\% (86.1\%) for recall/completeness. In terms of star classification, we achieved a precision of 99.44\% (95.59\%) and recall of 98.22\% (97.97\%). Finally for galaxies, a precision of 98.04\% (97.26\%) and a recall of 98.8\% (98.19\%) were obtained. These metrics demonstrate the efficacy of both our classifiers.  {For RF\_12S+4M$\oplus$, we also calculated the performances on objects without WISE counterpart, for which we obtained an overall F-score of 95.04\%; 52.47\% precision and 92.24\% recall for quasars; 98.17\% precision and 78.56\% recall for stars; 81.39\% precision and 85.37\% recall for galaxies. A total of 546 stars in the faint-end of $r$ distribution are being misclassified as quasars, which reflects in the bad quasar precision and star recall.}
    
\end{itemize}

We trained two classifiers using the complete spectroscopic sample from SDSS:
\begin{itemize}
    \item  Classifier RF\_12S+4M$\oplus$: 12 S-PLUS bands + Morphology parameters in a Random Forest algorithm
    \item Classifier RF\_12S+2W+4M: 12 S-PLUS bands + 2 WISE bands + Morphology parameters in a Random Forest algorithm (trained only with objects that have WISE counterparts)
 \end{itemize}

In our catalogue, we provided two classifications (using RF\_12S+4M$\oplus$ and RF\_12S+2W+4M) and their corresponding estimated probabilities of a source being a star, quasar or galaxy.
 {The classifier RF\_12S+2W+4M achieved a 2.79\% increase in macro-averaged F-score over classifier RF\_12S+4M$\oplus$ within objects with WISE counterpart, dominated by a 11.36\% increase in recall for quasars}. Thus we recommend that one uses the classification from RF\_12S+2W+4M  {whenever the WISE information is available}. For sources with no WISE counterpart, we can only rely on the classifier RF\_12S+4M$\oplus$ that has a  {reasonable performance down to magnitude $r=21$, for which all metrics are higher than 90\% (except for quasar precision that is 83.59\%). Moreover, we have shown that using the 12 S-PLUS bands provides better performances than using only 5 broad bands, improving macro-averaged F-score in 2.19\% (see \S\ref{ssec:feat_e}).} For any of the classifiers, we recommend that one uses Fig. \ref{fig:prob_wise} and \ref{fig:prob_threshold} shown in \S\ref{ssec:final} in order to select a probability threshold that gives the best star, quasar or galaxy sample that suits one's science. { One should also consider Fig. \ref{fig:miss_mag} for selecting an upper limit magnitude. }

Another useful approach for the star/quasar/galaxy separation are methods that directly process the broad- and narrow-band images, instead of using  {catalogued datasets created from source-finding algorithms}. Our group is currently investigating the use of  {deep learning techniques for such image-based classifications}. Preliminary processing of RGB images (derived from the 12-band images) by convolutional neural networks trained with transfer learning and self-supervised learning strategies presented promising classification results (\citealt{martinazzo2}, \citealt{martinazzo}). The present work will then serve as a baseline for future evaluation of classifications based on deep learning approaches. Moreover, this work can be easily applied to upcoming S-PLUS data releases or other multi-wavelength surveys that use narrow-band filters. With the full coverage of $\sim$9\,300 deg$
^2$, a classification catalogue for S-PLUS will have synergies with other surveys, such as LSST. Planned future work involves the estimation of photometric redshifts for all galaxies and quasars from our catalogue using machine learning and spectral energy distribution (SED) fitting.


\section*{Data Availability}
The S-PLUS DR2 can be obtained from \url{https://splus.cloud/query}.The AllWISE catalogue used in this work can be obtained at \url{https://datalab.noao.edu/}. The SDSS DR16 catalogue can be obtained at \url{http://skyserver.sdss.org/CasJobs}. The SDSS DR14Q catalogue can be obtained at \url{https://www.sdss.org/dr14/algorithms/qso_catalog/}.

The Stripe 82 S-PLUS DR2 catalogue containing classification and probabilities for all sources with $\texttt{PhotoFlag}\_r=0$ and $13<r\leq22$ is available for download at \url{https://splus.cloud/files/SPLUS_SQGTool_DR2_S82_v1.0.0.csv}. Classification for all sources in DR2 can also be downloaded from \url{https://splus.cloud/query} under dr2\_vacs schema as star\_galaxy\_quasar table. We made available the spectroscopic sample used in this paper through \url{https://splus.cloud/files/Spectroscopic_sample.csv}, where we use the flag 1 to identify sources that were used for training, and 0, otherwise (for both cross-validation and final model performance contexts).

The codes for the analyses presented in this paper are available at \url{https://github.com/marixko/MNRAS_nakazono_2021}, where the ADQL queries for downloading the above mentioned datasets can also be found.


\section*{acknowledgements}

At the time this work was accepted for publication, Brazil reached 496,172 deaths by COVID-19. I dedicate my first first-author paper to all Brazilians who lost your beloved ones. I would like to thank my family, my therapist and my close friends who have been of great support for my mental health throughout this political chaos. 

This work has been supported by a PhD fellowship to the lead author from Funda\c{c}\~ao de Amparo \`a Pesquisa do Estado de S\~ao Paulo (FAPESP), 2019/01312-2. LN also acknowledges the support of Coordena\c{c}\~ao de Aperfei\c{c}oamento de Pessoal de N\'ivel Superior - Brasil (CAPES) - Finance Code 001 and FAPESP through process number 2014/10566-4. LN also thanks the staff of the Astronomy department from the University of Florida, where part of this work was done; Marco Antonio dos Santos and Ulisses Manzo Castello for the technical support; Luis Manrique for the technical support, feedback and discussions about Machine Learning; Christian Massao Tsujiguchi Takagi and Vin\'icius Amaral Haga for the feedback on the accessibility of the figures in this paper; Gustavo Schwarz for the help with the database access. NSTH acknowledges FAPESP (grants 2017/25835-9 and 2015/22308-2). CQ acknowledges support from FAPESP (grants 2015/11442-0 and 2019/06766-1). AM acknowledges FAPESP scholarship grant 2018/25671-9. CEB acknowledges FAPESP, grant 2016/12331-0.  FA-F acknowledges funding for this work from FAPESP grant 2018/20977-2. LSJ acknowledges support from Brazilian agencies FAPESP (2019/10923-5) and CNPq (304819/201794). AAC acknowledges support from FAPERJ (grant E26/203.186/2016)  and  CNPq  (grants  304971/2016-2 and 401669/2016-5). EVL acknowledges funding for this work from CNPq grant 169181/2017-0 and CAPES grant 88887.470064/2019-00.
MLB acknowledges FAPESP, grants 2018/09165-6 and 2019/23388-0.
KMD acknowledges support from FAPERJ (grant E-26/203.184/2017), CNPq (grant 312702/2017-5) and the Serrapilheira Institute (grant Serra-1709-17357).  AAC acknowledges support from FAPERJ (grant E26/203.186/2016), CNPq (grants 304971/2016-2 and 401669/2016-5), from the Universidad de Alicante under contract UATALENTO18-02, and from the State Agency for Research of the Spanish MCIU through the "Center of Excellence Severo Ochoa" award to the Instituto de Astrof\'isica de Andaluc\'ia (SEV-2017-0709). ARL acknowledges the financial support from CNPq through the PCI fellowship. The authors would also like to thank all fellow colleagues from the S-PLUS collaboration who have been adding value to this work throughout those years with great suggestions and discussions. In special: Dr Jo\~ao E. Steiner (in memoriam), Dr Marcus V. Costa Duarte, Dr Alberto Molino, Dr Rodrigo C. Thom de Souza, Dr Raimundo Lopes, Dr Maria L. Linhares Dantas, Carlos A. Galarza, Dr Jose L. Nilo Castell\'on, Leonildo J. de Melo de Azevedo, and  Geferson Lucatelli. We would also like to deeply thank Dr Alex Clarke who has carefully read our paper and gave very valuable feedback.

This research made use of Scikit-learn \citep{scikit-learn}; Numpy \citep{numpy}; seaborn \citep{waskom2020seaborn}; matplotlib, a Python library for publication quality graphics \citep{Hunter:2007}; Astropy, a community-developed core Python package for Astronomy \citep{2013A&A...558A..33A}; TOPCAT, an interactive graphical viewer and editor for tabular data \citep{2005ASPC..347...29T}.  

The S-PLUS project, including the T80-South robotic telescope and the S-PLUS scientific survey, was founded as a partnership between the Funda\c{c}\~{a}o de Amparo \`{a} Pesquisa do Estado de S\~{a}o Paulo (FAPESP), the Observat\'{o}rio Nacional (ON), the Federal University of Sergipe (UFS), and the Federal University of Santa Catarina (UFSC), with important financial and practical contributions from other collaborating institutes in Brazil, Chile (Universidad de La Serena), and Spain (Centro de Estudios de F\'{\i}sica del Cosmos de Arag\'{o}n, CEFCA). We further acknowledge financial support from the S\~{a}o Paulo Research Foundation (FAPESP), the Brazilian National Research Council (CNPq), the Coordination for the Improvement of Higher Education Personnel (CAPES), the Carlos Chagas Filho Rio de Janeiro State Research Foundation (FAPERJ), and the Brazilian Innovation Agency (FINEP).

The authors are grateful for the contributions from CTIO staff in helping in the construction, commissioning and maintenance of the T80-South telescope and camera. We are also indebted to Rene Laporte and INPE, as well as Keith Taylor, for their important contributions to the project. 
We also thank CEFCA staff for their help with T80-South, specifically we thank Antonio Mar\'{i}n-Franch for his invaluable contributions in the early phases of the project, David Crist{\'o}bal-Hornillos and his team for their help with the installation of the data reduction package \textsc{jype} version 0.9.9, C\'{e}sar \'{I}\~{n}iguez for providing 2D measurements of the filter transmissions, and all other staff members for their support.

Funding for the Sloan Digital Sky Survey IV has been provided by the Alfred P. Sloan Foundation, the U.S. Department of Energy Office of Science, and the Participating Institutions. SDSS-IV acknowledges support and resources from the Center for High-Performance Computing at the University of Utah. The SDSS web site is www.sdss.org. SDSS-IV is managed by the Astrophysical Research Consortium for the Participating Institutions of the SDSS Collaboration including the Brazilian Participation Group, the Carnegie Institution for Science, Carnegie Mellon University, the Chilean Participation Group, the French Participation Group, Harvard-Smithsonian Center for Astrophysics, Instituto de Astrof\'isica de Canarias, The Johns Hopkins University, Kavli Institute for the Physics and Mathematics of the Universe (IPMU) / University of Tokyo, Lawrence Berkeley National Laboratory, Leibniz Institut f\"ur Astrophysik Potsdam (AIP), Max-Planck-Institut f\"ur Astronomie (MPIA Heidelberg), Max-Planck-Institut f\"ur Astrophysik (MPA Garching), Max-Planck-Institut f\"ur Extraterrestrische Physik (MPE), National Astronomical Observatories of China, New Mexico State University, New York University, University of Notre Dame, Observat\'ario Nacional / MCTI, The Ohio State University, Pennsylvania State University, Shanghai Astronomical Observatory, United Kingdom Participation Group, Universidad Nacional Aut\'onoma de M\'exico, University of Arizona, University of Colorado Boulder, University of Oxford, University of Portsmouth, University of Utah, University of Virginia, University of Washington, University of Wisconsin, Vanderbilt University, and Yale University. This publication makes use of data products from the Wide-field Infrared Survey Explorer, which is a joint project of the University of California, Los Angeles, and the Jet Propulsion Laboratory/California Institute of Technology, and NEOWISE, which is a project of the Jet Propulsion Laboratory/California Institute of Technology. WISE and NEOWISE are funded by the National Aeronautics and Space Administration.

\bibliographystyle{mnras} 
\bibliography{qso}

\newcommand{\noop}[1]{}
\begin{thebibliography}{}
\makeatletter
\relax
\def\mn@urlcharsother{\let\do\@makeother \do\$\do\&\do\#\do\^\do\_\do\%\do\~}
\def\mn@doi{\begingroup\mn@urlcharsother \@ifnextchar [ {\mn@doi@}
  {\mn@doi@[]}}
\def\mn@doi@[#1]#2{\def\@tempa{#1}\ifx\@tempa\@empty \href
  {http://dx.doi.org/#2} {doi:#2}\else \href {http://dx.doi.org/#2} {#1}\fi
  \endgroup}
\def\mn@eprint#1#2{\mn@eprint@#1:#2::\@nil}
\def\mn@eprint@arXiv#1{\href {http://arxiv.org/abs/#1} {{\tt arXiv:#1}}}
\def\mn@eprint@dblp#1{\href {http://dblp.uni-trier.de/rec/bibtex/#1.xml}
  {dblp:#1}}
\def\mn@eprint@#1:#2:#3:#4\@nil{\def\@tempa {#1}\def\@tempb {#2}\def\@tempc
  {#3}\ifx \@tempc \@empty \let \@tempc \@tempb \let \@tempb \@tempa \fi \ifx
  \@tempb \@empty \def\@tempb {arXiv}\fi \@ifundefined
  {mn@eprint@\@tempb}{\@tempb:\@tempc}{\expandafter \expandafter \csname
  mn@eprint@\@tempb\endcsname \expandafter{\@tempc}}}

\bibitem[\protect\citeauthoryear{{Almeida-Fernandes}
  et~al.,}{{Almeida-Fernandes} et~al.}{2021}]{2021arXiv210400020A}
{Almeida-Fernandes} F.,  et~al., 2021, arXiv e-prints, \href
  {https://ui.adsabs.harvard.edu/abs/2021arXiv210400020A} {p. arXiv:2104.00020}

\bibitem[\protect\citeauthoryear{{Astropy Collaboration} et~al.,}{{Astropy
  Collaboration} et~al.}{2013}]{2013A&A...558A..33A}
{Astropy Collaboration} et~al., 2013, \mn@doi [\aap]
  {10.1051/0004-6361/201322068}, 558, A33

\bibitem[\protect\citeauthoryear{Ball, Brunner, Myers  \& Tcheng}{Ball
  et~al.}{2006}]{Ball_2006}
Ball N.~M.,  Brunner R.~J.,  Myers A.~D.,   Tcheng D.,  2006, \mn@doi [The
  Astrophysical Journal] {10.1086/507440}, 650, 497

\bibitem[\protect\citeauthoryear{{Baqui} et~al.,}{{Baqui}
  et~al.}{2021}]{Baqui2021}
{Baqui} P.~O.,  et~al., 2021, \mn@doi [\aap] {10.1051/0004-6361/202038986},
  \href {https://ui.adsabs.harvard.edu/abs/2021A&A...645A..87B} {645, A87}

\bibitem[\protect\citeauthoryear{{Benitez} et~al.,}{{Benitez}
  et~al.}{2014}]{2014arXiv1403.5237B}
{Benitez} N.,  et~al., 2014, arXiv e-prints, \href
  {https://ui.adsabs.harvard.edu/abs/2014arXiv1403.5237B} {p. arXiv:1403.5237}

\bibitem[\protect\citeauthoryear{{Bertin} \& {Arnouts}}{{Bertin} \&
  {Arnouts}}{1996}]{1996A&AS..117..393B}
{Bertin} E.,  {Arnouts} S.,  1996, \mn@doi [\aaps] {10.1051/aas:1996164}, \href
  {https://ui.adsabs.harvard.edu/abs/1996A%26AS..117..393B} {117, 393}

\bibitem[\protect\citeauthoryear{{Bonoli} et~al.,}{{Bonoli}
  et~al.}{2020}]{miniJPAS}
{Bonoli} S.,  et~al., 2020, arXiv e-prints, \href
  {https://ui.adsabs.harvard.edu/abs/2020arXiv200701910B} {p. arXiv:2007.01910}

\bibitem[\protect\citeauthoryear{Bovy et~al.,}{Bovy et~al.}{2012}]{Bovy_2012}
Bovy J.,  et~al., 2012, \mn@doi [The Astrophysical Journal]
  {10.1088/0004-637x/749/1/41}, 749, 41

\bibitem[\protect\citeauthoryear{Breiman}{Breiman}{2001}]{Breiman2001}
Breiman L.,  2001, \mn@doi [Machine Learning] {10.1023/A:1010933404324}, 45, 5

\bibitem[\protect\citeauthoryear{Breiman, Friedman, Olshen  \& Stone}{Breiman
  et~al.}{1984}]{reason:BreFriOlsSto84a}
Breiman L.,  Friedman J.~H.,  Olshen R.~A.,   Stone C.~J.,  1984,
  Classification and Regression Trees.
Wadsworth and Brooks, Monterey, CA

\bibitem[\protect\citeauthoryear{Brescia, Cavuoti  \& Longo}{Brescia
  et~al.}{2015}]{Brescia_2015}
Brescia M.,  Cavuoti S.,   Longo G.,  2015, \mn@doi [\mnras]
  {10.1093/mnras/stv854}, 450, 3893

\bibitem[\protect\citeauthoryear{Burman}{Burman}{1989}]{10.2307/2336116}
Burman P.,  1989, Biometrika, 76, 503

\bibitem[\protect\citeauthoryear{Carrasco et~al.,}{Carrasco
  et~al.}{2015}]{Carrasco_2015}
Carrasco D.,  et~al., 2015, \mn@doi [Astronomy & Astrophysics]
  {10.1051/0004-6361/201525752}, 584, A44

\bibitem[\protect\citeauthoryear{{Cenarro} et~al.,}{{Cenarro}
  et~al.}{2019}]{2019A&A...622A.176C}
{Cenarro} A.~J.,  et~al., 2019, \mn@doi [\aap] {10.1051/0004-6361/201833036},
  \href {https://ui.adsabs.harvard.edu/abs/2019A&A...622A.176C} {622, A176}

\bibitem[\protect\citeauthoryear{{Clarke}, {Scaife}, {Greenhalgh}  \&
  {Griguta}}{{Clarke} et~al.}{2020}]{2019arXiv190910963C}
{Clarke} A.~O.,  {Scaife} A.~M.~M.,  {Greenhalgh} R.,   {Griguta} V.,  2020,
  \mn@doi [\aap] {10.1051/0004-6361/201936770}, \href
  {https://ui.adsabs.harvard.edu/abs/2020A&A...639A..84C} {639, A84}

\bibitem[\protect\citeauthoryear{{Coelho}}{{Coelho}}{2014}]{PaulaCoelho}
{Coelho} P.~R.~T.,  2014, \mn@doi [\mnras] {10.1093/mnras/stu365}, \href
  {http://adsabs.harvard.edu/abs/2014MNRAS.440.1027C} {440, 1027}

\bibitem[\protect\citeauthoryear{Cortes \& Vapnik}{Cortes \&
  Vapnik}{1995}]{Cortes1995}
Cortes C.,  Vapnik V.,  1995, \mn@doi [Machine Learning]
  {10.1023/A:1022627411411}, 20, 273

\bibitem[\protect\citeauthoryear{Costa-Duarte et~al.,}{Costa-Duarte
  et~al.}{2019}]{costaduarte2019splus}
Costa-Duarte M.~V.,  et~al., 2019, submitted

\bibitem[\protect\citeauthoryear{{Cutri} et~al.}{{Cutri}
  et~al.}{2013}]{2013yCat.2328....0C}
{Cutri} R.~M.,  et~al., 2013, VizieR Online Data Catalog, \href
  {https://ui.adsabs.harvard.edu/abs/2013yCat.2328....0C} {p. II/328}

\bibitem[\protect\citeauthoryear{{Dawson} et~al.,}{{Dawson}
  et~al.}{2016}]{2016AJ....151...44D}
{Dawson} K.~S.,  et~al., 2016, \mn@doi [\aj] {10.3847/0004-6256/151/2/44},
  \href {https://ui.adsabs.harvard.edu/abs/2016AJ....151...44D} {151, 44}

\bibitem[\protect\citeauthoryear{{Fluke} \& {Jacobs}}{{Fluke} \&
  {Jacobs}}{2020}]{2020WDMKD..10.1349F}
{Fluke} C.~J.,  {Jacobs} C.,  2020, \mn@doi [WIREs Data Mining and Knowledge
  Discovery] {10.1002/widm.1349}, \href
  {https://ui.adsabs.harvard.edu/abs/2020WDMKD..10.1349F} {10, e1349}

\bibitem[\protect\citeauthoryear{{Gaia Collaboration} et~al.,}{{Gaia
  Collaboration} et~al.}{2018}]{2018A&A...616A...1G}
{Gaia Collaboration} et~al., 2018, \mn@doi [\aap]
  {10.1051/0004-6361/201833051}, \href
  {https://ui.adsabs.harvard.edu/\#abs/2018A&A...616A...1G} {616, A1}

\bibitem[\protect\citeauthoryear{Guo, Qi, Liao, Cao, Lattanzi, Bucciarelli,
  Tang  \& Yan}{Guo et~al.}{2018}]{Guo_2018}
Guo S.,  Qi Z.,  Liao S.,  Cao Z.,  Lattanzi M.~G.,  Bucciarelli B.,  Tang Z.,
   Yan Q.-Z.,  2018, \mn@doi [Astronomy & Astrophysics]
  {10.1051/0004-6361/201833135}, 618, A144

\bibitem[\protect\citeauthoryear{Harris et~al.,}{Harris et~al.}{2020}]{numpy}
Harris C.~R.,  et~al., 2020, \mn@doi [Nature] {10.1038/s41586-020-2649-2}, 585,
  357

\bibitem[\protect\citeauthoryear{{Heintz}, {Fynbo}, {H{\o}g}, {M{\o}ller},
  {Krogager}, {Geier}, {Jakobsson}  \& {Christensen}}{{Heintz}
  et~al.}{2018}]{2018A&A...615L...8H}
{Heintz} K.~E.,  {Fynbo} J.~P.~U.,  {H{\o}g} E.,  {M{\o}ller} P.,  {Krogager}
  J.~K.,  {Geier} S.,  {Jakobsson} P.,   {Christensen} L.,  2018, \mn@doi
  [\aap] {10.1051/0004-6361/201833396}, \href
  {https://ui.adsabs.harvard.edu/\#abs/2018A&A...615L...8H} {615, L8}

\bibitem[\protect\citeauthoryear{{Hern{\'a}n-Caballero}, {Hatziminaoglou},
  {Alonso-Herrero}  \& {Mateos}}{{Hern{\'a}n-Caballero} et~al.}{2016}]{Hernan}
{Hern{\'a}n-Caballero} A.,  {Hatziminaoglou} E.,  {Alonso-Herrero} A.,
  {Mateos} S.,  2016, \mn@doi [\mnras] {10.1093/mnras/stw2107}, \href
  {http://adsabs.harvard.edu/abs/2016MNRAS.463.2064H} {463, 2064}

\bibitem[\protect\citeauthoryear{Hunter}{Hunter}{2007}]{Hunter:2007}
Hunter J.~D.,  2007, Computing In Science \& Engineering, 9, 90

\bibitem[\protect\citeauthoryear{Ivezi\'c et~al.,}{Ivezi\'c
  et~al.}{2019}]{Ivezi__2019}
Ivezi\'c Å.,  et~al., 2019, \mn@doi [The Astrophysical Journal]
  {10.3847/1538-4357/ab042c}, 873, 111

\bibitem[\protect\citeauthoryear{{Jin}, {Zhang}, {Zhang}, {Zhao}, {Wu}  \&
  {Fan}}{{Jin} et~al.}{2019}]{2019MNRAS.485.4539J}
{Jin} X.,  {Zhang} Y.,  {Zhang} J.,  {Zhao} Y.,  {Wu} X.-b.,   {Fan} D.,  2019,
  \mn@doi [\mnras] {10.1093/mnras/stz680}, \href
  {https://ui.adsabs.harvard.edu/abs/2019MNRAS.485.4539J} {485, 4539}

\bibitem[\protect\citeauthoryear{Kirkpatrick, Schlegel, Ross, Myers, Hennawi,
  Sheldon, Schneider  \& Weaver}{Kirkpatrick et~al.}{2011}]{Kirkpatrick_2011}
Kirkpatrick J.~A.,  Schlegel D.~J.,  Ross N.~P.,  Myers A.~D.,  Hennawi J.~F.,
  Sheldon E.~S.,  Schneider D.~P.,   Weaver B.~A.,  2011, \mn@doi [The
  Astrophysical Journal] {10.1088/0004-637x/743/2/125}, 743, 125

\bibitem[\protect\citeauthoryear{{Kron}}{{Kron}}{1980}]{1980ApJS...43..305K}
{Kron} R.~G.,  1980, \mn@doi [\apjs] {10.1086/190669}, \href
  {https://ui.adsabs.harvard.edu/abs/1980ApJS...43..305K} {43, 305}

\bibitem[\protect\citeauthoryear{Kurcz, Bilicki, Solarz, Krupa, Pollo  \&
  Ma\l{}ek}{Kurcz et~al.}{2016}]{Kurcz_2016}
Kurcz A.,  Bilicki M.,  Solarz A.,  Krupa M.,  Pollo A.,   Ma\l{}ek K.,  2016,
  \mn@doi [Astronomy & Astrophysics] {10.1051/0004-6361/201628142}, 592, A25

\bibitem[\protect\citeauthoryear{L\'opez-Sanjuan et~al.,}{L\'opez-Sanjuan
  et~al.}{2019}]{L_pez_Sanjuan_2019}
L\'opez-Sanjuan C.,  et~al., 2019, \mn@doi [Astronomy & Astrophysics]
  {10.1051/0004-6361/201732480}, 622, A177

\bibitem[\protect\citeauthoryear{Martinazzo, Espadoto  \& Hirata}{Martinazzo
  et~al.}{2020a}]{martinazzo2}
Martinazzo A.,  Espadoto M.,   Hirata N.,  2020a, in International Conference
  on Pattern Recognition (ICPR), in press.

\bibitem[\protect\citeauthoryear{Martinazzo, Espadoto  \& Hirata}{Martinazzo
  et~al.}{2020b}]{martinazzo}
Martinazzo A.,  Espadoto M.,   Hirata N.,  2020b, in In Proceedings of the 15th
  International Joint Conference on Computer Vision, Imaging and Computer
  Graphics Theory and Applications, INSTICC, SciTePress. pp 87--95

\bibitem[\protect\citeauthoryear{{Mendes de Oliveira} et~al.,}{{Mendes de
  Oliveira} et~al.}{2019}]{2019arXiv190701567M}
{Mendes de Oliveira} C.,  et~al., 2019, \mn@doi [\mnras]
  {10.1093/mnras/stz1985}, \href
  {https://ui.adsabs.harvard.edu/abs/2019MNRAS.489..241M} {489, 241}

\bibitem[\protect\citeauthoryear{Moore, Pimbblet  \& Drinkwater}{Moore
  et~al.}{2006}]{Moore_2006}
Moore J.~A.,  Pimbblet K.~A.,   Drinkwater M.~J.,  2006, \mn@doi [Publications
  of the Astronomical Society of Australia] {10.1071/as06010}, 23, 135

\bibitem[\protect\citeauthoryear{Oshiro, Perez  \& Baranauskas}{Oshiro
  et~al.}{2012}]{oshiro_2012}
Oshiro T.,  Perez P.,   Baranauskas J.,  2012. ,
  \mn@doi{10.1007/978-3-642-31537-4_13}

\bibitem[\protect\citeauthoryear{{P{\^a}ris} et~al.,}{{P{\^a}ris}
  et~al.}{2017}]{2017A&A...597A..79P}
{P{\^a}ris} I.,  et~al., 2017, \mn@doi [\aap] {10.1051/0004-6361/201527999},
  \href {https://ui.adsabs.harvard.edu/abs/2017A%26A...597A..79P} {597, A79}

\bibitem[\protect\citeauthoryear{P\^aris et~al.,}{P\^aris
  et~al.}{2018}]{Paris2018}
P\^aris I.,  et~al., 2018, \mn@doi [Astronomy & Astrophysics]
  {10.1051/0004-6361/201732445}, 613, A51

\bibitem[\protect\citeauthoryear{Pedregosa et~al.,}{Pedregosa
  et~al.}{2011}]{scikit-learn}
Pedregosa F.,  et~al., 2011, Journal of Machine Learning Research, 12, 2825

\bibitem[\protect\citeauthoryear{{Peng}, {Zhang}, {Zhao}  \& {Wu}}{{Peng}
  et~al.}{2012}]{2012MNRAS.425.2599P}
{Peng} N.,  {Zhang} Y.,  {Zhao} Y.,   {Wu} X.-b.,  2012, \mn@doi [\mnras]
  {10.1111/j.1365-2966.2012.21191.x}, \href
  {https://ui.adsabs.harvard.edu/abs/2012MNRAS.425.2599P} {425, 2599}

\bibitem[\protect\citeauthoryear{{Peters} et~al.,}{{Peters}
  et~al.}{2015}]{2015ApJ...811...95P}
{Peters} C.~M.,  et~al., 2015, \mn@doi [\apj] {10.1088/0004-637X/811/2/95},
  \href {https://ui.adsabs.harvard.edu/abs/2015ApJ...811...95P} {811, 95}

\bibitem[\protect\citeauthoryear{Pimbblet, Smail, Edge, Couch, O'Hely  \&
  Zabludoff}{Pimbblet et~al.}{2001}]{Pimbblet_2001}
Pimbblet K.~A.,  Smail I.,  Edge A.~C.,  Couch W.~J.,  O'Hely E.,   Zabludoff
  A.~I.,  2001, \mn@doi [\mnras] {10.1046/j.1365-8711.2001.04759.x}, 327, 588

\bibitem[\protect\citeauthoryear{{Richards} et~al.,}{{Richards}
  et~al.}{2002}]{Richards2002}
{Richards} G.~T.,  et~al., 2002, \mn@doi [\aj] {10.1086/340187}, \href
  {https://ui.adsabs.harvard.edu/abs/2002AJ....123.2945R} {123, 2945}

\bibitem[\protect\citeauthoryear{{Richards} et~al.,}{{Richards}
  et~al.}{2006}]{2006AJ....131.2766R}
{Richards} G.~T.,  et~al., 2006, \mn@doi [\aj] {10.1086/503559}, \href
  {https://ui.adsabs.harvard.edu/abs/2006AJ....131.2766R} {131, 2766}

\bibitem[\protect\citeauthoryear{{Ross} et~al.,}{{Ross}
  et~al.}{2013}]{2013ApJ...773...14R}
{Ross} N.~P.,  et~al., 2013, \mn@doi [\apj] {10.1088/0004-637X/773/1/14}, \href
  {https://ui.adsabs.harvard.edu/abs/2013ApJ...773...14R} {773, 14}

\bibitem[\protect\citeauthoryear{{Schindler}, {Fan}, {McGreer}, {Yang}, {Wu},
  {Jiang}  \& {Green}}{{Schindler} et~al.}{2017}]{2017ApJ...851...13S}
{Schindler} J.-T.,  {Fan} X.,  {McGreer} I.~D.,  {Yang} Q.,  {Wu} J.,  {Jiang}
  L.,   {Green} R.,  2017, \mn@doi [\apj] {10.3847/1538-4357/aa9929}, \href
  {https://ui.adsabs.harvard.edu/\#abs/2017ApJ...851...13S} {851, 13}

\bibitem[\protect\citeauthoryear{{Schneider} et~al.,}{{Schneider}
  et~al.}{2010}]{2010AJ....139.2360S}
{Schneider} D.~P.,  et~al., 2010, \mn@doi [\aj] {10.1088/0004-6256/139/6/2360},
  \href {https://ui.adsabs.harvard.edu/abs/2010AJ....139.2360S} {139, 2360}

\bibitem[\protect\citeauthoryear{Scholkopf \& Smola}{Scholkopf \&
  Smola}{2001}]{10.5555/559923}
Scholkopf B.,  Smola A.~J.,  2001, Learning with Kernels: Support Vector
  Machines, Regularization, Optimization, and Beyond.
MIT Press, Cambridge, MA, USA

\bibitem[\protect\citeauthoryear{{Stoughton} et~al.,}{{Stoughton}
  et~al.}{2002}]{2002AJ....123..485S}
{Stoughton} C.,  et~al., 2002, \mn@doi [\aj] {10.1086/324741}, \href
  {https://ui.adsabs.harvard.edu/abs/2002AJ....123..485S} {123, 485}

\bibitem[\protect\citeauthoryear{Strateva et~al.,}{Strateva
  et~al.}{2001}]{Strateva_2001}
Strateva I.,  et~al., 2001, \mn@doi [The Astronomical Journal]
  {10.1086/323301}, 122, 1861

\bibitem[\protect\citeauthoryear{{Taylor}}{{Taylor}}{2005}]{2005ASPC..347...29T}
{Taylor} M.~B.,  2005, in {Shopbell} P.,  {Britton} M.,   {Ebert} R.,  eds,
  Astronomical Society of the Pacific Conference Series Vol. 347, Astronomical
  Data Analysis Software and Systems XIV. p.~29

\bibitem[\protect\citeauthoryear{Walmsley et~al.,}{Walmsley
  et~al.}{2021}]{walmsley2021galaxy}
Walmsley M.,  et~al., 2021, Galaxy Zoo DECaLS: Detailed Visual Morphology
  Measurements from Volunteers and Deep Learning for 314,000 Galaxies
  (\mn@eprint {arXiv} {2102.08414})

\bibitem[\protect\citeauthoryear{Waskom \& the seaborn~development team}{Waskom
  \& the seaborn~development team}{2020}]{waskom2020seaborn}
Waskom M.,  the seaborn~development team 2020, mwaskom/seaborn,
  \mn@doi{10.5281/zenodo.592845}

\bibitem[\protect\citeauthoryear{Wilcoxon}{Wilcoxon}{1945}]{Wilcoxon}
Wilcoxon F.,  1945, Biometrics Bulletin, 1, 80

\bibitem[\protect\citeauthoryear{{Wright} et~al.,}{{Wright}
  et~al.}{2010}]{2010AJ....140.1868W}
{Wright} E.~L.,  et~al., 2010, \mn@doi [\aj] {10.1088/0004-6256/140/6/1868},
  \href {https://ui.adsabs.harvard.edu/abs/2010AJ....140.1868W} {140, 1868}

\bibitem[\protect\citeauthoryear{{Wu} \& {Jia}}{{Wu} \& {Jia}}{2010}]{Wu_2010}
{Wu} X.-B.,  {Jia} Z.,  2010, \mn@doi [\mnras]
  {10.1111/j.1365-2966.2010.16807.x}, \href
  {https://ui.adsabs.harvard.edu/abs/2010MNRAS.406.1583W} {406, 1583}

\bibitem[\protect\citeauthoryear{{Wu}, {Hao}, {Jia}, {Zhang}  \& {Peng}}{{Wu}
  et~al.}{2012}]{2012AJ....144...49W}
{Wu} X.-B.,  {Hao} G.,  {Jia} Z.,  {Zhang} Y.,   {Peng} N.,  2012, \mn@doi
  [\aj] {10.1088/0004-6256/144/2/49}, \href
  {https://ui.adsabs.harvard.edu/\#abs/2012AJ....144...49W} {144, 49}

\bibitem[\protect\citeauthoryear{{Yang} et~al.,}{{Yang}
  et~al.}{2017a}]{2017AJ....154..269Y}
{Yang} Q.,  et~al., 2017a, \mn@doi [\aj] {10.3847/1538-3881/aa943c}, \href
  {https://ui.adsabs.harvard.edu/\#abs/2017AJ....154..269Y} {154, 269}

\bibitem[\protect\citeauthoryear{Yang et~al.,}{Yang et~al.}{2017b}]{Yang_2017}
Yang Q.,  et~al., 2017b, \mn@doi [The Astronomical Journal]
  {10.3847/1538-3881/aa943c}, 154, 269

\bibitem[\protect\citeauthoryear{van~der Maaten \& Hinton}{van~der Maaten \&
  Hinton}{2008}]{vanDerMaaten2008}
van~der Maaten L.,  Hinton G.,  2008, Journal of Machine Learning Research, 9,
  2579

\makeatother
\end{thebibliography}

\appendix
\clearpage
\section{Affiliations}

$^{6}$ INAF - Osservatorio Astronomico di Padova, Vicolo Osservatorio 5, 35122 Padova, Italy \\
$^{7}$ Observat\'orio do Valongo, Universidade Federal do Rio de Janeiro, Ladeira Pedro Ant\^onio 43, Sa\'ude, Rio de Janeiro, RJ, Brasil, CEP 20080-090 \\
$^{8}$
Institute for Astronomy, Astrophysics, Space Applications and Remote Sensing, National Observatory of Athens, Penteli GR 15236, Greece \\
$^{9}$ Instituto de Astrof\'isica de Andaluc\'ia, CSIC, Apt 3004, E18080 Granada, Spain\\
$^{10}$ IUFACyT, Universidad de Alicante, San Vicent del Raspeig, E03080, Alicante, Spain\\
$^{11}$ GMTO Corporation 465 N. Halstead Street, Suite 250 Pasadena, CA 91107 \\
$^{12}$
Departamento de F\'isica, Universidade Federal de Santa Catarina, Florian\'opolis, SC, 88040-900, Brazil \\
$^{13}$
NOAO, P.O. Box 26732, Tucson, AZ 85726 \\

\section{Statistical Tests}
\label{sec:stat}

 {We performed some statistical tests to be able to conclude which model had the best performance compared to another with a given significance level, as discussed in Section \ref{sec:comparison}. We used the one-tailed Wilcoxon signed-rank test \citep{Wilcoxon}, which is a non-parametric method for testing two hypothesis: 
}
\begin{itemize}
    \item $H_0$: the median of the differences in any considered metric is negative,
    \item $H_1$: the median of the differences in any considered metric is positive
\end{itemize}

 {As we are using five folds for the cross validation, we have five independent measurements for each metric. For each classifier/metric, these measurements are ordered from highest to lowest and paired. The test statistics W is given by:
}
\begin{equation}
    W = \sum_{i=1}^{N_r}[sgn(x_{2,i}-x_{1,i}) \times R_i],
\end{equation}

 {where sgn refers to the sign function, N$_r$ is the total number of pairs (excluding pairs that $|x_{2,i} - x_{1,i}| = 0$), R$_i$ is the rank for each pair, and $i = \{1,...,5\}$.
To calculate the statistics W, we used the \texttt{wilcoxon} module from \texttt{scipy.stats} in Python, with \texttt{mode} = "exact" and \texttt{alternative} = "greater". We always put in the parameter \texttt{x} of \texttt{wilcoxon} the set of measurements with highest average between the two. Thus, if the null hypothesis H$_0$ is rejected, there is statistical evidence to accept that the corresponding classifier with highest average in a specific metric had performed better than the other. The null hypothesis H$_0$ is rejected if the p-value is below 10\%, i.e., considering a confidence level of 90\%.}

 {We show in Table \ref{tab:stat_W} the statistics W and the corresponding p-value for each of the comparisons we have done in Sec. \ref{sec:comparison}}

\section{Model settings}
\label{sec:hyper}

 {In this section we show the hyper-parameters we used for t-SNE, Random Forest, and Support Vector Machines with \texttt{sklearn} (version 0.24.1) package in Python. }

\begin{table}
\caption{ {Hyperparameters of the t-SNE implementation.}}
\label{tab:hyper_tsne}
\begin{tabular}{ll}
\toprule
Hyperparameter             & Value \\
\midrule
n\_components              &   2    \\
perplexity                 &   30    \\
early\_exaggeration        &  12     \\
learning\_rate             &   200    \\
n\_iter                    &  1000     \\
n\_iter\_without\_progress &    300   \\
min\_grad\_norm            &  10$^{-7}$     \\
metric                     &  euclidian     \\
init                       &random   \\
verbose                    &  0     \\
random\_state              &  2     \\
method                     &   barnes\_hut    \\
angle                      &   0.5    \\
n\_jobs                    &  None  \\
square\_distances          & legacy \\
\bottomrule
\end{tabular}
\end{table}

\begin{table}
\caption{ {Hyperparameters of the Random Forest implementation.}}
\label{tab:hyper_rf}
\begin{tabular}{ll}
\toprule
Hyperparameter             & Value \\
\midrule
n\_estimators             &  100     \\
criterion                &   gini    \\
max\_depth        &    None   \\
min\_samples\_split             &  2     \\
min\_samples\_leaf                 &  1     \\
min\_weigth\_fraction\_leaf  &  0     \\
max\_features            &  auto     \\
max\_leaf\_nodes                     &    None   \\
min\_impurity\_decrease                   &    0   \\
min\_impurity\_split                  &   None    \\
bootstrap              &   False    \\
oob\_score                    &  False     \\
n\_jobs                      &   None    \\
random\_state                    &  2  \\
warm\_start                    &  False  \\
class\_weight                   &  None  \\
ccp\_alpha                   &  0  \\
max\_samples                   &  None  \\
\bottomrule
\end{tabular}
\end{table}

\begin{table}
\caption{ {Hyperparameters of the Support-Vector Machine implementation.}}
\label{tab:hyper_svm}
\begin{tabular}{ll}
\toprule
Hyperparameter             & Value \\
\midrule
C                         & 1      \\
kernel                    & rbf   \\
degree                    &  3     \\
gamma                     &  scale     \\
shrinking                 &  True     \\
tol                       &   10$^{-3}$    \\
cache\_size               &   200    \\
class\_weight             &   None    \\
max\_iter                 &  -1     \\
decision\_function\_shape &   ovo    \\
break\_ties               &   False    \\
random\_state             &   2       \\
\bottomrule
\end{tabular}
\end{table}

\section{Missing-band values analyses}
\label{sec:mb}
 {In this section we describe the analyses we made to investigate any bias effect due to missing-band values. For that, we separate our entire spectroscopic sample into two complementary datasets containing objects that:}

\begin{enumerate}
    \item have information in all 12 S-PLUS bands
    \item have missing value in at least one S-PLUS band
\end{enumerate}

\begin{figure}
    \includegraphics[width=0.45\textwidth]{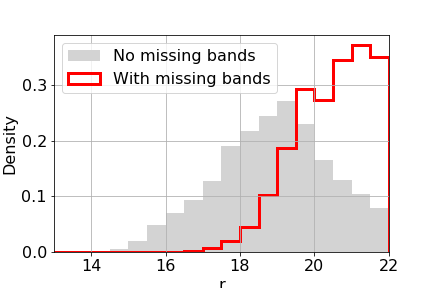}
  \caption{Distribution of magnitude in r band for the two subsets of our spectroscopic sample: one containing objects that have information in all 12 S-PLUS bands (grey), and another containing objects that have at least one missing-band value (red).}
    \label{fig:mb_dist}
\end{figure}

 {The magnitude distribution from each of these subsets are shown in Fig. \ref{fig:mb_dist}. By using subset (i), we are able to investigate how missing-band values can possibly affect the performance of a random forest modelling. We use subset (ii) to empirically calculate the proportion of bands with missing values for each class. This proportion is calculated considering each combination of bands, totaled 12282 cases. We randomly sample objects from subset (i) following these empirical proportions. Then we replace their observed magnitudes in the corresponding bands to missing values (i.e. 99). We run 4-fold cross validation for which we want to compare the performances of a random forest algorithm trained either with subset (i) or with the same subset but with imputed missing values. If there is any bias effect due to missing-band values, we should see a significant decrease in performance for the latter compared to the former. The cross-validation results are shown in Table \ref{tab:cv_bias}, for which we do not see any great difference for the performances between the two datasets. 
}


\begin{table*}
\centering
\caption{ {Cross-validation results for a random forest algorithm. Two different dataset were used within each considered model features. The "original" dataset refers to the subset of our spectroscopic sample for which all sources have information in the twelve S-PLUS bands. The "original with simulations" refers to the same dataset but with imputed missing values in some bands. The performance are described in terms of precision, recall and F-score (F) for each class. The overall performance is shown in terms of macro-averaged F-score ($\bar{\text{F}}$). } }
\label{tab:cv_bias}
\begin{tabular}{@{}lcccccc@{}}
\arrayrulecolor{gray}
\toprule
Model features & Train/validation data &  $\bar{\text{F}}$ (\%) & CLASS  & Precision (\%) & Recall (\%) & F (\%) \\ 
 
\midrule

\multirow{6}{*}{\begin{tabular}[c]{@{}l@{}}12 S-PLUS bands + \\ 4 morphological features\end{tabular}} & \multirow{3}{*}{original}  & \multirow{3}{*}{ 93.09 $\pm$ 0.2 } & QSO & 85.57 $\pm$ 0.78 & 89.68 $\pm$ 0.53 & 87.57 $\pm$ 0.46 \\
&&& STAR & 96.21 $\pm$ 0.3& 94.31 $\pm$ 0.32 & 95.25 $\pm$ 0.21 \\
&&& GAL & 96.42 $\pm$ 0.19 & 96.48 $\pm$ 0.08 & 96.45 $\pm$ 0.11 \\

\cmidrule(l){2-7} 
& \multirow{3}{*}{original with simulations} & \multirow{3}{*}{92.92 $\pm$ 0.19 } & QSO & 85.53 $\pm$ 0.62 & 89.03 $\pm$ 0.63 &87.24 $\pm$ 0.38 \\
&&& STAR & 95.87 $\pm$ 0.38 & 94.12 $\pm$ 0.34 & 94.99 $\pm$ 0.21 \\
&&& GAL & 96.42 $\pm$ 0.24 &96.62 $\pm$ 0.07 &96.52 $\pm$ 0.15 \\

\midrule

\multirow{6}{*}{\begin{tabular}[c]{@{}l@{}}12 S-PLUS bands + 2 WISE bands \\ + 4 morphological features\end{tabular}}   & \multirow{3}{*}{original} & \multirow{3}{*}{ 97.95 $\pm$ 0.15 }  & QSO & 96.12 $\pm$ 0.58 & 97.38 $\pm$ 0.15 & 96.75 $\pm$ 0.3 \\
&&& STAR & 99.41 $\pm$ 0.1 & 98.53 $\pm$ 0.11 & 98.97 $\pm$ 0.07 \\
&&& GAL & 98.01 $\pm$ 0.12& 98.23 $\pm$ 0.22 & 98.12 $\pm$ 0.14 \\

\cmidrule(l){2-7}

  &\multirow{3}{*}{original with simulations}  &   \multirow{3}{*}{ 98.01 $\pm$ 0.2 } & QSO & 96.2 $\pm$ 0.61 & 97.54 $\pm$ 0.34& 96.87 $\pm$ 0.4 \\
&&& STAR & 99.39 $\pm$ 0.06 & 98.55 $\pm$ 0.11 & 98.97 $\pm$ 0.05 \\
&&& GAL & 98.11 $\pm$ 0.18 & 98.26 $\pm$ 0.2 &98.18 $\pm$ 0.17 \\

\bottomrule
\end{tabular}
\end{table*}

\begin{table*}
\label{tab:test_bias}
\caption{{Performance on the subset of our spectroscopic sample that have missing-band value in at least one S-PLUS band for the model trained with the subset with no missing-band values. In parenthesis we show the performances for the model trained with the same subset but with simulated missing-band values.}}
\begin{tabular}{@{}lccccc@{}}
\toprule
Model Features                                                       & \multicolumn{1}{c}{$\bar{\text{F}}$ (\%)} & CLASS & \multicolumn{1}{c}{Precision (\%)} & \multicolumn{1}{c}{Recall (\%)} & \multicolumn{1}{c}{F (\%)} \\ \midrule
\multirow{3}{*}{\begin{tabular}[c]{@{}c@{}}12 S-PLUS bands + \\ 4 morphological features\end{tabular}}                & \multirow{3}{*}{85.31 (85.19)}               & QSO   & 64.70 (65.76)                         & 72.71 (71.00)                      & 68.47 (68.28)                 \\
                                                                                                                      &                                           & STAR  & 95.13 (95.82)                         & 87.88 (87.14)                      & 91.36 (91.27)                 \\
                                                                                                                      &                                           & GAL   & 95.00 (94.41)                         & 97.19 (97.66)                       & 96.08 (96.01)                 \\
                                                                                                                      \midrule
\multirow{3}{*}{\begin{tabular}[c]{@{}l@{}}12 S-PLUS bands + 2 WISE bands \\ + 4 morphological features\end{tabular}} & \multirow{3}{*}{93.71 (93.39)}               & QSO   & 84.58 (83.64)                         & 86.44 (86.11)                       & 85.50 (84.86)                 \\
                                                                                                                      &                                           & STAR  & 99.44 (99.53)                         & 95.74 (95.30)                      & 97.55 (97.37)                 \\
                                                                                                                      &                                           & GAL   & 97.43 (97.22)                         & 98.76 (98.69)                      & 98.09 (97.95)                 \\ \bottomrule
\end{tabular}
\end{table*}

\section{Catalogue description}
\label{sec:appendix_catalogue}

\begin{table*}
\centering
\caption{ {Description of the columns in our catalogue.}}
\label{tab:columns}
\begin{threeparttable}
\begin{tabular}{@{}llll@{}}
\toprule
Column & Name             & Format & Description \\
\midrule
 1      & RA               & float64      &    Right ascension in degrees (J2000)         \\
2       & DEC              & float64       &    Declination in degrees (J2000)         \\
3       & FWHM\_n          &  float64      &   Full width at half maximum normalized per field          \\
4       & A                & float64       &   Major semi-axis          \\
5       & B                & float64       &   Minor semi-axis          \\
6       & KRON\_RADIUS     & float64       &   Kron Radius           \\
7       & u\_iso           & float64       &   Magnitude in AB system for $u$ band          \\
8       & J0378\_iso        & float64       &   Magnitude in AB system for J0378 band          \\
9       & J0395\_iso        & float64       &  Magnitude in AB system for J0395 band           \\
10       & J0410\_iso        & float64       & Magnitude in AB system for J0410 band            \\
11       & J0430\_iso        & float64       &  Magnitude in AB system for J0430 band           \\
12       & g\_iso           & float64       &  Magnitude in AB system for $g$ band           \\
13       & J0515\_iso        & float64       &  Magnitude in AB system for J0515 band           \\
14       & r\_iso           & float64       &  Magnitude in AB system for $r$ band           \\
15       & J0660\_iso        & float64       &  Magnitude in AB system for J0660 band           \\
16       & i\_iso           & float64       &  Magnitude in AB system for $i$ band           \\
17       & J0861\_iso        & float64       &  Magnitude in AB system for J0861 band           \\
18       & z\_iso           & float64       &  Magnitude in AB system for $z$ band           \\
19       & w1mpro           &  float64      & Magnitude in Vega system for W1 band taken from ALLWISE catalogue           \\
20       & w2mpro           & float64       &  Magnitude in Vega system for W2 band taken from ALLWISE catalogue          \\
21       & w1snr       & float64       &  W1 signal-to-noise ratio taken from ALLWISE catalogue            \\
22       & w2snr         & float64       & W2 signal-to-noise ratio taken from ALLWISE catalogue            \\
23       & w1sigmpro        & float64       & W1 uncertainty in mag units taken from ALLWISE catalogue            \\
24       & w2sigmpro        & float64       &  W2 uncertainty in mag units taken from ALLWISE catalogue             \\
25       & CLASS\_SPEC      & object       &  Spectroscopic classification taken from SDSS DR14Q/DR16           \\
26       & spec\_flag\tnote{*}       & int64       &  0, if the object was in our traning sample and 1, otherwise.      \\
27       & CLASS            & float64       &   Object classification from RF\_12S+4M$\oplus$ for sources with no WISE counterpart         \\
28       & PROB\_QSO        & float64       &  Probability of the source being a quasar from RF\_12S+4M$\oplus$           \\
29       & PROB\_STAR       & float64       & Probability of the source being a star from RF\_12S+4M$\oplus$             \\
30       & PROB\_GAL        & float64       &  Probability of the source being a galaxy from RF\_12S+4M$\oplus$            \\
31       & CLASS\_WISE      & float64       &   Object classification from RF\_12S+2W+4M for sources with WISE counterpart         \\
 32      & PROB\_QSO\_WISE  & float64       &   Probability of the source being a quasar from RF\_12S+2W+4M           \\
33       & PROB\_STAR\_WISE & float64       &  Probability of the source being a star from RF\_12S+2W+4M           \\
34      & PROB\_GAL\_WISE  & float64       & Probability of the source being a galaxy from RF\_12S+2W+4M   \\
       \bottomrule
\end{tabular}

\begin{tablenotes}[para,flushleft]
\item[*] Not all training objects are in this catalogue due to our selection criteria. 
\end{tablenotes}
\end{threeparttable}
\end{table*}

\begin{figure*}
\subfloat[Distribution of FWHM.]{\includegraphics[width=1\textwidth]{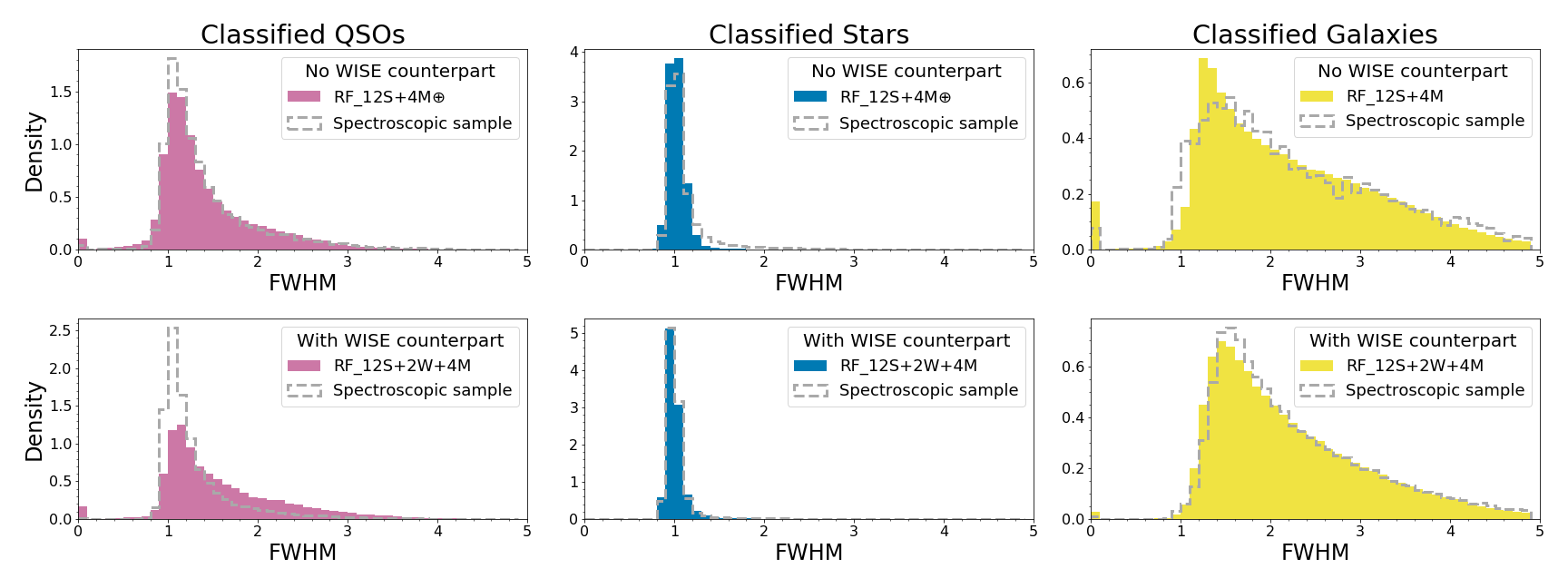}}\\
\subfloat[Distribution of magnitude in $r$ band.]{\includegraphics[width=1\textwidth]{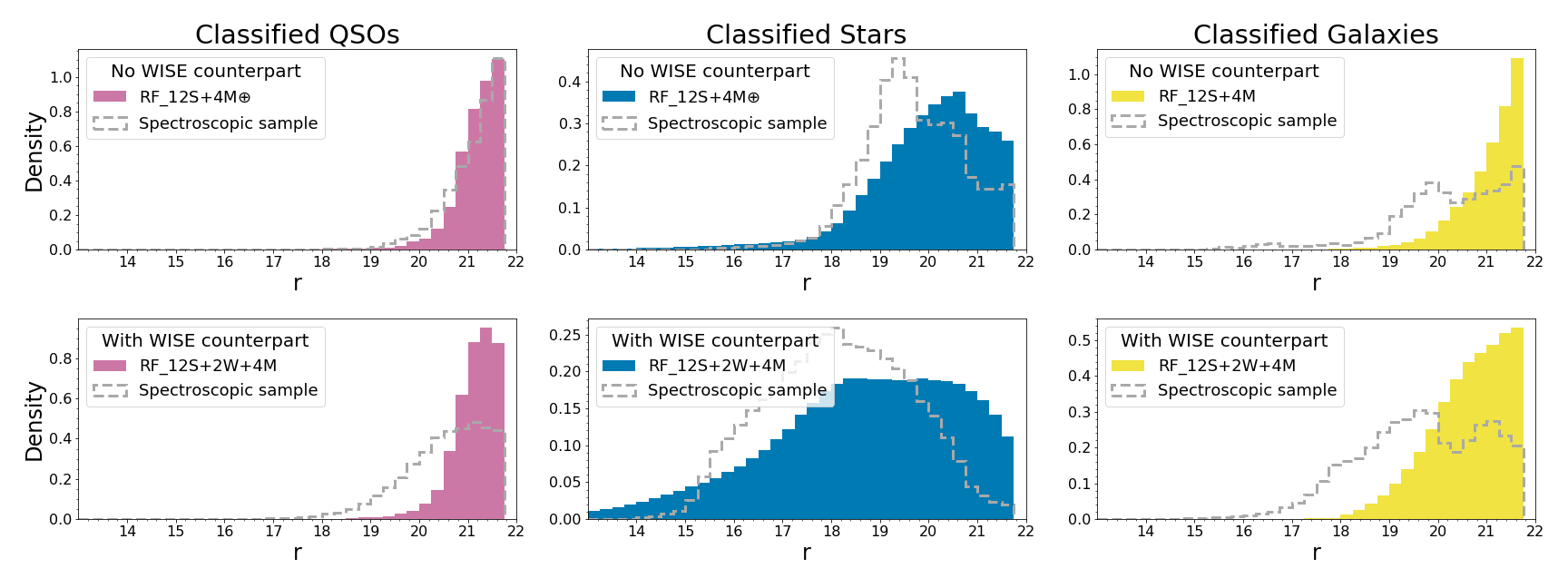}} \\
    \caption{Distribution  of (a) FWHM and (b) magnitude in $r$ band for the sources we classify as quasars (pink), stars (blue), or galaxies (yellow). Top (bottom) panels show objects without (with) WISE counterpart that were classified by RF\_12S+4M$\oplus$ (RF\_12S+2W+4M). We also show in gray the distribution for the spectroscopically confirmed objects. }
    \label{fig:dist_feat_classification}

\end{figure*}

\begin{table}
\caption{P-values for the comparisons from Section \ref{sec:comparison}. P-values below 0.1 indicate evidence that Classifier 1 performs statistically better than Classifier 2.}
\label{tab:stat_W}
\begin{tabular}{@{}lllcc@{}}
\toprule
Classifier 1                & Classifier 2                 & Metric             & W  & p-value \\ \midrule
RF\_5S                      & SVM\_5S                      & All                & 15 & 0.03    \\
\midrule
RF\_12S                     & SVM\_12S                     & All                & 15 & 0.03    \\
\midrule
RF\_5S+2W                   & SVM\_5S+2W                   & All                & 15 & 0.03    \\
\midrule
\multirow{2}{*}{RF\_12S+2W} & \multirow{2}{*}{SVM\_12S+2W} & P\_QSO & 8 & 0.5   \\
                            &                              & Others             & 15  & 0.03     \\
                            \midrule
RF\_12S                     & RF\_5S                       & All                & 15 & 0.03    \\
\midrule
\multirow{6}{*}{RF\_12S+2W}                  & \multirow{6}{*}{RF\_5S+2W}                    & P\_QSO             & 14 & 0.06    \\
                            &                              & R\_QSO             & 1  & 0.97    \\
                            &                              & F\_QSO             & 12 & 0.16    \\
                            &                              & R\_STAR            & 13 & 0.09    \\
                            &                              & P\_GAL             & 11 & 0.22    \\
                            && Others & 15 & 0.03 \\

                            \midrule
RF\_5S+2W                   & RF\_5S                       & All                & 15 & 0.03    \\
\midrule
RF\_12S+2W                  & RF\_12S                      & All                & 15 & 0.03    \\ 
\midrule
\multirow{4}{*}{RF\_12S+2W+4M }                  & \multirow{4}{*}{RF\_12S+2W}                    & P\_QSO             & 10 & 0.31    \\
                            &                              & R\_QSO             & 14  & 0.06    \\
                            &                              & F\_QSO             & 14 & 0.06    \\
                            &                              & Others             & 15 & 0.03    \\
                            \midrule
RF\_12S+4M                  & RF\_12S                      & All                & 15 & 0.03    \\ 
\midrule
RF\_12S+2W+4M                  & RF\_12S+4M                     & All                & 15 & 0.03    \\ 

\bottomrule
\end{tabular}
\end{table}

\bsp	
\label{lastpage}
\end{document}